\documentclass[aps,prl,reprint,superscriptaddress]{revtex4-2}

\usepackage[T1]{fontenc}
\usepackage[utf8]{inputenc}
\usepackage{lmodern}
\usepackage{microtype}
\usepackage{amsmath,amssymb,mathtools,bm}
\usepackage{graphicx}
\usepackage{xcolor}

\definecolor{darkblue}{rgb}{0,0,0.6}
\definecolor{darkred}{rgb}{0.6,0,0}
\usepackage[colorlinks=true, urlcolor=darkblue, anchorcolor=blue, citecolor=darkblue,filecolor=blue,linkcolor=darkred,menucolor=blue]{hyperref}

\begin{document}

\title{Universal fluctuations of first discoveries in competitive exploration}

\author{Arthur Plaud}
\affiliation{Sorbonne Universit\'e, CNRS, Laboratoire de Physique Th\'eorique de la Mati\`ere Condens\'ee (LPTMC), 4 Place Jussieu, 75005 Paris, France}

\author{P.L. Krapivsky}
\affiliation{Department of Physics, Boston University, Boston, Massachusetts 02215, USA}

\author{S. Redner}
\affiliation{Santa Fe Institute, Santa Fe, New Mexico 87501, USA}

\author{Olivier B\'enichou}
\affiliation{Sorbonne Universit\'e, CNRS, Laboratoire de Physique Th\'eorique de la Mati\`ere Condens\'ee (LPTMC), 4 Place Jussieu, 75005 Paris, France}

\begin{abstract}

Random exploration is usually quantified by how fast new space is found, from
 the range of a single walker to the territory collectively covered by many
 walkers. In competitive exploration, first arrival secures an exclusive resource, as
when foragers compete for food items or agents capture distributed targets. It
is then no longer enough to know which sites have been discovered: one must
determine, for each discovered site, which searcher reached it first. We introduce the discovery
 share \(X_n\), the fraction of the first \(n\) collective discoveries secured by
 a tagged searcher. For two identical competitors, exchange symmetry fixes
 \(\langle X_n\rangle=1/2\), but the central question is whether this equal
 split emerges in each long exploration history or only on average, \emph{i.e.} whether early
 competitive advantages are erased or persist. Here we show that the answer is
 controlled by the spectral dimension \(d_s\), defined by the large-time decay of the probability that a single searcher is at
 its starting point after \(t\) steps, \(p_0(t)\sim t^{-d_s/2}\).
 Across ordinary diffusion, long-range superdiffusion and subdiffusion induced
 by crowding or memory, \(d_s\) separates persistent randomness in
 recurrent exploration \((d_s<2)\), anomalously slow non-Gaussian concentration
 for \(2\le d_s<3\), and Gaussian concentration, logarithmically corrected at
 \(d_s=3\), for \(d_s\ge3\). For \(d_s\ge2\), we derive exact asymptotic
 variances, including prefactors, and the discovery scale on which competitive
 imbalances are erased. Two-point correlations of first-discovery labels identify the memory mechanism
behind these regimes.
 The same phase structure persists under changes in geometry, competitor
 heterogeneity, number of competitors and memory, revealing a general fluctuation
 theory of first-arrival inequalities.

\end{abstract}

\maketitle

At the single-walker level, random exploration is classically framed as a
problem of search efficiency: finding a target~\cite{Redner2001,SlutskyMirny2004,Condamin2007,Guerin2016},
growing explored ranges~\cite{Hughes1995,Regnier2023}, characterizing visited
sets~\cite{wijland_statistical_1997,mariz_statistics_2001}, or exhausting a
domain~\cite{Dembo2004,Chupeau2015}.
%At the single-walker level, random exploration is classically framed as a
%problem of search efficiency: finding a target~\cite{Redner2001,SlutskyMirny2004,Condamin2007,Guerin2016},
%growing an explored range~\cite{wijland_statistical_1997,mariz_statistics_2001,Regnier2023}, or exhausting a
%domain~\cite{Chupeau2015,Dembo2004}. 
With several walkers, this viewpoint
usually becomes a question of collective coverage: how much territory the
population discovers~\cite{DonskerVaradhan1975,Larralde1992}, or how long it
takes to cover space~\cite{Chupeau2015}. Yet collective coverage is
silent about how discoveries are allocated when first arrival removes the
resource from further competition. This form of exclusivity is widespread: it
arises whenever a resource is consumed, claimed or captured by the first
searcher to reach it. A food item reached by one forager is no longer available
to the others, and a target captured by one
searcher can no longer be claimed by another.
In such settings, exploration is not merely collective but competitive.

The competition becomes especially consequential when first discoveries reshape
the subsequent dynamics. In starving exploration~\cite{BenichouRedner2014,Regnier2024},
each newly discovered site provides a discrete life-extending reward, so sites
claimed by others are irrevocably lost and directly lengthen starvation
intervals. In reinforced and self-interacting
exploration~\cite{Davis1990,Pemantle1992,Pemantle2007,PhysRevLett.130.227101},
first discoveries can instead alter mobility itself, so early imbalances feed
back into faster or slower motion. More generally, whenever first discoveries
affect later access to space or resources, collective coverage is not enough:
the allocation of discoveries becomes central.

Despite decades of work on multiparticle random walks~\cite{AcedoYuste2002},
first-discovery allocation lies largely outside the standard theory. Existing
results describe collective ranges~\cite{Larralde1992}, distinct and commonly visited
sites~\cite{MajumdarTamm2012,KunduMajumdarSchehr2013}, and non-intersection probabilities in two
dimensions~\cite{DuplantierKwon1988,LawlerSchrammWerner2001II}. They determine
which sites are covered or jointly visited, not how covered sites are allocated
by first arrival. Related colouring problems were studied previously, first for
two one-dimensional random walkers on a ring~\cite{GomesJuniorLucenaDaSilvaHilhorst1996}
and later on more general \emph{finite} graphs~\cite{MillerPainting}, but they concern
final colouring after full coverage rather than the asymptotic division of
first discoveries considered here.

\begin{figure*}[ht]
    \centering
    \includegraphics[width=\linewidth]{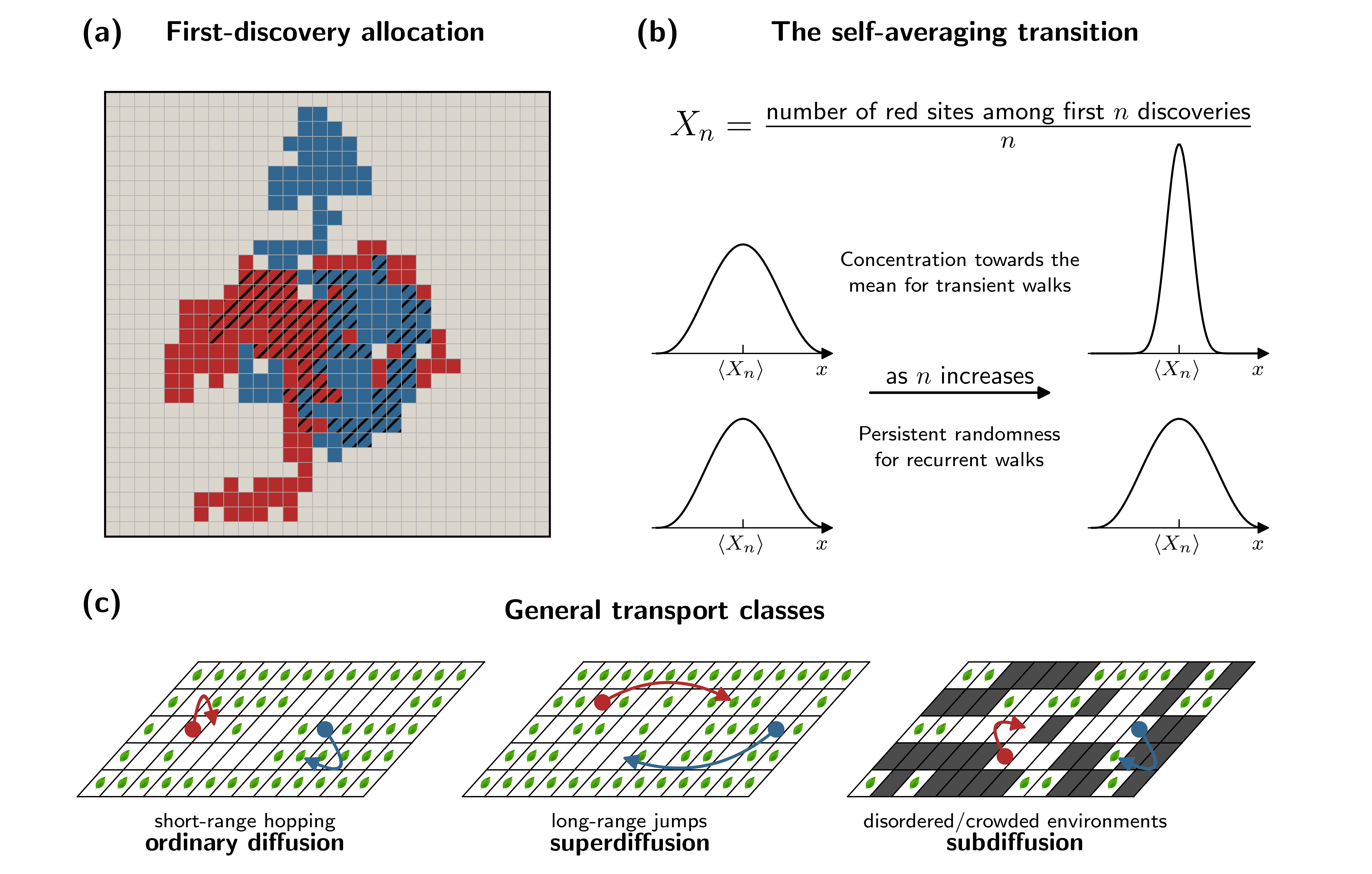}
    \caption{\textbf{Competitive exploration requires a missing observable.}
\textbf{a}, First-discovery allocation. Two identical searchers explore a
landscape in which each site initially carries one unclaimed resource. Grey
sites are undiscovered; coloured sites have been discovered and are labelled by
the searcher that reached them first; hatched sites have been visited by both
searchers. Classical observables measure collective coverage or geometric
interference, but not which searcher secured each discovery.
\textbf{b}, Self-averaging transition. We tag one searcher and set \(I_k=1\)
if the \(k\)-th collective discovery is secured by this tagged searcher, and
\(I_k=0\) otherwise. The discovery share
\(X_n=n^{-1}\sum_{k=1}^n I_k\) has mean \(1/2\) for two identical competitors.
It remains broadly random in the recurrent regime, but self-averages towards
the equal split in the marginal and transient regimes, with threshold
\(d_s=2\).
\textbf{c}, General transport classes. Beyond ordinary diffusion with local
steps, competitive exploration can involve superdiffusive motion driven by
long relocations or subdiffusive motion induced by geometry, crowding, disorder
or memory. In the crowded/disordered schematic, grey sites represent
inaccessible regions.}
\label{fig:competitive_exploration_missing_observable}
\end{figure*}

In competitive exploration, each discovered site carries a label: the searcher
that reached it first. Here we ask how these labels are split globally between
competitors. We tag one searcher and denote by \(I_k\) the indicator that the
\(k\)-th collective discovery, ordered by discovery time, is secured by this
tagged searcher. After \(n\) distinct sites have been collectively discovered,
we define the discovery share
\[
X_n=\frac{1}{n}\sum_{k=1}^n I_k,
\]
as illustrated in Fig.~\ref{fig:competitive_exploration_missing_observable}.
We first focus on two identical competitors. In this symmetric case, exchange
symmetry fixes the mean allocation: $\langle X_n\rangle=1/2$ for every $n$. The central question is whether
  individual
realizations concentrate around it,
$X_n\longrightarrow 1/2$ 
in probability as $n\to\infty$, or whether the discovery share remains broadly
random. We refer to this concentration towards the ensemble mean as
self-averaging. In physical terms, the question is whether competition
eventually erases early asymmetries, or whether inequalities in access to first
discoveries can persist indefinitely~\cite{Note1}. More broadly, can the same question be answered beyond ordinary diffusion,
for superdiffusive motion driven by  long relocations or subdiffusive motion
induced by crowding or memory~\cite{Sokolov2012,HoeflingFranosch2013}?

Here we show that the discovery share is governed by a universal fluctuation
theory controlled by the spectral dimension \(d_s\), defined by the large-\(t\)
decay of the probability that a single searcher is at its starting point after
\(t\) steps, \(p_0(t)\sim t^{-d_s/2}\). For ordinary diffusion in \(d\)
Euclidean dimensions, \(d_s=d\). More generally, when the number of accessible
sites within distance \(r\) scales as \(r^{d_f}\), defining the fractal
dimension \(d_f\), and the typical displacement after \(t\) steps scales as
\(r(t)\sim t^{1/d_w}\), defining the walk dimension \(d_w\) (\(d_w>2\) for
subdiffusive spreading and \(d_w<2\) for superdiffusive spreading), one has
\(d_s=2d_f/d_w\)~\cite{HavlinBenAvraham1987}.
The spectral dimension first determines whether equal splitting is reached:
in recurrent exploration \((d_s<2)\), \(X_n\) remains persistently random,
whereas in the marginal and transient regimes \((d_s\ge2)\), \(X_n\to1/2\)
in probability. Within the self-averaging phase, \(d_s\) also fixes how equal
splitting is approached: \(2\le d_s<3\) gives anomalously slow non-Gaussian
concentration, while \(d_s\ge 3\) gives  Gaussian concentration, standard for $d_s>3$ and  logarithmically corrected for $d_s=3$. For \(d_s\ge2\), we
obtain exact asymptotic variances, including prefactors. Beyond this global
fluctuation law, we determine   the first-discovery label
correlations along the discovery sequence, which provides a finer probe of
first-discovery allocation.

We first analyse two identical Markovian searchers on a \(d\)-dimensional
regular lattice with a general symmetric jump law \(w({\bf r})\), where
\({\bf r}\) is the jump vector. Finite-variance laws, including isotropic tails
\(w({\bf r})\sim r^{-(d+\alpha)}\) with \(r=|{\bf r}|\) and \(\alpha>2\) at
large \(r\), have \(d_w=2\) and therefore \(d_s=d\). Long-range laws with the
same tail and \(\alpha<2\) have \(d_w=\alpha\) and therefore
\(d_s=2d/\alpha\). We return below to subdiffusive transport generated by
geometry or memory.

We now rewrite the deviation of the discovery share from equal splitting in
terms of two elementary quantities: the imbalance between the individual explored
ranges and the allocation of sites visited by both walkers. In this setting, after
\(n\) distinct sites have been collectively discovered, \(X_n\) is the fraction of
these first discoveries secured by a tagged walker. Denote by \(N_1(t)\) and \(N_2(t)\) the individual
ranges of walkers 1 and 2 up to time \(t\), regardless of which walker reached each
site first, and let \(N_{\rm comm}(t)\) be the overlap, defined as the number of
sites visited by both walkers. Denote by \(t_n\) the time of the
\(n\)-th collective discovery. At this random time \(t_n\), inclusion-exclusion 
gives
\begin{equation}
\label{eq:inclusion_exclusion}
N_1+N_2-N_{\rm comm}=n.
\end{equation}
We suppress \(t_n\) to avoid cluttering formulas; e.g., in \eqref{eq:inclusion_exclusion}, we write $N_j$ instead of $N_j(t_n)$. 

Let \(Q_{n}\) be the fraction of common sites that were first discovered by
walker 1 at time $t_n$.  The individual range \(N_1\) counts all sites visited
by walker 1, including common sites that may have been first discovered by the other
walker.  The numbers of first discoveries by the two walkers  are thus
\begin{align*}
&F_1=nX_n=N_1-N_{\rm comm}(1-Q_{n}),\\
&F_2=n(1-X_n)=N_2-N_{\rm comm}Q_{n}.\\
\end{align*}
Subtracting these two relations and using Eq.~\eqref{eq:inclusion_exclusion} gives
\begin{equation}
\label{eq:Xn_symmetric_decomposition}
X_n-\frac12 = \frac{N_1-N_2}{2n}
+ \frac{N_{\rm comm}}{n} \left(Q_{n}-\frac12\right),
\end{equation}
showing that deviations from equal
splitting are the result of a range-imbalance contribution and a
contested-allocation contribution. The self-averaging problem is therefore reduced to determining whether the
right-hand side of Eq.~\eqref{eq:Xn_symmetric_decomposition} vanishes as
\(n\) grows, and at what rate.

The first quantity to estimate in this decomposition is the size of this common,
or contested, set. Starting from the general identity
\begin{equation}
\label{eq:common_sites_identity}
\langle N_{\rm comm}(t)\rangle
=
\sum_x \mathbb P(x\ {\rm reached\ before}\ t)^2,
\end{equation}
and using the standard renewal relation between the hitting probability
\(\mathbb P(x\ {\rm reached\ before}\ t)\) and the on-site probability
\(p_0(t)\), together
with the large-time scaling \(p_0(t)\sim t^{-d_s/2}\) defining the spectral
dimension, we obtain the asymptotic size of the contested set. Transferring this
estimate to the random discovery time \(t_n\) (SI), we find that
\(N_{\rm comm}\) is of order \(n\) for \(d_s<2\), as are the individual ranges
\(N_1\) and \(N_2\), but \(N_{\rm comm}\) is asymptotically
subleading for \(d_s\ge 2\). Thus \(d_s=2\) is the overlap threshold: below it,
competition for the same resources remains macroscopic; above it, overlap becomes
subextensive, opening the way to concentration towards an equal split of discovered
sites.

Overlap identifies the threshold for self-averaging, but it does not by itself
determine the size or nature of the fluctuations. To obtain the full fluctuation
theory, one must go beyond the mean size of the contested set and determine the
fluctuations of the two terms in Eq.~\eqref{eq:Xn_symmetric_decomposition}. This
technically more demanding step is carried out in the SI and leads to the central
fluctuation law of competitive exploration:
\begin{equation}
\label{eq:VarXn_main}
{\rm Var}(X_n)\sim
\begin{cases}
{\rm Var}(X_\infty)>0, & d_s<2,\\
(\log n)^{-2}, & d_s=2,\\
n^{-(d_s-2)}, & 2<d_s<3,\\
(\log n)/n, & d_s=3,\\
1/n, & d_s>3.
\end{cases}
\end{equation}

\begin{figure*}[t]
    \centering
    \includegraphics[width=\linewidth]{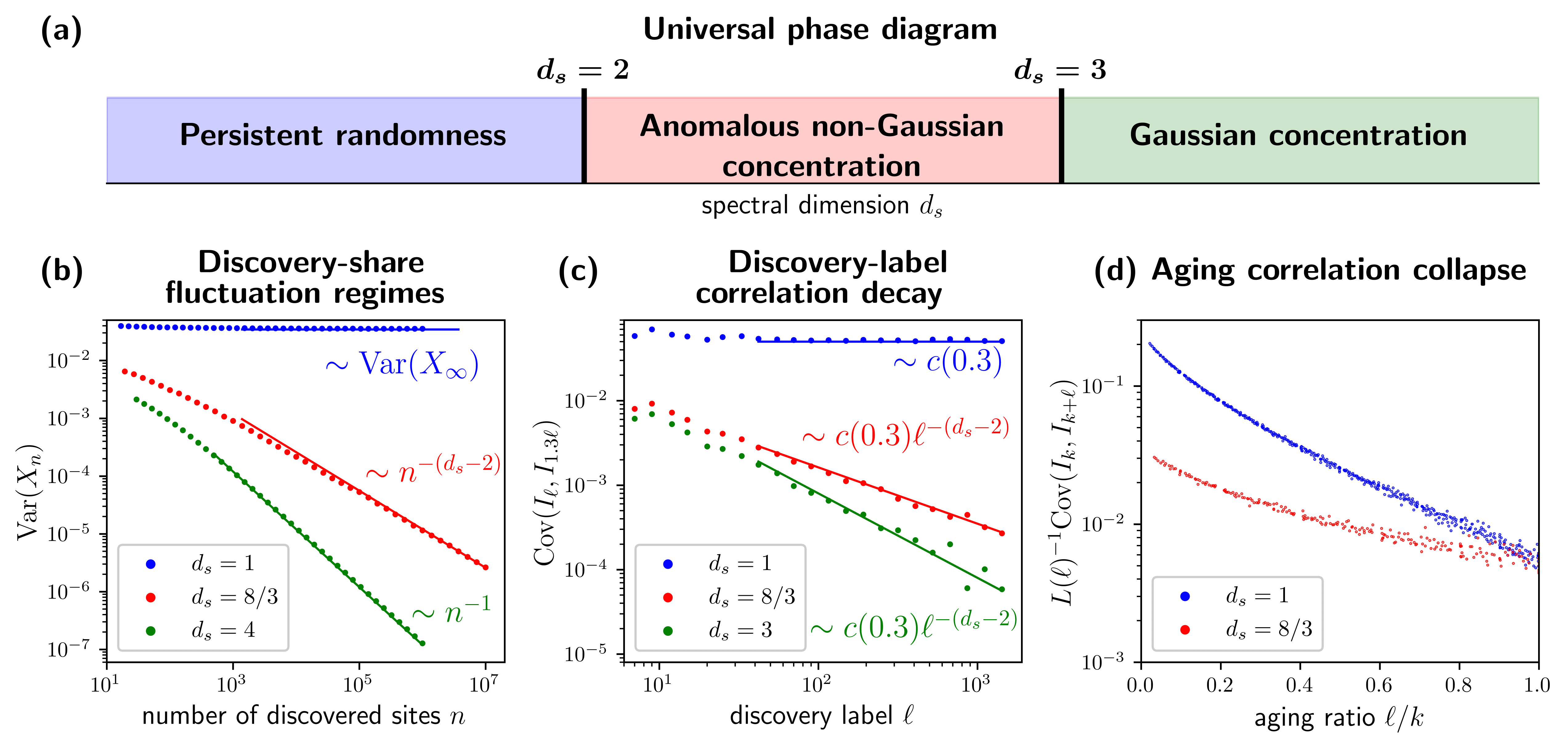}
\caption{\textbf{First-discovery label correlations underlie the discovery-share
fluctuation phase diagram.}
\textbf{a}, Discovery-share fluctuation regimes. The spectral dimension \(d_s\),
defined by the on-site decay \(p_0(t)\sim t^{-d_s/2}\), separates persistent
randomness \((d_s<2)\), anomalously slow non-Gaussian concentration
\((2\le d_s<3)\), and Gaussian concentration \((d_s\ge3)\), with logarithmic
corrections at \(d_s=3\).
\textbf{b}, Discovery-share variance. Simulations for two identical searchers on a
one-dimensional lattice with a general symmetric jump law (simulation details given in the SI) confirm
Eq.~\eqref{eq:VarXn_main}: non-zero limiting variance for \(d_s<2\), anomalous
decay for \(2<d_s<3\), and \(1/n\) Gaussian decay for \(d_s>3\). 
\textbf{c}, Discovery-label correlation decay. The panel shows
\(\mathrm{Cov}(I_k,I_{k+\ell})\) at the fixed aging ratio \(\ell/k=0.3\), as a
function of the separation \(\ell\) along the discovery sequence. The amplitude
\(c(0.3)\) denotes the value of the scaling function \(\mathcal C_{d,d_s}\) at
this ratio. The predicted decay is of order one for \(d_s<2\), \(\ell^{-2/3}\)
for \(d_s=8/3\), and \(\ell^{-1}\) at \(d_s=3\). Thus \(d_s=8/3\) and \(d_s=3\)
belong to the same two-point aging regime, with
\(\mathrm{Cov}(I_k,I_{k+\ell})\propto \ell^{-(d_s-2)}\) at fixed \(\ell/k\),
although \(d_s=3\) is the marginal threshold separating anomalously slow
non-Gaussian concentration from Gaussian concentration in the global
discovery-share variance.
\textbf{d}, Aging correlation collapse. The plotted quantity
\(L_{d_s}(\ell)^{-1}\mathrm{Cov}(I_k,I_{k+\ell})\) collapses as a function of the
aging ratio \(\ell/k\). This shows that, after removal of the lag-dependent
factor \(L_{d_s}\), the remaining dependence is the non-stationary scaling
function \(\mathcal C_{d,d_s}(\ell/k)\).}
\label{fig:variance+correlations_decay}
\end{figure*}

\begin{figure*}[t]
    \centering
    \includegraphics[width=\linewidth]{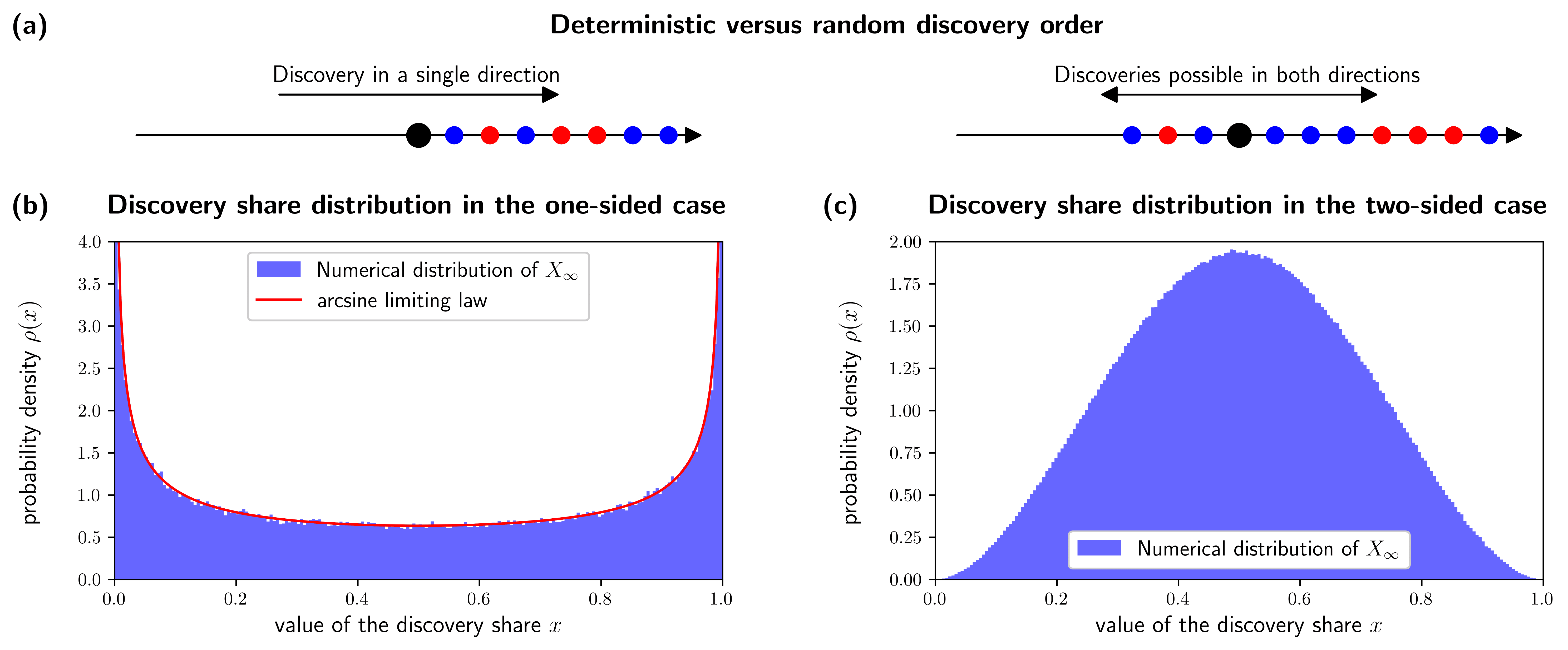}
   \caption{\textbf{Site-discovery order shapes broad recurrent fluctuations.}
In recurrent exploration, discovery-label correlations prevent self-averaging,
but the limiting law also depends on the spatial order in which sites are
discovered. \textbf{a}, Deterministic versus random site-discovery order. The black dot
marks the starting point. Red and blue sites are first discovered by the two
competitors, respectively. In the one-sided geometry, the first \(n\)
collective discoveries are necessarily the sites \(1,\ldots,n\), so the spatial
order of discovery is deterministic. In the symmetric geometry, discoveries can
occur on either side of the origin, and  the left-right order of discovered sites is random.
\textbf{b}, Deterministic order: beta distribution. Blue bars show simulations of the
discovery-share distribution; the red curve is the
\({\rm Beta}(1/2,1/2)\) arcsine law.
\textbf{c}, Random order: non-beta distribution. The limiting distribution remains broad
but differs from the parameter-free beta benchmark; the exclusion of this
benchmark is documented in the SI.}
\label{fig:illu_1D_limiting_distributions}
\end{figure*}

\begin{figure*}[ht]
    \centering
    \includegraphics[width=\linewidth,height=0.6\textheight]{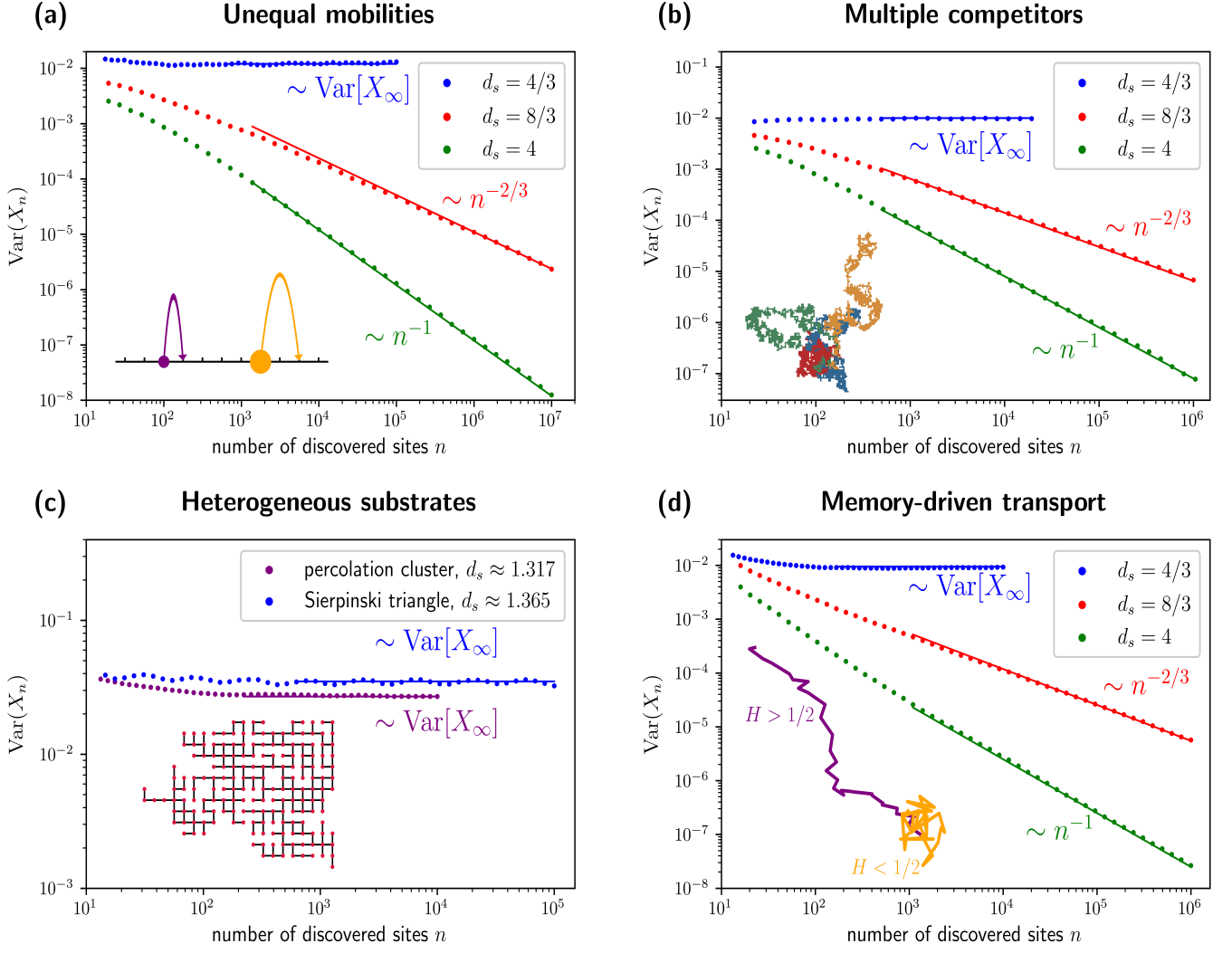}
    \caption{\textbf{Discovery-share fluctuation regimes persist beyond the minimal model.}
Numerical simulations of representative extensions show that the fluctuation
structure predicted by the analytical framework is robust to changes in
microscopic dynamics, geometry and competitive setting. The simulations use
one-dimensional heavy-tailed jump processes in panels \textbf{a} and \textbf{b}, simple
diffusion on fractal substrates in panel \textbf{c}, and fractional Brownian
motion in panel \textbf{d}. Small schematics in the panels illustrate the
corresponding extension.
\textbf{a}, Unequal mobilities. When the two competitors have unequal
mobilities, here implemented by different jump scales, the mean discovery share
is shifted away from \(1/2\); fluctuations around this biased mean retain the
same scaling structure.
\textbf{b}, Multiple competitors.  For \(M\) identical competitors, the mean
discovery share of a tagged competitor is \(1/M\). Fluctuations around this
mean retain the same scaling structure, showing that the phenomenology is not
restricted to pairwise competition.
\textbf{c}, Heterogeneous substrates. On deterministic or random geometrically heterogeneous substrates, including
Sierpinski-type fractals and critical percolation clusters, the recurrent regime remains broadly
fluctuating. Persistent competitive randomness is therefore not restricted to
simple Euclidean geometries.
\textbf{d}, Memory-driven transport. Fractional Brownian exploration provides a
non-Markovian test case with memory, spanning both subdiffusive ($H<1/2$) and
superdiffusive ($H>1/2$) regimes. Across these extensions, the theory retains the same three regimes:
persistent randomness, anomalously slow non-Gaussian concentration and standard
Gaussian concentration, with the \(n\)-dependence governed by \(d_s\).
%Across these extensions, the same three regimes
%persist---persistent randomness, anomalously slow non-Gaussian concentration and standard Gaussian concentration--- \art{ici aussi} with the \(n\)-dependence governed by \(d_s\).
The curves shown
correspond to two-dimensional fBm with Hurst exponents \(H=1/3\) (blue) and
\(H=2/3\) (red), and three-dimensional fBm with Hurst exponent \(H=2/3\)
(green).}
\label{fig:illu_decay_robustness}
\end{figure*}

Because the two competitors are identical, decay of \({\rm Var}(X_n)\) means
concentration of \(X_n\) around the equal split \(1/2\).
Equation~\eqref{eq:VarXn_main} therefore identifies \(d_s=2\) as the threshold
between two long-time behaviours: for \(d_s\ge2\), the discovery share
concentrates towards equal splitting, \(X_n\to1/2\) in probability, whereas for
\(d_s<2\), it remains broad at long times and is described by a non-degenerate
limiting random variable \(X_\infty\). Within the self-averaging regimes, the
threshold at \(d_s=3\) separates anomalously slow non-Gaussian concentration
from standard Gaussian concentration, with the logarithmic correction displayed
in Eq.~\eqref{eq:VarXn_main}. Beyond the scaling exponents, the theory also
gives exact amplitudes whenever the variance decays  (see SI). These prefactors have a
regime-dependent degree of universality: at the overlap marginal point
\(d_s=2\), they are fixed solely by the large-distance universality class, whereas away
from marginality some short-distance information  enters. The phase
diagram and representative variance tests are shown in
Fig.~\ref{fig:variance+correlations_decay}a,b. All numerical protocols and
parameter values used in
Figs.~\ref{fig:variance+correlations_decay}--\ref{fig:illu_decay_robustness}
are specified in the SI.

For instance, at the marginal value \(d_s=2\), the one-dimensional Cauchy
universality class, represented by symmetric jump processes with jump distribution \(w(r)\sim |r|^{-2}\), gives
\begin{equation}
    {\rm Var}(X_n)\sim \frac{\log 2}{2(\log n)^2}\,,
\end{equation}
whereas the two-dimensional diffusive universality class, represented by
finite-variance symmetric jump processes such as the simple random walk on
\(\mathbb Z^2\), gives
\begin{equation}
    {\rm Var}(X_n)\sim
\frac{{\rm Cl}_2(\pi/2)-3^{-1/2}{\rm Cl}_2(\pi/3)}{(\log n)^2}\,.
\end{equation}
Here
\({\rm Cl}_2(\theta)=\sum_{m\geq 1}\sin(m\theta)/m^2\)
denotes the Clausen function.

The variance law also sets the discovery scale at which equal splitting is
reached. The typical distance of $X_n$ from $1/2$ is measured by
$\sqrt{{\rm Var}(X_n)}$. Let $n_\varepsilon$ denote the number of discoveries
needed for this typical deviation to become of order $\varepsilon$, that is,
$\sqrt{{\rm Var}(X_n)}\sim \varepsilon$. Equation~\eqref{eq:VarXn_main} then
gives, in the strongly transient regime,
\begin{equation}
n_\varepsilon \sim \varepsilon^{-2},
\qquad d_s>3,
\label{eq:neps_gaussian}
\end{equation}
up to the logarithmic correction at $d_s=3$. By contrast, in the weakly
transient regime,
\begin{equation}
n_\varepsilon \sim \varepsilon^{-2/(d_s-2)},
\qquad 2<d_s<3.
\label{eq:neps_nongaussian}
\end{equation}
Thus weak transience produces a genuine regime of anomalously slow convergence
towards equal splitting.

The variance law describes the global fluctuations of the discovery share. We
next ask how the labels of two discoveries are correlated along the discovery
sequence. These label correlations form the natural two-point structure of
first-discovery allocation: they measure how the identity of the searcher
securing one discovery is related to the identity of the searcher securing a
later discovery. Writing
\[
\mathrm{Cov}(I_k,I_{k+\ell})=
\langle I_k I_{k+\ell}\rangle-
\langle I_k\rangle\langle I_{k+\ell}\rangle,
\]
the variance of \(X_n\) is obtained by summing these non-stationary
correlations, but before this summation the two-point correlations retain the
memory and aging structure of the discovery history. As shown in the SI, in the scaling limit of
large \(k\) and \(\ell\),
\begin{equation}
\mathrm{Cov}(I_k,I_{k+\ell})
\sim
\mathcal C_{d,d_s}\!\left(\frac{\ell}{k}\right)
L_{d_s}(\ell),
\label{eq:cov_all}
\end{equation}
where
\(L_{d_s}(\ell)=1\) for \(d_s<2\),
\(L_{d_s}(\ell)=(\log \ell)^{-2}\) for \(d_s=2\), and
\(L_{d_s}(\ell)=\ell^{-(d_s-2)}\) for \(2<d_s<4\). The factor \(L_{d_s}\) controls the decay with the separation \(\ell\) along
the discovery sequence, whereas
the scaling function \(\mathcal C_{d,d_s}(\ell/k)\) captures the aging
structure of the discovery-label sequence: correlations are not stationary functions of
the lag alone, but depend on the relative position within the discovery
history. For \(d_s\ge2\), these scaling functions are computed exactly in the SI for the
corresponding large-distance universality classes and they determine the exact
variance amplitudes together with the lag factor \(L_{d_s}\).

Summing Eq.~\eqref{eq:cov_all} over discovery indices recovers the variance
law~\eqref{eq:VarXn_main}. The two-point correlations explain why this law has two thresholds. For \(d_s>3\), the label correlations are summable, placing the
system in a short-memory regime where Gaussian fluctuations are expected on
central-limit grounds~\cite{BouchaudGeorges1990} and observed numerically. For
\(2<d_s<3\), they are non-summable, producing a long-memory regime in which
\(X_n\) self-averages but with anomalously large, non-Gaussian fluctuations. At
\(d_s=2\) and \(d_s=3\), the marginal decays generate the logarithmic
corrections displayed in Eq.~\eqref{eq:VarXn_main}. In the recurrent regime
\(d_s<2\), correlations remain of order one across macroscopic portions of the
discovery history, preventing self-averaging altogether. This two-point
structure of the discovery-label sequence is shown in
Fig.~\ref{fig:variance+correlations_decay}c,d.

The recurrent phase requires a separate treatment because the discovery share
does not collapse to its mean. The one-dimensional nearest-neighbour random walk provides the most transparent
paradigm. In the SI, we derive an exact expression for the limiting
variance, \(\mathrm{Var}(X_\infty)\simeq 0.0343\), showing that fluctuations
remain of order one. Unlike the self-averaging regimes, where the large-\(n\)
behaviour is expressed in terms of one-walker first-passage and survival
quantities, the recurrent case keeps the random discovery time and the two
discovery histories strongly coupled.

Inhomogeneous environments provide a natural direction beyond the homogeneous
setting considered so far. An analytically tractable extreme is the one-sided
geometry, in which resources are located at each lattice site on the positive
axis while sites on the negative axis are empty.
%A useful reference case is the one-sided geometry, in which resources are
%located at each lattice site on the positive axis while sites on the negative
%axis are empty. 
Here, the first \(n\) collective discoveries are necessarily the
sites \(1,\ldots,n\). Writing \(T_i(k)\) for the first-passage time of walker
\(i\) to site \(k\), and \(Z_k=T_2(k)-T_1(k)\) for the inter-walker delay, the
discovery share is rewritten as
\[
X_n=\frac{1}{n}\sum_{k=1}^n \mathbf{1}_{\{Z_k>0\}}.
\]
By the Markov property and translational invariance, the increments
\(Z_{k+1}-Z_k\) are independent and identically distributed differences of
unit-distance first-passage times~\cite{randon-furling_markovian_2015}. These \(t^{-3/2}\) first-passage-time tails place \(Z_k\) in the symmetric
\(1/2\)-stable universality class, and \(X_n\) is its occupation
fraction above zero. The generalized arcsine theorem~\cite{spitzer_combinatorial_1956, Lamperti1958} gives
\(X_n\Rightarrow {\rm Beta}(1/2,1/2)\), the arcsine law on \([0,1]\) with
density \((\pi\sqrt{x(1-x)})^{-1}\).

This reduction relies on the deterministic order in which sites are discovered.
In the symmetric geometry, resources are present on both sides of the origin,
and this order becomes random. If the limiting law nevertheless remained beta,
exchange symmetry would restrict it to the one-parameter family
\({\rm Beta}(a,a)\), and the exact variance would fix \(a\). This remaining
possibility is ruled out in the SI. Thus the departure from the arcsine/beta distribution in this symmetric geometry 
has a direct interpretation: random site-discovery order adds an extra source of
correlations to first-discovery allocation. This mechanism is illustrated in
Fig.~\ref{fig:illu_1D_limiting_distributions}.

We finally show that the theory extends well beyond the minimal analytical
setting. Scaling arguments developed in the SI predict that the same asymptotic
law in the number of discoveries \(n\), Eq.~\eqref{eq:VarXn_main}, persists
across a broad class of competitive exploration processes, and simulations
confirm this prediction in representative extensions
(Fig.~\ref{fig:illu_decay_robustness}). These extensions probe distinct
physical ingredients of competitive exploration. Unequal mobilities (panel ~\ref{fig:illu_decay_robustness}a) represent
the simplest source of intrinsic competitive advantage: changing the mobility
scale of one competitor shifts the mean discovery share away from \(1/2\). For
ordinary diffusive competitors with diffusivities \(D_1\) and \(D_2\), this
mean is \(D_1/(D_1+D_2)\). Multiple competitors (panel~\ref{fig:illu_decay_robustness}b) move the mean discovery share
of a tagged competitor from \(1/2\) to \(1/M\) for \(M\) identical competitors,
and show that the fluctuation structure is not restricted to pairwise
competition. Heterogeneous substrates (panel~\ref{fig:illu_decay_robustness}c), including deterministic fractals and
critical percolation clusters~\cite{RammalToulouse1983,HavlinBenAvraham1987,HoeflingFranoschFrey2006}, provide minimal models of geometrically
constrained, crowded or disordered environments~\cite{Sokolov2012,HoeflingFranosch2013}. Memory-driven transport (panel~\ref{fig:illu_decay_robustness}d) is
represented by fractional Brownian motion~\cite{MetzlerKlafter2000}, a
paradigmatic non-Markovian Gaussian process with stationary increments and
mean-square displacement
\(\langle [Y(t)-Y(0)]^2\rangle\sim t^{2H}\), where \(H\) is the Hurst exponent.
Across these extensions, simulations show quantitative agreement with
Eq.~\eqref{eq:VarXn_main}: persistent randomness, anomalously slow non-Gaussian
concentration and standard Gaussian concentration all occur with the predicted
\(n\)-dependence governed by \(d_s\).
 The
very different nature of these examples shows the broad applicability of the
approach.

Together, these results enlarge the focus of random-exploration theory. When
first arrival grants exclusive access to resources or targets, the explored set
carries, beyond its geometry, history-dependent allocation information: each
site is labelled by the competitor that discovered it first. The discovery
share is the simplest global probe of this allocation. Its label correlations
give a finer, two-point description along the discovery sequence. These
observables answer a question that is invisible to range, overlap and cover-time
observables: whether the fair mean allocation of two identical competitors
becomes fair in a single long exploration history.

The answer is universal. The spectral dimension determines whether
first-arrival competition leaves a persistent random imbalance or washes it out,
and, when it is washed out, on which discovery scale. Trajectory overlap, discovery-label correlations and site-discovery order
provide the mechanisms by which this universality emerges across representative
random-walk models.
%The mechanisms identified
%here---trajectory overlap, discovery-label correlations and site-discovery
%order--- \art{tirets} show how this universality emerges across representative random-walk models.
First-discovery allocation therefore defines a robust statistical structure of
competitive exploration, present across ordinary diffusion, long-range hopping,
heterogeneous substrates, unequal mobilities, many competitors and random walks
with memory.

\end{document}

\title{Supplementary Information - Universal fluctuations of first discoveries in competitive exploration}

\author{Arthur Plaud}
\affiliation{Sorbonne Universit\'e, CNRS, Laboratoire de Physique Th\'eorique de la Mati\`ere Condens\'ee (LPTMC), 4 Place Jussieu, 75005 Paris, France}

\author{P.L. Krapivsky}
\affiliation{Department of Physics, Boston University, Boston, Massachusetts 02215, USA}

\author{S. Redner}
\affiliation{Santa Fe Institute, Santa Fe, New Mexico 87501, USA}

\author{Olivier B\'enichou}
\affiliation{Sorbonne Universit\'e, CNRS, Laboratoire de Physique Th\'eorique de la Mati\`ere Condens\'ee (LPTMC), 4 Place Jussieu, 75005 Paris, France}

\maketitle

\tableofcontents

\section{Fluctuation law of competitive exploration}\label{sec:fluc}

\subsection{Definitions}

We first define the relevant observables of competitive exploration that are used when proving results of the main text. These are either random variables whose fluctuations we aim to quantify or simply quantities relevant at intermediate steps in the proof.

We first define the random time $t_n$. It is the time at which the $2$ walkers collectively discovered $n$ sites in total. Because of our definition of the discovery-share $X_n$, obtaining the distribution of $X_n$ a priori requires to integrate over the fluctuations of $t_n$.

We also use notations for number of discoveries, including or not competitive aspects. Let $\mathcal{N}_1(t)$ and $\mathcal{N}_2(t)$ be the number of distinct sites visited by walkers $1$ and $2$ up until time $t$, regardless of whether they were the first one to discover these sites. They are known as the ranges of the walkers, and are among the most classic quantities in random exploration. We also introduce the overlap or number of common sites, $\mathcal{N}_{\text{comm}}(t)$, defined as the number of sites visited by both walkers up until time $t$.

Until now, we only introduced classic quantities in random exploration, namely the range of a single walker, its inverse $t_n$, and the number of common sites visited. But the competitive nature of our problem suggests introducing new observables that quantify the allocation of sites among the $2$ walkers. For this purpose, we denote by $F_i(t)$ the number of first discoveries secured by walker $i$ up until time $t$. Because some of the sites reached by walker $i$ were discovered by the other walker, we have:
\begin{equation}
    F_i(t) \leq \mathcal{N}_i(t).
\end{equation}

We can now define the discovery-share $X_n$, measuring the fraction of total discoveries made by a tagged walker. At a given fixed time, this fraction of discoveries is:
\begin{equation}
    X_t = \dfrac{F_1(t)}{F_1(t)+F_2(t)}.
\end{equation}
The discovery-share corresponds to this fraction measured at the random time $t_n$, where the walkers collectively discovered $n$ sites and thus $F_1(t_n)+F_2(t_n) = n$. It therefore writes as:
\begin{equation}
    X_n = \dfrac{F_1(t_n)}{n}.
\end{equation}
Denoting by $I_k$ the indicator that the $k$-th collective discovery is secured by walker $1$, we have the alternative definition:
\begin{equation}
    X_n = \dfrac{1}{n}\sum_{k=1}^n I_k.
\end{equation}
We also introduce the fraction of contested discoveries $Q_i(t)$. This is a quantity analogous to the discovery-share $X$, but defined only on the sites that were reached by both walkers, where they effectively competed for first discovery. It thus counts the fraction of common sites discovered by walker $i$.

Lastly, we denote as $d_s$ the spectral dimension of the walk, defined by the asymptotic behavior:
\begin{equation}
    \mathbb{P}(\text{the walker stands at its starting point at time }t) \underset{t \sim +\infty}{\sim} t^{-\frac{d_s}{2}}.
\end{equation}
For ordinary diffusion in \(d\) Euclidean dimensions, $d_s=d$. More generally,
if the number of accessible sites within distance $r$ scales as $r^{d_f}$
and the typical displacement after time $t$ scales as $r(t)\sim t^{1/d_w}$,
then:
\begin{equation}
    d_s=\frac{2d_f}{d_w}.
\end{equation}
In most of the calculations below we work on regular $d$-dimensional
lattices, so $d_f=d$, and therefore $d_s=2d/d_w$. For a finite-variance symmetric jump law, $d_w=2$ (normal diffusion), hence $d_s=d$. For a long-range symmetric jump law with tail:
\begin{equation}
p(\ell)\sim |\ell|^{-(d+\alpha)},
\end{equation}
one has $d_w=\alpha$ when $\alpha<2$, and thus:
\begin{equation}
d_s=\frac{2d}{\alpha}.
\end{equation}

The spectral dimension gauges the speed at which the random walk spreads away from the starting site and characterizes the recurrence/transience transition. If $d_s < 2$, the walk is recurrent: every site of the lattice is visited an infinite number of times. On the other hand, walks with $d_s > 2$ are transient and visit each point on the lattice a (random) finite number of times. Walks with $d_s=2$ are called marginal. The spectral dimension $d_s$ also enters into scaling forms of several first-passage observables, e.g., first-passage time densities and survival probabilities.

\subsection{Scaling of the random stopping time $t_n$}

We start by obtaining relevant scalings for $t_n$ the time at which the walkers collectively discovered $n$ sites.  The scaling of $t_n$ is the same as for a single walker, a quantity widely studied in random exploration. Indeed, $t^{1W}_{\frac{n}{2}}<t_n<t^{1W}_n$ (where the superscript denotes the inverse range time for a single walker) and the $\frac{n}{2}$ does not change the scaling. 

We can therefore directly transfer known \cite{mariz_statistics_2001, gall_range_1991} scalings for $t^{1W}_n$ to $t_n$. These results are obtained by inverting the scaling of the mean range and are expressed in term of the spectral dimension. Distinguishing between recurrent and transient walks, we have:
\begin{equation}
\begin{split}
    \begin{cases}
    & d_s < 2 : \langle \mathcal{N}(t) \rangle \sim t^{\frac{d_s}{2}} \Rightarrow \langle t_n \rangle \sim n^{\frac{2}{d_s}}\\
    & d_s = 2 : \langle \mathcal{N}(t) \rangle \sim \dfrac{t}{\log{t}} \Rightarrow \langle t_n \rangle \sim n \log{n}\\
    & d_s > 2 : \langle \mathcal{N}(t) \rangle \sim t \Rightarrow \langle t_n \rangle \sim n.
    \end{cases}
\end{split}
\end{equation}
We also use the fluctuations of $t_n$, that are directly related to fluctuations of the range:
\begin{equation}
\begin{split}
    \begin{cases}
    & d_s < 2 : \Var[\mathcal{N}(t)] \sim t^{d_s} \Rightarrow \Var[t_n] \sim n^{\frac{4}{d_s}}\\
    & d_s = 2 : \Var[\mathcal{N}(t)] \sim \dfrac{t^2}{(\log{t})^4} \Rightarrow \Var[t_n] \sim n^2\\
    & 2 < d_s < 3 : \Var[\mathcal{N}(t)] \sim t^{4-d_s} \Rightarrow \Var[t_n] \sim n^{4-d_s}\\
    & d_s = 3 : \Var[\mathcal{N}(t)] \sim t \log{t} \Rightarrow \Var[t_n] \sim n \log{n}\\
    & d_s > 3 : \Var[\mathcal{N}(t)] \sim t \Rightarrow \Var[t_n] \sim n.
    \end{cases}
\end{split}
\end{equation}
From this information, one can prove a self-averaging transition for the range. A random variable is said to be self-averaging if its fluctuations are subdominant compared to its mean, namely:
\begin{equation}
    \sqrt{\Var[X]} \ll \langle X \rangle.
\end{equation}
Using the scalings above, one sees easily that the range is self-averaging for marginal and transient walks ($d_s \geq 2$), whereas it shows fluctuations of the same order as the mean for recurrent walks.

In \ref{sec:quant}, more precise information will be needed about range scalings, both for $\mathcal{N}_i(t)$ and for $t_n$, as prefactors play a role in the variance decay prefactor for the discovery-share. But for now such precise information is not relevant.

\begin{figure}[ht]
    \centering
    \includegraphics[width=1\textwidth]{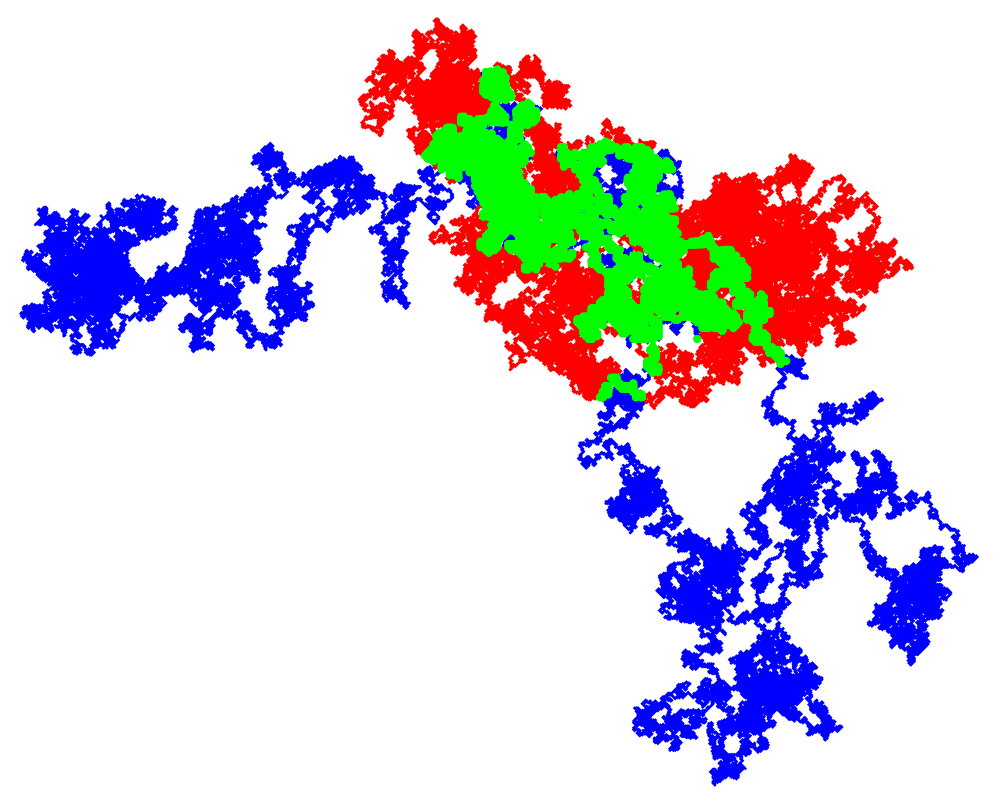}
    \caption{An illustration of sites visited by two independent random walkers on the square lattice at time $t = 2^{18}$. Sites visited only by the first walker are shown in red, sites visited only by the second walker are shown in blue, and sites visited by both walkers are shown in green. The common sites constitute an asymptotically negligible part of the range of each walker: the average number of common sites visited increases as $t/(\log{t})^2$, while the average single walker range scales as $t/\log{t}$. This negligibility property is generalized to the case of the random time $t_n$.}
    \label{fig:overlap_2D}
\end{figure}

\subsection{Scaling of the number of common sites}

In the context of competitive exploration, it is important to quantify the overlap between walker trajectories by computing the scaling of the number of sites visited by both walkers. In Fig.~\ref{fig:overlap_2D}, we illustrate the ranges and the overlap of two independent random walkers on the square lattice at a fixed time. In our case, we instead consider the situation at time $t_n$, where the walkers collectively discovered $n$ sites.

Computing the scaling of the mean is sufficient. We start from the exact expression at fixed $t$ for the mean number of common sites visited:
\begin{equation}
    \langle \mathcal{N}_{\text{comm}}(t) \rangle = \sum_{x \in \mathbf{R}^d} \mathbb{P}(x \text{ reached before } t)^2.
    \label{eq 6}
\end{equation}
This general formula is valid for any two independent identical random walkers, on hypercubic lattices $\mathbb{Z}^d$. Whereas we focus on the competition between two searchers, many foundational formulas can be extended to arbitrary number of searchers; e.g., in Eq.~\eqref{eq 6}, the power of two should be replaced by the number of searchers.

Specializing Eq.~\eqref{eq 6} to the case of our random stopping time $t_n$ gives:
\begin{equation} \begin{split}
    \langle \mathcal{N}_{\text{comm}}(t_n) \rangle & = \sum_{x \in \mathbf{R}^d} \mathbb{P}(x \text{ reached before } t_n)^2 \\
     & = \sum_{x \in \mathbf{R}^d} \left( \int_0^{+\infty} \dd t \, \mathbb{P}(x \text{ reached before } t, t_n = t) \right)^2.
\end{split} \end{equation}
We now insert the scaling form of the joint probability. We already described the scaling of $t_n$ (in this case we do not track fluctuations as we are interested in the mean number of common sites visited). The scaling form of the survival probability, on the other hand, is well-known for Markovian random walks and depends on the spectral dimension \cite{majumdar_number_2012}:
\begin{equation}
    \mathbb{P}(x \text{ reached before } t) \sim \begin{cases}
    d_s < 2 \;: \; & f\left(\dfrac{x}{t^{\frac{d_s}{2 d}}}\right), \\
    d_s = 2 \;: \; & \dfrac{1}{\log{t}}f\left(\dfrac{x}{t^{\frac{1}{d}}}\right), \\
    d_s > 2 \;: \; & \dfrac{1}{t^{\frac{d_s}{2}-1}}f\left(\dfrac{x}{t^{\frac{d_s}{2 d}}}\right).
    \end{cases}
\end{equation}
This means we can write, first in the case $d_s<2$:
\begin{equation} \begin{split}
    \sum_{x \in \mathbf{R}^d} \left( \int_0^{+\infty} \dd t \, \mathbb{P}(x \text{ reached before } t, t_n = t) \right)^2 & \sim \sum_{x \in \mathbf{R}^d} \left[ \int_0^{+\infty} \dd t \, n^{-\frac{2}{d_s}} f\left(\dfrac{x}{t^{\frac{d_s}{2 d}}},\dfrac{t}{n^{\frac{2}{d_s}}}\right) \right]^2 \\
     & \sim \sum_{x \in \mathbf{R}^d} \left[ \int_0^{+\infty} \dd u \, f\left(\dfrac{x}{n^{\frac{1}{d}} u^{\frac{d_s}{2 d}}},u\right) \right]^2 \\
     & \sim \int_{\mathbf{R}^d} \dd^d x F\left(\dfrac{x}{n^{\frac{1}{d}}}\right)^2 \\
     & \sim n \int_{\mathbf{R}^d} \dd^d u F\left(u\right)^2,
\end{split} \end{equation}
where in the last steps, we replaced summation by integration to obtain the leading behavior in the $n\rightarrow +\infty$ limit. For the recurrent case, we obtain that $\mathcal{N}_{\text{comm}}(t_n) \sim n$. Because the territory explored by the walkers is typically compact around the starting point, both walkers tend to explore the same sites, and these common sites represent a finite portion of all the discoveries.

For $d_s = 2$, we simply add in the extra logarithmic factor and adjust the scaling of $t_n \sim n \log{n}$:
\begin{equation} \begin{split}
    \sum_{x \in \mathbf{R}^d} \left( \int_0^{+\infty} \dd t \, \mathbb{P}(x \text{ reached before } t, t_n = t) \right)^2 & \sim \sum_{x \in \mathbf{R}^d} \left[ \int_0^{+\infty} \dd t \, \dfrac{1}{n \log{n}} \dfrac{1}{\log{t}}f\left(\dfrac{x}{t^{\frac{1}{d}}},\dfrac{t}{n \log{n}}\right) \right]^2 \\
     & \sim \sum_{x \in \mathbf{R}^d} \left[ \int_0^{+\infty} \dd u \dfrac{1}{\log{n}}\, f\left(\dfrac{x}{(n \log{n})^{\frac{1}{d}} u^{\frac{1}{d}}},u\right) \right]^2 \\
     & \sim \dfrac{1}{(\log{n})^2}\int_{\mathbf{R}^d} \dd^d x F\left(\dfrac{x}{(n \log{n})^{\frac{1}{d}}}\right)^2 \\
     & \sim \dfrac{n}{\log{n}} \int_{\mathbf{R}^d} \dd^d u F\left(u\right)^2,
\end{split} \end{equation}
and thus $\mathcal{N}_{\text{comm}}(t_n) \sim \dfrac{n}{\log{n}}$. The common sites become a negligible part of the territory explored in the $n \rightarrow +\infty$ limit. This decay is even stronger for $d_s >2$ (now $t_n \sim n$):
\begin{equation} \begin{split}
    \sum_{x \in \mathbf{R}^d} \left( \int_0^{+\infty} \dd t \, \mathbb{P}(x \text{ reached before } t, t_n = t) \right)^2 & \sim \sum_{x \in \mathbf{R}^d} \left[ \int_0^{+\infty} \dfrac{\dd t}{n} \,  \dfrac{1}{t^{\frac{d_s}{2}-1}}f\left(\dfrac{x}{t^{\frac{d_s}{2 d}}},\dfrac{t}{n}\right) \right]^2 \\
     & \sim \sum_{x \in \mathbf{R}^d} \left[ \int_0^{+\infty} \dd u \dfrac{1}{n^{\frac{d_s}{2}-1}u^{\frac{d_s}{2}-1}}\, f\left(\dfrac{x}{n^{\frac{d_s}{2d}} u^{\frac{d_s}{2 d}}},u\right) \right]^2 \\
     & \sim \dfrac{1}{n^{d_s-2}}\int_{\mathbf{R}^d} \dd^d x F\left(\dfrac{x}{n^{\frac{d_s}{2 d}}}\right)^2 \\
     & \sim n^{2-\frac{d_s}{2}} \int_{\mathbf{R}^d} \dd^d u F\left(u\right)^2.
\end{split} \end{equation}
Since $d_s>2$, $2 - \frac{d_s}{2} < 1$, the mean number of common sites grows sublinearly with $n$. The vanishing of the exponent for $d_s = 4$ signals the breakdown of the scaling approximation of the discrete sum, with terms $x \sim 1$ dominating the sum. The number of common sites will then stop growing with $n$ and remain of order $1$. In this work, we do not need a fine description of the scaling of $\mathcal{N}_{\text{comm}}(t_n)$ in the regime $d_s >3$, only that it is negligible compared to $\sqrt{n}$. But we still mention that $\mathcal{N}_{\text{comm}}(t_n) \sim \log{n}$ when $d_s=4$, and remains of order one in the $n \rightarrow +\infty$ limit for $d_s >4$.

This transition from $\mathcal{N}_{\text{comm}}(t_n) \sim n$ in the recurrent case to $\mathcal{N}_{\text{comm}}(t_n) \ll n$ for transient and marginal walks drives the discovery-share to a self-averaging state. In the transient case, the competition decays with time, as the walkers start to explore space in different directions. This is explained with more details in the following section.

\subsection{Fraction decomposition and range fluctuations}

We start by expressing the discovery-share in terms of the number of common sites and the individual ranges of the walkers, whose scalings were studied in the previous subsection. By definition of $t_n$, we have:
\begin{equation}
    \mathcal{N}_1(t_n) + \mathcal{N}_2(t_n) - \mathcal{N}_{\text{comm}}(t_n) = n,
    \label{eq 19}
\end{equation}
as common sites are counted twice, once for each walker.
We can also express the number of first discoveries for walker $1$ in term of these variables:
\begin{equation}
    F_1(t_n) = n X_n = \mathcal{N}_1(t_n) - \mathcal{N}_{\text{comm}}(t_n) (1-Q_1(t_n)).
    \label{eq 20}
\end{equation}
Similarly,
\begin{equation}
F_2(t_n)=n(1-X_n) =\mathcal N_2(t_n)-\mathcal N_{\rm comm}(t_n)Q_1(t_n).
\end{equation}
Combining these two identities with Eq.~\eqref{eq 19} gives:
\begin{equation}
X_n-\frac12 = \frac{\mathcal N_1(t_n)-\mathcal N_2(t_n)}{2n} + \frac{\mathcal N_{\rm comm}(t_n)}{n} \left(Q_1(t_n)-\frac12\right),
\label{eq 16}
\end{equation}
which is exactly Eq.~\eqref{eq:Xn_symmetric_decomposition} in the main text. This equation clarifies the physics behind self-averaging for the discovery-share: if $\mathcal{N}_{\text{comm}}(t_n) \ll \mathcal{N}_i(t_n) \sim n$ (because $t_n$ scales exactly in the same way as the corresponding time for a single random walker), the second term in Eq.~\eqref{eq 16} is negligible compared to the first term. The discovery-share is thus self-averaging if $\mathcal{N}_1(t_n)$ self-averages. The only technicality in quantifying fluctuations of $\mathcal{N}_1(t_n)$ involves going from the known results for the fluctuations of the range at fixed time to the case at random time $t_n$. This is done in exactly the same spirit as for the mean number of common sites, but now, since we track fluctuations, we introduce the scaling forms given by the central limit theorem (CLT) for the range, in the case where the law of large number holds (transient case).

We start with the recurrent case $d_s < 2$. Here, the range is not self-averaging as $\Var[\mathcal{N}(t)] \sim \langle \mathcal{N}(t) \rangle^2$, and we have $\mathbb{P}(\mathcal{N}^1(t) = n) \sim t^{-\frac{d_s}{2}} \varphi(n t^{-\frac{d_s}{2}})$, with non-trivial scaling function $\varphi$, i.e. not equal to the Dirac delta function. We therefore write:
\begin{equation} \begin{split}
    \mathbb{P}(\mathcal{N}_1(t_n) = n') & = \int_0^{+\infty} \dd t \, \mathbb{P}(\mathcal{N}_1(t) = n',t_n = t) \\
    & \sim  \int_0^{+\infty} \dd t \, n^{-\frac{2}{d_s}} t^{-\frac{d_s}{2}} \varphi(n' t^{-\frac{d_s}{2}},t n^{-\frac{2}{d_s}}) \\
    & \sim \int_0^{+\infty} \dd u \dfrac{u^{-\frac{d_s}{2}}}{n} \varphi\left(\frac{n'}{n}u^{-\frac{d_s}{2}},u\right) \\
    & \sim \dfrac{1}{n}\Phi\left(\frac{n'}{n}\right).
\end{split} \end{equation}
This proves that in the recurrent case $\mathcal{N}_1(t_n)$ has fluctuations of order $n$ (the same order as its mean), as expected because of the non-self-averaging property.

For $d_s \geq 2$, we replace the scaling form by the one induced by the CLT, both for the random variable $t_n$ and for $\mathcal{N}_1(t)$.In the marginal case:
\begin{equation} \begin{split}
    \mathbb{P}(\mathcal{N}_1(t_n) = n') & = \int_0^{+\infty} \dd t \, \mathbb{P}(\mathcal{N}_1(t) = n',t_n = t) \\
    & \sim  \int_0^{+\infty} \dd t \, \dfrac{1}{n} \dfrac{(\log{t})^2}{t} \varphi\left[ \left(n' - \frac{c t}{\log{t}}\right)\frac{(\log{t})^2}{t},\left(t - \frac{n \log{n}}{2 c}\right)\frac{1}{n} \right] \\
    & \sim \int_{-\infty}^{+\infty} \dd u \dfrac{(\log{n})^2}{n \log{n}} \varphi\left[ \left(n' - \frac{c }{\log{n}}\frac{n \log{n}}{2 c}\right)\frac{(\log{n})^2}{n \log{n}},u \right] \\
    & \sim \dfrac{\log{n}}{n}\Phi\left[\left(n' - \frac{n}{2}\right) \frac{\log{n}}{n}\right],
\end{split} \end{equation}
where the prefactors of the means are such that $\langle \mathcal{N}_1(t_n) \rangle = \frac{n}{2}$ (equal share on average). This proves the CLT for $\mathcal{N}_1(t_n)$, with the fluctuations being of order $\frac{n}{\log{n}}$ in this case. 

The proof in the case $d_s > 2$ is exactly the same, adjusting for the correct order of fluctuations of $\mathcal{N}_1(t)$ and $t_n$.
\begin{equation} \begin{split}
    \mathbb{P}(\mathcal{N}_1(t_n) = n') & = \int_0^{+\infty} \dd t \, \mathbb{P}(\mathcal{N}_1(t) = n',t_n = t) \\
    & \sim \int_0^{+\infty} \dd t \dfrac{1}{\sqrt{t\,n}^{\max\{4-d_s,1\}}} \varphi\left(\dfrac{n' - c\, t}{\sqrt{t}^{\max\{4-d_s,1\}}},\dfrac{t - \frac{n}{2c}}{\sqrt{n}^{\max\{4-d_s,1\}}} \right) \\
    & \sim \int_{-\infty}^{+\infty} \dd u \dfrac{1}{\sqrt{n}^{\max\{4-d_s,1\}}} \varphi\left( \frac{n'-\frac{n}{2}}{\sqrt{n}^{\max\{4-d_s,1\}}},u \right)\\
    & \sim \dfrac{1}{\sqrt{n}^{\max\{4-d_s,1\}}} \Phi\left(\frac{n'-\frac{n}{2}}{\sqrt{n}^{\max\{4-d_s,1\}}}\right).
\end{split} \end{equation}
Since $t_n \sim n$, going from $\mathcal{N}_1(t)$ to $\mathcal{N}_1(t_n)$ does not change the magnitude of fluctuations appearing in the CLT in this case.

Using these results for the fluctuations of $\mathcal{N}_1(t_n)$ and what we previously obtained for the scaling of the number of common sites visited, we can obtain the scaling of the fraction variance for all values of the spectral dimension $d_s$, showing first a self-averaging transition at $d_s=2$ and then a transition from anomalously slow non-Gaussian concentration to Gaussian concentration at $d_s=3$. These results are collected in Table \ref{tab:tab 1}, which shows the scaling of fluctuations of all quantities involved in Eq.~\eqref{eq 20} and the resulting scaling of the variance of the $X_n$.

\begin{table}[h]
\caption{Summary of relevant scalings}
\label{tab:tab 1}
\begin{tabular}{c c c c c}
\toprule
Spectral dimension & $\sqrt{\Var[\mathcal{N}_1(t_n)]}$ & $\langle \mathcal{N}_{\text{comm}}(t_n) \rangle$ & $\Var[X_n]$ & Limiting fluctuations \\
\midrule
$d_s<2$ & $\mathcal{O}(n)$ & $\mathcal{O}(n)$ & $\mathcal{O}(1)$ & Persistent randomness \\
\addlinespace[3ex]
$d_s=2$ & $\mathcal{O}\left(n / \log n\right)$ & $\mathcal{O}\left(n / \log n\right)$ & $\mathcal{O}\left(1 / \log^2 n\right)$ & non-Gaussian vanishing fluctuations \\
$2<d_s<3$ & $\mathcal{O}\left(n^{2-\frac{d_s}{2}}\right)$ & $\mathcal{O}\left(n^{2-\frac{d_s}{2}}\right)$ &  $\mathcal{O}\left(n^{2-d_s}\right)$ & non-Gaussian vanishing fluctuations \\
\addlinespace[3ex]
$d_s=3$ & $\mathcal{O}\left(\sqrt{n \log n}\right)$ & $\mathcal{O}\left(\sqrt{n}\right)$ & $\mathcal{O}\left( \log n / n\right)$ & Gaussian vanishing fluctuations \\
$d_s > 3$ & $\mathcal{O}\left(\sqrt{n}\right)$ & $\mathcal{O}\left(n^{2-\frac{d_s}{2}}\right)$\footnote{This scaling is only valid for $d_s < 4$ and has to be replaced by $\mathcal{O}(\log n)$ for $d_s=4$ and $\mathcal{O}(1)$ for $d_s>4$. In all cases, it is negligible compared to $\mathcal{O}(\sqrt{n})$.} & $\mathcal{O}\left(1 / n\right)$ & Gaussian vanishing fluctuations \\
\bottomrule
\end{tabular}
\end{table}

This table explains the transitions shown in equation Eq.~\eqref{eq:VarXn_main} in the main text. For recurrent walks, both fluctuations of individual ranges and number of common sites scale similarly to the mean individual range, explaining the non-self-averaging nature of the fraction and the persistence of fluctuations in the large time limit. In the intermediate region ($2\leq d_s <3$), encompassing both marginal and weakly transient walks, these two scalings are still the same but are now negligible compared to $n$, which is the scaling of the mean individual range. They both contribute to leading order fluctuations of the fraction. Lastly, in the strongly transient case $d_s\geq 3$, the number of common sites become negligible compared to fluctuations of the individual ranges and leading order fluctuations of the fraction are now fully explained by individual range fluctuations, leading to quantitative predictions for the prefactor of the decay of the variance and to a Gaussian central limit theorem for the fraction as the corresponding result for the individual range has been proven for $d_s \geq 3$. This table also shows the two transitions for discovery-share fluctuations, with the discovery-share becoming self-averaging above for $d_s \geq 2$ and having asymptotically Gaussian fluctuations for $d_s \geq 3$.

\section{Quantitative theory for the self-averaging case}\label{sec:quant}

Here, we compute variance prefactors in the self-averaging case. We first consider the intermediate case where common sites play a role and leading order fluctuations of the fraction are no longer fully quantified by individual range fluctuations. Therefore, we need to describe precisely correlations between walker discoveries, including two-walkers effects. We distinguish between the marginal and the weakly transient case, although they are treated in a very similar manner.

\subsection{Fixed-time formulation and linearization}

The approach in this section (and in the next one) will be the following. We first compute the variance of the discovery-share at fixed time using a combinatorial description using first passage times, and then we transfer this result to the random time $t_n$ using similar arguments as those used to obtain Table \ref{tab:tab 1} in the previous section. 

We start by "linearizing" the variance of the fraction at fixed time. Still denoting by $F_i(t)$ the amount of sites discovered by walker $i$ until time $t$, the share is expressed by definition via the ratio:
\begin{equation}
    X_t = \dfrac{F_1(t)}{F_1(t) + F_2(t)}.
\end{equation}
Written in this way, the computation of the variance seems intractable: $F_1(t)$ and $F_2(t)$ are correlated, and a ratio of random variables is not well suited for indicator decomposition computations of the variance. But self-averaging considerably simplifies the problem. We write $F_i(t) = \langle F_i(t) \rangle + \Delta F_i(t)$ and expand to first order in deviations (where the first term is actually the same for $i=1,2$ since the walkers are identical). This yields:
\begin{equation} \begin{split}
    \dfrac{F_1(t)}{F_1(t) + F_2(t)} & \sim \dfrac{ \langle F_1(t) \rangle + \Delta F_1(t)}{ \langle F_1(t) \rangle + \Delta F_1(t) +  \langle F_2(t) \rangle + \Delta F_2(t)} \\
    & \sim \dfrac{ \langle F_1(t) \rangle}{\langle F_1(t) \rangle + \langle F_1(t) \rangle}\left( 1+ \dfrac{\Delta F_1(t)}{\langle F_1(t) \rangle}\right)\left(1-\dfrac{\Delta F_1(t) + \Delta F_2(t)}{\langle F_1(t) \rangle + \langle F_1(t) \rangle}\right) \\
    & \sim \dfrac{1}{2} + \dfrac{\Delta F_1(t)}{2\langle F_1(t) \rangle} - \dfrac{\Delta F_1(t) + \Delta F_2(t)}{4\langle F_1(t) \rangle}\\
    & \sim \dfrac{1}{2} + \dfrac{\Delta F_1(t) - \Delta F_2(t)}{4 \langle F_1(t) \rangle}.
\end{split} \end{equation}
Since $\Delta F_1(t) - \Delta F_2(t) = F_1(t) - F_2(t)$, we can write:
\begin{equation}
    \Var[X_t] \sim \dfrac{1}{16 \langle F_1(t) \rangle^2}\langle [F_1(t) - F_2(t)]^2 \rangle.
    \label{eq 36}
\end{equation}
We cannot expand the square since the two random variables are not independent, $\langle (F_1(t) - F_2(t))^2 \rangle > 2 \Var[F_1(t)]$. As the mean involved at the denominator is a single walker quantity (overlap is negligible in the self-averaging case at the level of the mean), the main task is to compute the second moment $(F_1(t) - F_2(t))^2$ and we can replace $F_1$ with $\mathcal{N}_1$ in the denominator:
\begin{equation}
\tcbhighmath[
colback=white,
colframe=black,
arc=2mm]{
    \Var[X_t] \sim \dfrac{1}{16 \langle \mathcal{N}_1(t) \rangle^2}\langle [F_1(t) - F_2(t)]^2 \rangle.
    }
    \label{eq 36 prime}
\end{equation}

\subsection{Indicator decomposition and exact formula for the variance}

Expressing the ranges as sums of discoveries of given sites, we write:
\begin{equation}
    \langle [F_1(t) - F_2(t)]^2 \rangle = \sum_{x,y} \langle I_x I_y \rangle,
\end{equation} 
where $I_x = \mathds{1}_{1 \text{ discovered } x \text{ before time } t} - \mathds{1}_{2 \text{ discovered } x \text{ before time } t}$.
We now aim to express the correlation $\langle I_x I_y \rangle$. Counting the cases $I_x I_y = 1$ and $I_x I_y = -1$ and using symmetry of the walkers, we get:
\begin{eqnarray}
\langle I_x I_y \rangle &=& 2 \mathbb{P}(1 \text{ discovered } x \text{ before time } t,1 \text{ discovered } y \text{ before time } t) \nonumber \\
   & - & 2 \mathbb{P}(1 \text{ discovered } x \text{ before time } t,2 \text{ discovered } y \text{ before time } t).
\end{eqnarray}
Since we will sum these correlations over $x$ and $y$, we can restrict ourselves to the cases where $x$ was discovered before $y$, adding a factor of $2$:
\begin{equation}
    \langle I_x I_y \rangle + \langle I_y I_x \rangle = 2 \langle I_x I_y \rangle = 2 (\langle I_x I_y, x \text{ before } y \rangle + \langle I_x I_y, y \text{ before } x \rangle ) =  2 \langle I_x I_y, x \text{ before } y \rangle + 2 \langle I_x I_y, y \text{ before } x \rangle.
\end{equation}
We associate the first term to $\langle I_x I_y \rangle$ and the second to $\langle I_y I_x \rangle$. This breaks the symmetry but makes calculations shorter. With this decomposition, we get:
\begin{eqnarray}
    \langle I_x I_y \rangle & = &4 \sum_{0\leq i<j \leq t} F^{\{y\}}_i(x) F_{j-i}(y-x) \sum_{z \neq x,y} P^{\{x,y\}}_i(z)S^{\{y-z\}}_{j-i} \nonumber \\
   & - & 4 \sum_{0\leq i<j \leq t} F^{\{y\}}_i(x) S_{j-i}^{\{y-x\}} \sum_{z \neq x,y} P^{\{x,y\}}_i(z)F_{j-i}(y-z).
\end{eqnarray}
In this equation, $F$ is the first passage time distribution starting from $0$, $F^{\{y\}}$ is the first passage time distribution without hitting $y$, $P^{\{x,y\}}$ is the propagator without hitting $x$ and $y$, and $S^{\{y\}}$ is the probability starting from $0$ not to hit $y$. Summing over $x$ and $y$, we therefore write:
\begin{tcolorbox}[
colback=white,
colframe=black,
arc=2mm]
\begin{varwidth}{\textwidth}
\begin{multline}
    \langle (F_1(t) - F_2(t))^2 \rangle = 4 \sum_{x,y} \sum_{0\leq i<j \leq t} F^{\{y\}}_i(x) F_{j-i}(y-x) \sum_{z \neq x,y} P^{\{x,y\}}_i(z)S^{\{y-z\}}_{j-i} \\
    -  4 \sum_{x,y} \sum_{0\leq i<j \leq t} F^{\{y\}}_i(x) S_{j-i}^{\{y-x\}} \sum_{z \neq x,y} P^{\{x,y\}}_i(z)F_{j-i}(y-z).
    \label{eq 43}
\end{multline}
\end{varwidth}
\end{tcolorbox}
We included terms with $x=y$, but we can discard them if needed, as they lead to a subdominant contribution to the variance. Eq.~\eqref{eq 43} is exact and the basis of all the expansions performed in the following.

\subsection{Green function identities}

Now we compute the \emph{detailed} scaling expansions of the random walk quantities involved in Eq.~\eqref{eq 43}. This is done by relating these quantities to the propagator using renewal equations and then using the scaling form of the propagator. We will go by order of difficulty.

For the first-passage distribution, we write:
\begin{equation}
    P_i(x) = \sum_{k=0}^i F_k(x)P_{i-k}(0).
    \label{eq 26}
\end{equation}
Introducing the generating functions:
\begin{equation}
G(x,\xi) = \displaystyle \sum_{n\geq 0} \xi^n P_n(x), \quad F(x,\xi) = \displaystyle \sum_{n\geq 0} \xi^n F_n(x),
\end{equation}
we recast Eq.~\eqref{eq 26} into:
\begin{equation}
    F(x,\xi) = \dfrac{G(x,\xi)}{G_0(\xi)}\,, \qquad G_0(\xi) = G(x=0,\xi).
\end{equation}
The survival is obtained by simple summation:
\begin{equation}
    S_i(x) = 1 - \sum_{k=0}^i F_k(x),
\end{equation}
so that:
\begin{equation}
    S(x,\xi) = \dfrac{1-F(x,\xi)}{1-\xi} = \dfrac{G_0(\xi) - G(x,\xi)}{(1-\xi)G_0(\xi)}.
\end{equation}
We also write a renewal equation for $F^{\{y\}}$:
\begin{equation}
    F_i(x) = F_i^{\{y\}}(x) + \sum_{j=0}^i F_j^{\{x\}}(y) F_{i-j}(y,x),
\end{equation}
which translates into:
\begin{equation}
    F^{\{y\}}(x,\xi) = \dfrac{F(x,\xi) - F(x-y,\xi)F(y,\xi)}{1-F(x-y,\xi)F(y-x,\xi)} = \dfrac{G(x,\xi) G_0(\xi) - G(x-y,\xi)G(y,\xi)}{G_0(\xi)^2 - G(x-y,\xi)G(y-x,\xi)}.
\end{equation}
Lastly for $P^{\{x,y\}}$, we use:
\begin{equation}
    P_i^{\{x\}}(z) = P_i^{\{x,y\}}(z) + \sum_{k=0}^i F^{\{x\}}_k(y) P_{i-k}^{\{x-y\}}(z-y),
\end{equation}
leading to:
\begin{equation}
    P^{\{x,y\}}(z,\xi) = P^{\{x\}}(z,\xi) - F^{\{x\}}(y,\xi)P^{\{x-y\}}(z-y,\xi).
\end{equation}
We also write the renewal for $P^{\{x\}}$:
\begin{equation}
    P^{\{x\}}(z,\xi) = G(z,\xi) - G(z-x,\xi) F(x,\xi).
\end{equation}
Combining everything, we obtain:
\begin{eqnarray}
    P^{\{x,y\}}(z,\xi) &=& G(z,\xi) - G(z-x,\xi) \dfrac{G(x,\xi)}{G_0(\xi)} \nonumber \\
     &-& \left[G(z-y,\xi) - G(z-x,\xi) \dfrac{G(x-y,\xi)}{G_0(\xi)}\right] \dfrac{G(y,\xi) G_0(\xi) - G(y-x,\xi)G(x,\xi)}{G_0(\xi)^2 - G(x-y,\xi)G(y-x,\xi)}.
\end{eqnarray}

\subsection{Weakly transient regime $2<d_s<3$}

\subsubsection{Leading asymptotic contribution}

We now perform the scaling expansions of all these quantities for $\xi \rightarrow 1$ for $2<d_s<3$. As a building block, we only need the behavior of $G_0$ and $G$ as $\xi \rightarrow 1$. From the definition of $d_s$, it is clear that:
\begin{equation}
    G_0(1) = \sum_{n=0}^{+\infty} P_n(0) < +\infty
\end{equation} when $d_s >2$. We denote this constant $g_0$. Introducing the scaling form of the propagator, the Green function becomes:
\begin{equation} \begin{split}
    G(x,\xi) & = \sum_{n=0}^{+\infty} \xi^n P_n(x)\\
    & = \sum_{n=0}^{+\infty} \xi^n \dfrac{1}{n^{\frac{d_s}{2}}} p\left(\dfrac{x}{n^{\frac{d_s}{2d}}}\right) \\
    & \sim \int_0^{+\infty} \dd n \,  e^{-n(1-\xi)} \dfrac{1}{n^{\frac{d_s}{2}}} p\left(\dfrac{x}{n^{\frac{d_s}{2d}}}\right) \\
    & \sim (1-\xi)^{\frac{d_s}{2}-1} \int_0^{+\infty} \dd u \, e^{-u} \dfrac{1}{u^{\frac{d_s}{2}}} p\left(\dfrac{x (1-\xi)^{\frac{d_s}{2d}}}{u^{\frac{d_s}{2d}}}\right) \\
    & = (1-\xi)^{\frac{d_s}{2}-1} \mathcal{G}(x (1-\xi)^{\frac{d_s}{2d}}).
\end{split} \end{equation}
To lighten notations, we denote the small parameter in front of the scaling function $\epsilon$ and the rescaled positions with a star superscript (ex. $x (1-\xi)^{\frac{d_s}{2d}} = x^*$).Expanding the $4$ quantities appearing in Eq.~\eqref{eq 43} up to order $\epsilon^3$ gives:
\begin{equation}
    F(x,\xi) \sim \dfrac{\epsilon}{g_0} \mathcal{G}(x^*),\quad    S(x,\xi) \sim \dfrac{1}{1-\xi}\left[1 - \dfrac{\epsilon}{g_0} \mathcal{G}(x^*)\right],
\end{equation}
\begin{equation}
    F^{\{y\}}(x,\xi) \sim \dfrac{\epsilon}{g_0}\mathcal{G}(x^*) -  \dfrac{\epsilon^2}{g_0^2}\mathcal{G}(x^*-y^*)\mathcal{G}(y^*) +  \dfrac{\epsilon^3}{g_0^3}\mathcal{G}(x^*)\mathcal{G}(x^*-y^*)\mathcal{G}(y^*-x^*),
\end{equation}
\begin{eqnarray}
    P^{\{x,y\}}(z,\xi) &\sim& \epsilon \mathcal{G}(z^*) - \dfrac{\epsilon^2}{g_0}[\mathcal{G}(x^*)\mathcal{G}(z^*-x^*) 
    + \mathcal{G}(y^*)\mathcal{G}(z^*-y^*)] \nonumber \\
    &+& \dfrac{\epsilon^3}{g_0^2}[\mathcal{G}(x^*)\mathcal{G}(y^*-x^*)\mathcal{G}(z^*-y^*) + \mathcal{G}(y^*)\mathcal{G}(x^*-y^*)\mathcal{G}(z^*-x^*)]
\end{eqnarray}

To obtain the scaling expansions of these constrained propagators, we perform the Laplace inversion of the above expressions. This is done using $\mathcal{L}^{-1}\{\epsilon \mathcal{G}(x^*) \} = \dfrac{1}{t^{\frac{d_s}{2}}} p\left(\dfrac{x}{t^{\frac{d_s}{2d}}}\right)$, and converting products in Laplace space to convolutions. We obtain:
\begin{equation}
    F_t(x) \sim \dfrac{1}{g_0 t^{\frac{d_s}{2}}} p\left(x t^{-\frac{d_s}{2d}}\right).
\end{equation}
The inversion for $S$ involves a slightly different scaling function:
\begin{equation}
    S_t(x) \sim 1 - \dfrac{1}{g_0 t^{\frac{d_s}{2}-1}} g_2\left(x t^{-\frac{d_s}{2d}}\right),\quad    g_2(x) = \int_0^1 \dfrac{\dd u}{u^{\frac{d_s}{2}}}\, p\left(x u^{-\frac{d_s}{2d}}\right).
\end{equation}
In total, the expansions of the relevant quantities involve $4$ different scaling functions, all defined as inverse Laplace transforms of products of propagators. We get:
\begin{eqnarray}
    F^{\{y\}}_t(x) &\sim& \dfrac{1}{ g_0 t^{\frac{d_s}{2}}} p\left(x t^{-\frac{d_s}{2d}}\right) - \dfrac{1}{ g_0^2 t^{d_s-1}} g_3\left(y t^{-\frac{d_s}{2d}},(x-y) t^{-\frac{d_s}{2d}}\right) \nonumber \\
    & + & \dfrac{1}{ g_0^3 t^{\frac{3}{2}d_s-2}}\, g_4\left(x t^{-\frac{d_s}{2d}},(y-x) t^{-\frac{d_s}{2d}}, (x-y) t^{-\frac{d_s}{2d}}\right),
\end{eqnarray}
for $F^{\{y\}}$, and:
\begin{multline}
    \dfrac{P^{\{x,y\}}_t(z)}{g_0} \sim \dfrac{1}{  t^{\frac{d_s}{2}}} p\left(z t^{-\frac{d_s}{2d}}\right) - \dfrac{1}{ g_0 t^{d_s-1}} \left\{ g_3\left(x t^{-\frac{d_s}{2d}},(z-x) t^{-\frac{d_s}{2d}}\right) + g_3\left(y t^{-\frac{d_s}{2d}},(z-y) t^{-\frac{d_s}{2d}}\right) \right\} \\
    + \dfrac{1}{ g_0^2 t^{\frac{3}{2}d_s-2}} \left\{ g_4\left(x t^{-\frac{d_s}{2d}},(y-x) t^{-\frac{d_s}{2d}}, (z-y) t^{-\frac{d_s}{2d}}\right) + g_4\left(y t^{-\frac{d_s}{2d}},(x-y) t^{-\frac{d_s}{2d}}, (z-x) t^{-\frac{d_s}{2d}}\right) \right\},
\end{multline}
for $P^{\{x,y\}}$. The scaling functions $g_3$ and $g_4$ are defined as:
\begin{equation}
    g_3(x,y) = \int_0^1 \dfrac{\dd u}{[u(1-u)]^{\frac{d_s}{2}}} p\left(x u^{-\frac{d_s}{2d}}\right) p\left(y (1-u)^{-\frac{d_s}{2d}}\right),
\end{equation}
and:
\begin{equation}
    g_4(x,y,z) = \int_0^1 \dd u \int_0^u \dd v \dfrac{1}{[v(u-v)(1-u)]^{\frac{d_s}{2}}} p\left(x v^{-\frac{d_s}{2d}}\right) p\left(y (u-v)^{-\frac{d_s}{2d}}\right) p\left(z (1-u)^{-\frac{d_s}{2d}}\right).
    \label{eq 47}
\end{equation}
Using all this, we can rewrite the correlation $\langle I_x I_y \rangle$ (and therefore the second moment) as:
\begin{multline}
    \langle [F_1(t) - F_2(t)]^2 \rangle \sim 4  \sum_{x,y,z} \sum_{0\leq i<j \leq t} \left[\dfrac{1}{ g_0 i^{\frac{d_s}{2}}} p\left(x i^{-\frac{d_s}{2d}}\right) - \dfrac{1}{g_0^2  i^{d_s-1}} g_3\left(y i^{-\frac{d_s}{2d}},(x-y) i^{-\frac{d_s}{2d}}\right) \right. \\
    \left. + \dfrac{1}{g_0^3 i^{\frac{3}{2}d_s-2}} g_4\left(x i^{-\frac{d_s}{2d}},(y-x) i^{-\frac{d_s}{2d}}, (x-y) i^{-\frac{d_s}{2d}}\right) \right] \left[\dfrac{1}{ i^{\frac{d_s}{2}}} p\left(z i^{-\frac{d_s}{2d}}\right) - \dfrac{1}{ g_0 i^{d_s-1}} \left\{ g_3\left(x i^{-\frac{d_s}{2d}},(z-x) i^{-\frac{d_s}{2d}}\right) \right. \right. \\
    \left. \left. + g_3\left(y i^{-\frac{d_s}{2d}},(z-y) i^{-\frac{d_s}{2d}}\right) \right\} + \dfrac{1}{g_0^2  i^{\frac{3}{2}d_s-2}} \left\{ g_4\left(x i^{-\frac{d_s}{2d}},(y-x) i^{-\frac{d_s}{2d}}, (z-y) i^{-\frac{d_s}{2d}}\right) + g_4\left(y i^{-\frac{d_s}{2d}},(x-y) i^{-\frac{d_s}{2d}}, (z-x) i^{-\frac{d_s}{2d}}\right) \right\} \right] \\
    \left[ \dfrac{1}{g_0 (j-i)^{\frac{d_s}{2}}} p\left((y-x) (j-i)^{-\frac{d_s}{2d}}\right) \left\{1 - \dfrac{1}{g_0 (j-i)^{\frac{d_s}{2}-1}} g_2\left((y-z) (j-i)^{-\frac{d_s}{2d}}\right) \right\} \right.
    \\ \left. - \dfrac{1}{g_0 (j-i)^{\frac{d_s}{2}}} p\left((y-z) (j-i)^{-\frac{d_s}{2d}}\right) \left\{1 - \dfrac{1}{g_0 (j-i)^{\frac{d_s}{2}-1}} g_2\left((y-x) (j-i)^{-\frac{d_s}{2d}}\right) \right\} \right],
\end{multline}
or using the time difference $j-i$ as a variable:
\begin{multline}
    \langle [F_1(t) - F_2(t)]^2 \rangle \sim 4  \sum_{x,y,z} \sum_{i=0}^t \sum_{j=1}^{t-i} \left[\dfrac{1}{ g_0 i^{\frac{d_s}{2}}} p\left(x i^{-\frac{d_s}{2d}}\right) - \dfrac{1}{ g_0^2 i^{d_s-1}} g_3\left(y i^{-\frac{d_s}{2d}},(x-y) i^{-\frac{d_s}{2d}}\right) \right. \\
    \left. + \dfrac{1}{ g_0^3i^{\frac{3}{2}d_s-2}} g_4\left(x i^{-\frac{d_s}{2d}},(y-x) i^{-\frac{d_s}{2d}}, (x-y) i^{-\frac{d_s}{2d}}\right) \right] \left[\dfrac{1}{ i^{\frac{d_s}{2}}} p\left(z i^{-\frac{d_s}{2d}}\right) - \dfrac{1}{ g_0 i^{d_s-1}} \left\{ g_3\left(x i^{-\frac{d_s}{2d}},(z-x) i^{-\frac{d_s}{2d}}\right) \right. \right. \\
    \left. \left. + g_3\left(y i^{-\frac{d_s}{2d}},(z-y) i^{-\frac{d_s}{2d}}\right) \right\} + \dfrac{1}{g_0^2  i^{\frac{3}{2}d_s-2}} \left\{ g_4\left(x i^{-\frac{d_s}{2d}},(y-x) i^{-\frac{d_s}{2d}}, (z-y) i^{-\frac{d_s}{2d}}\right) + g_4\left(y i^{-\frac{d_s}{2d}},(x-y) i^{-\frac{d_s}{2d}}, (z-x) i^{-\frac{d_s}{2d}}\right) \right\} \right] \\
    \left[ \dfrac{1}{g_0 j^{\frac{d_s}{2}}} p\left((y-x) j^{-\frac{d_s}{2d}}\right) \left\{1 - \dfrac{1}{g_0 j^{\frac{d_s}{2}-1}} g_2\left((y-z) j^{-\frac{d_s}{2d}}\right) \right\}  - \dfrac{1}{g_0 j^{\frac{d_s}{2}}} p\left((y-z) j^{-\frac{d_s}{2d}}\right) \left\{1 - \dfrac{1}{g_0 j^{\frac{d_s}{2}-1}} g_2\left((y-x) j^{-\frac{d_s}{2d}}\right) \right\} \right].
    \label{eq 70}
\end{multline}
Keeping only the lower order terms, we get:
\begin{equation}
    \langle (F_1(t) - F_2(t))^2 \rangle \sim 4  \sum_{x,y,z} \sum_{i=0}^t \sum_{j=1}^{t-i} \dfrac{1}{ g_0 i^{\frac{d_s}{2}}} p\left(x i^{-\frac{d_s}{2d}}\right)  \dfrac{1}{ i^{\frac{d_s}{2}}} p\left(z i^{-\frac{d_s}{2d}}\right) \left[ \dfrac{1}{g_0 j^{\frac{d_s}{2}}} p\left((y-x) j^{-\frac{d_s}{2d}}\right)  - \dfrac{1}{g_0 j^{\frac{d_s}{2}}} p\left((y-z) j^{-\frac{d_s}{2d}}\right) \right]
\end{equation}
at first order. This sum immediately vanishes, as swapping $x$ and $z$ changes the sign. The second order also vanishes, although the cancellation in this case is more involved. Because the first order has been proven to be zero, we now write:
\begin{multline}
    \langle [F_1(t) - F_2(t)]^2 \rangle \sim - 4 \sum_{x,y,z} \sum_{i=0}^t \sum_{j=1}^{t-i} \dfrac{1}{g_0^2 i^{d_s-1}} g_3\left(y i^{-\frac{d_s}{2d}},(x-y) i^{-\frac{d_s}{2d}}\right)  \dfrac{1}{ i^{\frac{d_s}{2}}} p\left(z i^{-\frac{d_s}{2d}}\right) \left[ \dfrac{1}{g_0 j^{\frac{d_s}{2}}} p\left((y-x) j^{-\frac{d_s}{2d}}\right) \right. \\
    \left. - \dfrac{1}{g_0 j^{\frac{d_s}{2}}} p\left((y-z) j^{-\frac{d_s}{2d}}\right) \right] + \dfrac{1}{g_0  i^{\frac{d_s}{2}}} p\left(x i^{-\frac{d_s}{2d}}\right) \left[ \dfrac{1}{g_0 i^{d_s-1}} g_3\left(x i^{-\frac{d_s}{2d}},(z-x) i^{-\frac{d_s}{2d}}\right) + \dfrac{1}{g_0 i^{d_s-1}} g_3\left(y i^{-\frac{d_s}{2d}},(z-y) i^{-\frac{d_s}{2d}}\right) \right]\\
    \left[ \dfrac{1}{g_0 j^{\frac{d_s}{2}}} p\left((y-x) j^{-\frac{d_s}{2d}}\right)  - \dfrac{1}{g_0 j^{\frac{d_s}{2}}} p\left((y-z) j^{-\frac{d_s}{2d}}\right) \right] + \dfrac{1}{g_0 i^{\frac{d_s}{2}}} p\left(x i^{-\frac{d_s}{2d}}\right)  \dfrac{1}{ i^{\frac{d_s}{2}}} p\left(z i^{-\frac{d_s}{2d}}\right) \\
    \left[ \dfrac{1}{g_0 j^{\frac{d_s}{2}}} p\left((y-x) j^{-\frac{d_s}{2d}}\right)  \dfrac{1}{g_0 j^{\frac{d_s}{2}-1}} g_2\left((y-z) j^{-\frac{d_s}{2d}}\right) - \dfrac{1}{g_0 j^{\frac{d_s}{2}}} p\left((y-z) j^{-\frac{d_s}{2d}}\right)\dfrac{1}{g_0 j^{\frac{d_s}{2}-1}} g_2\left((y-x) j^{-\frac{d_s}{2d}}\right) \right].
\end{multline}
The third term sums to $0$ for the same reason as at first order, as swapping $x$ and $z$ changes its sign. Therefore,
\begin{multline}
    \langle [F_1(t) - F_2(t)]^2 \rangle \sim - 4  \sum_{i=0}^t \dfrac{1}{ g_0^2 i^{\frac{3d_s}{2}-1}} \sum_{j=1}^{t-i} \dfrac{1}{g_0 j^{\frac{d_s}{2}}} \sum_{x,y,z} \left[p\left((y-x) j^{-\frac{d_s}{2d}}\right)  -  p\left((y-z) j^{-\frac{d_s}{2d}}\right) \right] \\
    \left[ g_3\left(y i^{-\frac{d_s}{2d}},(x-y) i^{-\frac{d_s}{2d}}\right)  p\left(z i^{-\frac{d_s}{2d}}\right) + p\left(x i^{-\frac{d_s}{2d}}\right)  g_3\left(x i^{-\frac{d_s}{2d}},(z-x) i^{-\frac{d_s}{2d}}\right) + p\left(x i^{-\frac{d_s}{2d}}\right)  g_3\left(y i^{-\frac{d_s}{2d}},(z-y) i^{-\frac{d_s}{2d}}\right) \right].
\end{multline}
We can swap $x$ and $z$ in all the terms involving $g_1\left((y-z) j^{-\frac{d_s}{2d}}\right)$ to yield:
\begin{multline}
    \langle [F_1(t) - F_2(t)]^2 \rangle \sim - 4 \sum_{i=0}^t \dfrac{1}{g_0^2 i^{\frac{3d_s}{2}-1}} \sum_{j=1}^{t-i} \dfrac{1}{g_0 j^{\frac{d_s}{2}}} \sum_{x,y,z} p\left((y-x) j^{-\frac{d_s}{2d}}\right) \left[g_3\left(y i^{-\frac{d_s}{2d}},(x-y) i^{-\frac{d_s}{2d}}\right)  p\left(z i^{-\frac{d_s}{2d}}\right) \right. \\
    \left. + p\left(x i^{-\frac{d_s}{2d}}\right)  g_3\left(x i^{-\frac{d_s}{2d}},(z-x) i^{-\frac{d_s}{2d}}\right) + p\left(x i^{-\frac{d_s}{2d}}\right)  g_3\left(y i^{-\frac{d_s}{2d}},(z-y) i^{-\frac{d_s}{2d}}\right) - p\left(x i^{-\frac{d_s}{2d}}\right)  g_3\left(y i^{-\frac{d_s}{2d}},(z-y) i^{-\frac{d_s}{2d}}\right) \right. \\
    \left. - p\left(z i^{-\frac{d_s}{2d}}\right)  g_3\left(z i^{-\frac{d_s}{2d}},(x-z) i^{-\frac{d_s}{2d}}\right) - p\left(z i^{-\frac{d_s}{2d}}\right)  g_3\left(y i^{-\frac{d_s}{2d}},(x-y) i^{-\frac{d_s}{2d}}\right) \right] \sim - 4  \sum_{i=0}^t \dfrac{1}{ g_0^2 i^{\frac{3d_s}{2}-1}} \sum_{j=1}^{t-i} \dfrac{1}{g_0 j^{\frac{d_s}{2}}} \\
    \sum_{x,y,z} p\left((y-x) j^{-\frac{d_s}{2d}}\right) \left[p\left(x i^{-\frac{d_s}{2d}}\right)  g_3\left(x i^{-\frac{d_s}{2d}},(z-x) i^{-\frac{d_s}{2d}}\right) -  p\left(z i^{-\frac{d_s}{2d}}\right)  g_3\left(z i^{-\frac{d_s}{2d}},(x-z) i^{-\frac{d_s}{2d}}\right) \right].
\end{multline}
It is now clear that the second order fully cancels out, as we can redefine $y$ as $y-x$, perform the sum over this new variable, and we are left with the difference of $2$ terms where $x$ and $z$ have just been swapped. 

We therefore go to third order, where we have $5$ terms. To compute them, we rely on two crucial identities for the propagator. The first is the normalization:
\begin{equation}
    \int \dd x \, p\left(x\right) = 1.
    \label{eq 74}
\end{equation}
The second identity expresses the spatial convolution of the propagator. Such an identity is quite natural and is simply the scaling version of the path decomposition identity:
\begin{equation} \begin{split}
    \sum_x P_n(x)P_{n'}(y-x) =  P_{n+n'}(y).
\end{split} \end{equation}
Taking the scaling limit on both sides, we obtain:
\begin{equation}
    \int \dd x \dfrac{1}{n^{\frac{d_s}{2}}} p\left(x n^{-\frac{d_s}{2d}}\right) \dfrac{1}{n'^{\frac{d_s}{2}}} p\left((y-x) n'^{-\frac{d_s}{2d}}\right) = \dfrac{1}{(n+n')^{\frac{d_s}{2}}} p\left(y (n+n')^{-\frac{d_s}{2d}}\right).
\end{equation}
In the following, we consider a random walk with symmetric scaling limit ($p(x) = p(-x)$). Using this symmetry and the integral identities shown above, we can compute all the integrals involved in the leading order of the variance, except for the spatial integral involving the propagator cubed. We denote this integral by $\Theta$:
\begin{equation}
    \Theta(u,v,w) = \int \dd x u^{-\frac{d_s}{2}} p\left(x u^{-\frac{d_s}{2d}}\right) v^{-\frac{d_s}{2}} p\left(x v^{-\frac{d_s}{2d}}\right) w^{-\frac{d_s}{2}} p\left(x w^{-\frac{d_s}{2d}}\right),
    \label{eq 57}
\end{equation}
and rewrite all the terms using this function. 

All these terms involve the multiplicative $4g_0^4/t^{4-d_s}$ factor which we temporarily suppress, and restore in the final formula. 
After replacing the temporal sums by integrals, the first ($3-1-1$) term becomes:
\begin{multline}
   \int_0^1 \dfrac{\dd u}{u^{2 d_s - 2}} \int_0^{1-u} \dfrac{\dd v}{v^{\frac{d_s}{2}}} \iiint \dd x^* \dd y^* \dd z^* p\left(z^* u^{-\frac{d_s}{2d}}\right) g_4\left(x^* u^{-\frac{d_s}{2d}},(y^*-x^*)u^{-\frac{d_s}{2d}},(x^*-y^*)u^{-\frac{d_s}{2d}}\right)\\
    \left[p\left((y^*-x^*)v^{-\frac{d_s}{2d}}\right)-p\left((y^*-z^*)v^{-\frac{d_s}{2d}}\right)\right] \\
    = 
    \int_0^1 \dfrac{\dd u}{u^{2 d_s - 2}} \int_0^{1-u} \dfrac{\dd v}{v^{\frac{d_s}{2}}} \iint \dd x^* \dd y^*  g_4\left(x^* u^{-\frac{d_s}{2d}},(y^*-x^*)u^{-\frac{d_s}{2d}},(x^*-y^*)u^{-\frac{d_s}{2d}}\right)\\
    \left[ p\left((y^*-x^*)v^{-\frac{d_s}{2d}}\right) \int \dd z^* p\left(z^* u^{-\frac{d_s}{2d}}\right) - \int \dd z^* p\left(z^* u^{-\frac{d_s}{2d}}\right) p\left((y^*-z^*)v^{-\frac{d_s}{2d}}\right) \right].
\end{multline}
Integrating over $z^*$ we obtain:
\begin{multline}
   \int_0^1 \dfrac{\dd u}{u^{\frac{3d_s}{2} - 2}} \int_0^{1-u} \dd v \iint \dd x^* \dd y^*  g_4\left(x^* u^{-\frac{d_s}{2d}},(y^*-x^*)u^{-\frac{d_s}{2d}},(x^*-y^*)u^{-\frac{d_s}{2d}}\right)\\
    \left[ \dfrac{1}{v^{\frac{d_s}{2}}} p\left((y^*-x^*)v^{-\frac{d_s}{2d}}\right)  - \dfrac{1}{(u+v)^{\frac{d_s}{2}}} p\left(y^* (u+v)^{-\frac{d_s}{2d}}\right) \right].
\end{multline}
Recalling the explicit formula Eq.~\eqref{eq 47} for $g_4$, and redefining $y^* = y^* -x^*$ gives:
\begin{multline}
    \int_0^1 \dfrac{\dd u}{u^{\frac{3d_s}{2} - 2}} \int_0^{1-u} \dd v \int_0^1 \dd a \int_0^a \dfrac{\dd b}{[b(a-b)(1-a)]^{\frac{d_s}{2}}} \int \dd y^* p\left(y^* (a-b)^{-\frac{d_s}{2d}}u^{-\frac{d_s}{2d}}\right) p\left(y^* (1-a)^{-\frac{d_s}{2d}}u^{-\frac{d_s}{2d}}\right) \\
    \left[ \dfrac{1}{v^{\frac{d_s}{2}}} p\left(y^*v^{-\frac{d_s}{2d}}\right) \int \dd x^* p\left(x^* b^{-\frac{d_s}{2d}}u^{-\frac{d_s}{2d}}\right) - \int \dd x^* p\left(x^* b^{-\frac{d_s}{2d}}u^{-\frac{d_s}{2d}}\right) \dfrac{1}{(u+v)^{\frac{d_s}{2}}} p\left((y^*+x^*) (u+v)^{-\frac{d_s}{2d}}\right) \right],
\end{multline}
which we integrate over $x^*$ to yield:
\begin{multline}
    \int_0^1 \dfrac{\dd u}{u^{d_s - 2}} \int_0^{1-u} \dd v \int_0^1 \dd a \int_0^a \dfrac{\dd b}{[(a-b)(1-a)]^{\frac{d_s}{2}}} \int \dd y^* p\left(y^* (a-b)^{-\frac{d_s}{2d}}u^{-\frac{d_s}{2d}}\right) p\left(y^* (1-a)^{-\frac{d_s}{2d}}u^{-\frac{d_s}{2d}}\right) \\
    \left[ \dfrac{1}{v^{\frac{d_s}{2}}} p\left(y^*v^{-\frac{d_s}{2d}}\right) -  \dfrac{1}{(u+v+ub)^{\frac{d_s}{2}}} p\left(y^* (u+v+ub)^{-\frac{d_s}{2d}}\right) \right].
\end{multline}
Rewriting this result via the function $\Theta$ defined by Eq.~\eqref{eq 57} we arrive at:
\begin{multline}
   \int_0^1 \dd u \, u^2 \int_0^{1-u} \dd v \int_0^1 \dd a \int_0^a \dd b \left[\Theta(u(a-b),u(1-a),v) - \Theta(u(a-b),u(1-a),u+v+ub)\right]\\
    =  \int_0^1 \dd u  \int_0^{1-u} \dd v \int_0^u \dd a \int_0^a \dd b \left[\Theta(a-b,u-a,v) - \Theta(a-b,u-a,u+v+b)\right]. 
\end{multline}

We rely on similar simplifications in the computations of the $4$ other terms. Often, we can even perform all the integrals when the spatial integrals are all first or second power of the propagator. This is the case for half of the second ($1-3-1$) term. The second term is written as:
\begin{multline}
   \int_0^1 \dfrac{\dd u}{u^{2 d_s - 2}} \int_0^{1-u} \dfrac{\dd v}{v^{\frac{d_s}{2}}} \iiint \dd x^* \dd y^* \dd z^* p\left(x^* u^{-\frac{d_s}{2d}}\right) \left[g_4\left(x^* u^{-\frac{d_s}{2d}},(y^*-x^*)u^{-\frac{d_s}{2d}},(z^*-y^*)u^{-\frac{d_s}{2d}}\right) \right.\\
    \left. + g_4\left(y^* u^{-\frac{d_s}{2d}},(x^*-y^*)u^{-\frac{d_s}{2d}},(z^*-x^*)u^{-\frac{d_s}{2d}}\right)\right] \left[p\left((y^*-x^*)v^{-\frac{d_s}{2d}}\right)-p\left((y^*-z^*)v^{-\frac{d_s}{2d}}\right)\right].
\end{multline}
We treat the two $g_4$ terms separately for clarity. After integrating over $z^*$, the first one simplifies to 
\begin{multline}
    \int_0^1 \dfrac{\dd u}{u^{\frac{3 d_s}{2}-2}}\int_0^{1-u} \dd v \iint \dd x^* \dd y^* p\left(x^* u^{-\frac{d_s}{2d}}\right) \int_0^1 \dd a \int_0^a  \dfrac{\dd b}{[b(a-b)]^{\frac{d_s}{2}}} p\left(x^* b^{-\frac{d_s}{2d}} u^{-\frac{d_s}{2d}}\right) \\
    p\left((y^*-x^*) (a-b)^{-\frac{d_s}{2d}} u^{-\frac{d_s}{2d}}\right) \left[ \dfrac{1}{v^{\frac{d_s}{2}}} p\left((y^*-x^*)v^{-\frac{d_s}{2d}}\right) - \dfrac{g_1(0)}{(v+u -a u)^{\frac{d_s}{2}}} \right].
\end{multline}
Integrating over the difference $y^* -z^*$, we arrive at 
\begin{multline}
    \int_0^1 \dfrac{\dd u}{u^{d_s-2}}\int_0^{1-u} \dd v \int \dd x^*  p\left(x^* u^{-\frac{d_s}{2d}}\right) \int_0^1 \dd a \int_0^a  \dfrac{\dd b}{b^{\frac{d_s}{2}}} p\left(x^* b^{-\frac{d_s}{2d}} u^{-\frac{d_s}{2d}}\right) \left[\dfrac{p(0)}{(v+u a -u b)^{\frac{d_s}{2}}} - \dfrac{p(0)}{(v+u -a u)^{\frac{d_s}{2}}} \right],
\end{multline}
which we integrate over $x^*$ and obtain:
\begin{equation}
    p(0)^2 \int_0^1 \dd u \, u^2 \int_0^{1-u} \dd v \int_0^1 \dd a \int_0^a \dd b \dfrac{1}{(u+b u)^{\frac{d_s}{2}}} \left[\dfrac{1}{(v+u a -u b)^{\frac{d_s}{2}}} - \dfrac{1}{(v+u -a u)^{\frac{d_s}{2}}} \right]. 
\end{equation}
Integrating over $a$ then gives $0$. This cancellation only exists because we restricted ourselves to symmetric scaling limits. For the term involving the other $g_4$ function, we get after integrating over $z^*$:
\begin{multline}
    \int_0^1 \dfrac{\dd u}{u^{\frac{3 d_s}{2}-2}}\int_0^{1-u} \dd v \iint \dd x^* \dd y^* p\left(x^* u^{-\frac{d_s}{2d}}\right) \int_0^1 \dd a \int_0^a \dfrac{\dd b}{[b(a-b)]^{\frac{d_s}{2}}} p\left(y^* b^{-\frac{d_s}{2d}} u^{-\frac{d_s}{2d}}\right) \\
    p\left((x^*-y^*) (a-b)^{-\frac{d_s}{2d}} u^{-\frac{d_s}{2d}}\right) \left[ \dfrac{1}{v^{\frac{d_s}{2}}} p\left((y^*-x^*)v^{-\frac{d_s}{2d}}\right) - \dfrac{1}{(v+u -a u)^{\frac{d_s}{2}}} p\left((y^*-x^*)(v+u-u a)^{-\frac{d_s}{2d}}\right) \right].
\end{multline}
Redefining $y^* = y^* -x^*$ and integrating over $x^*$ gives:
\begin{multline}
    \int_0^1 \dfrac{\dd u}{u^{\frac{d_s}{2}-2}}\int_0^{1-u} \dd v \int_0^1 \dd a \int_0^a \dfrac{\dd b}{(a-b)^{\frac{d_s}{2}}} \int \dd y^*  p\left(y^* u^{-\frac{d_s}{2d}} (a-b)^{\frac{d_s}{2}}\right)\dfrac{1}{(u+ub)^{\frac{d_s}{2}}}  p\left(y^* (u+ub)^{-\frac{d_s}{2d}}\right)\\
    \left[\dfrac{1}{v^{\frac{d_s}{2}}} p\left(y^* v^{-\frac{d_s}{2}}\right) - \dfrac{1}{(v+u-ua)^{\frac{d_s}{2}}} p\left(y^* (v+u-ua)^{-\frac{d_s}{2}}\right) \right].
\end{multline}
Rewriting this result via the function $\Theta$ defined by Eq.~\eqref{eq 57} we arrive at:
\begin{multline}
    \int_0^1 \dd u\, u^2 \int_0^{1-u} \dd v \int_0^1 \dd a \int_0^a \dd b \left[\Theta(ua -ub,u+ub,v) - \Theta(ua-ub,u+ub,v+u-ua)\right]\\
    =  \int_0^1 \dd u \int_0^{1-u} \dd v \int_0^u \dd a \int_0^a \dd b \left[\Theta(a -b,u+b,v) - \Theta(a-b,u+b,v+u-a)\right].
\end{multline}

We continue with the $2-2-1$ term, which takes a similar form:
\begin{multline}
   \int_0^1 \dfrac{\dd u}{u^{2 d_s - 2}} \int_0^{1-u} \dfrac{\dd v}{v^{\frac{d_s}{2}}} \iiint \dd x^* \dd y^* \dd z^* g_3\left(y^* u^{-\frac{d_s}{2d}}, (x^*-y^*) u^{-\frac{d_s}{2d}}\right) \left[g_3\left(x^* u^{-\frac{d_s}{2d}},(z^*-x^*)u^{-\frac{d_s}{2d}}\right) \right.\\
    \left. + g_3\left(y^* u^{-\frac{d_s}{2d}},(z^*-y^*)u^{-\frac{d_s}{2d}}\right)\right] \left[p\left((y^*-x^*)v^{-\frac{d_s}{2d}}\right)-p\left((y^*-z^*)v^{-\frac{d_s}{2d}}\right)\right].
\end{multline}
As for the second term, the product involving $g_3\left(y^* u^{-\frac{d_s}{2d}},(z^*-y^*)u^{-\frac{d_s}{2d}}\right)$ vanishes because it is antisymmetric with respect to the exchange $x^* \leftrightarrow z^*$. We are therefore left with:
\begin{multline}
     \int_0^1 \dfrac{\dd u}{u^{2 d_s - 2}} \int_0^{1-u} \dfrac{\dd v}{v^{\frac{d_s}{2}}} \iiint \dd x^* \dd y^* \dd z^* g_3\left(y^* u^{-\frac{d_s}{2d}}, (x^*-y^*) u^{-\frac{d_s}{2d}}\right) g_3\left(x^* u^{-\frac{d_s}{2d}},(z^*-x^*)u^{-\frac{d_s}{2d}}\right) \\
    \left[p\left((y^*-x^*)v^{-\frac{d_s}{2d}}\right)-p\left((y^*-z^*)v^{-\frac{d_s}{2d}}\right)\right].
\end{multline}
Inserting the definition of $g_3$ and integrating over $z^*$ gives
\begin{multline}
     \int_0^1 \dfrac{\dd u}{u^{\frac{3 d_s}{2}-2}}\int_0^{1-u} \dd v \iint \dd x^* \dd y^* \int_0^1 \dfrac{\dd a}{[a(1-a)]^{\frac{d_s}{2}}} \int_0^1 \dfrac{\dd b}{[b(1-b)]^{\frac{d_s}{2}}}  p\left(y^* u^{-\frac{d_s}{2d}} a^{-\frac{d_s}{2d}} \right) p\left((x^*-y^*) (1-a)^{-\frac{d_s}{2d}} u^{-\frac{d_s}{2d}}\right) \\
    p\left(x^* b^{-\frac{d_s}{2d}} u^{-\frac{d_s}{2d}}\right) \left[ \dfrac{1}{v^{\frac{d_s}{2}}} p\left((y^*-x^*)v^{-\frac{d_s}{2d}}\right) - \dfrac{1}{(v+u -b u)^{\frac{d_s}{2}}} p\left((y^*-x^*)(v+u-bu)^{-\frac{d_s}{2d}}\right)\right],
\end{multline}
which simplifies to 
\begin{multline}
     \int_0^1 \dfrac{\dd u}{u^{\frac{d_s}{2}-2}}\int_0^{1-u} \dd v \int_0^1 \dfrac{\dd a}{(1-a)^{\frac{d_s}{2}}} \int_0^1 \dd b \int \dd y^*  p\left(y^* u^{-\frac{d_s}{2d}} (1-a)^{\frac{d_s}{2}}\right)\dfrac{1}{(ua+ub)^{\frac{d_s}{2}}}  p\left(y^* (ua+ub)^{-\frac{d_s}{2d}}\right)\\
    \left[\dfrac{1}{v^{\frac{d_s}{2}}} p\left(y^* v^{-\frac{d_s}{2}}\right) - \dfrac{1}{(v+u-ub)^{\frac{d_s}{2}}} p\left(y^* (v+u-ub)^{-\frac{d_s}{2}}\right) \right]
\end{multline}
after integrating over $x^*$. Equivalently,
\begin{multline}
     \int_0^1 \dd u\, u^2 \int_0^{1-u} \dd v \int_0^1 \dd a \int_0^1 \dd b \left[\Theta(ua -ub,u+ub,v) - \Theta(ua-ub,u+ub,v+u-ua)\right]\\
    =  \int_0^1 \dd u \int_0^{1-u} \dd v \int_0^u \dd a \int_0^u \dd b \left[\Theta(u-a,a+b,v) - \Theta(u-a,a+b,v+u-b)\right],
\end{multline}
in terms of $\Theta$. 

The last two terms involve the $g_2$ corrections to the survival. The $1-2-2$ term reads:
\begin{multline}
    - \int_0^1 \dfrac{\dd u}{u^{\frac{3d_s}{2} - 1}} \int_0^{1-u} \dfrac{\dd v}{v^{d_s-1}} \iiint \dd x^* \dd y^* \dd z^* p\left(x^* u^{-\frac{d_s}{2d}}\right) \left[ g_3\left(x^* u^{-\frac{d_s}{2d}}, (z^*-x^*) u^{-\frac{d_s}{2d}}\right) \right. \\
    \left. + g_3\left(y^* u^{-\frac{d_s}{2d}},(z^*-y^*)u^{-\frac{d_s}{2d}}\right) \right] \left[p\left((y^*-z^*)v^{-\frac{d_s}{2d}}\right)g_2\left((y^*-x^*)v^{-\frac{d_s}{2d}}\right)-p\left((y^*-x^*)v^{-\frac{d_s}{2d}}\right)g_2\left((y^*-z^*)v^{-\frac{d_s}{2d}}\right)\right].
\end{multline}
Integrating over $z^*$ yields:
\begin{multline}
    - \int_0^1 \dfrac{\dd u}{u^{d_s-1}}\int_0^{1-u} \dfrac{\dd v}{v^{\frac{d_s}{2}-1}} \iint \dd x^* \dd y^* \int_0^1 \dfrac{\dd a}{a^{\frac{d_s}{2}}} \int_0^1 \dd b  p\left(x^* u^{-\frac{d_s}{2d}}\right) \\
    \left[ p\left(x^* (au)^{-\frac{d_s}{2d}}\right) \dfrac{1}{b^{\frac{d_s}{2d}}} p\left((y^*-x^*)(bv)^{-\frac{d_s}{2d}}\right) \dfrac{1}{(u-ua+v)^{\frac{d_s}{2}}} p\left((y^*-x^*)(u-ua+v)^{-\frac{d_s}{2d}}\right)\right. \\
    \left. - p\left(x^* (au)^{-\frac{d_s}{2d}}\right) p\left((y^*-x^*)v^{-\frac{d_s}{2d}}\right) \dfrac{1}{(u-ua+vb)^{\frac{d_s}{2}}} p\left((y^*-x^*)(u-ua+vb)^{-\frac{d_s}{2d}}\right) \right. \\
    \left. + p\left(y^* (au)^{-\frac{d_s}{2d}}\right) \dfrac{1}{b^{\frac{d_s}{2d}}} p\left((y^*-x^*)(bv)^{-\frac{d_s}{2d}}\right) \dfrac{p(0)}{(u-ua+v)^{\frac{d_s}{2}}} - p\left(y^* (au)^{-\frac{d_s}{2d}}\right)  p\left((y^*-x^*)v^{-\frac{d_s}{2d}}\right) \dfrac{p(0)}{(u-ua+vb)^{\frac{d_s}{2}}}\right].
\end{multline}
which we integrate over $x^*$ (we also redefine $y^* = y^* -x^*$ in the first $2$ terms) to give: 
\begin{multline}
    -\int_0^1 \dd u \, u \int_0^{1-u} \dd v\, v \int_0^1 \dd a \int_0^1 \dd b \int \dd y^* \left[ \dfrac{p(0)}{(u+a u)^{\frac{d_s}{2}}} \dfrac{1}{(u-ua+v)^{\frac{d_s}{2}}} p\left(y^*(u-ua+v)^{\frac{d_s}{2}}\right) \dfrac{1}{(v b)^{\frac{d_s}{2}}} p\left(y^*(v b)^{\frac{d_s}{2}}\right) \right.\\
    \left. - \dfrac{p(0)}{(u+a u)^{\frac{d_s}{2}}} \dfrac{1}{(u-ua+v b)^{\frac{d_s}{2}}} p\left(y^*(u-ua+v b)^{\frac{d_s}{2}}\right) \dfrac{1}{v^{\frac{d_s}{2}}} p\left(y^*v^{\frac{d_s}{2}}\right) + \dfrac{p(0)}{(u-a u+v)^{\frac{d_s}{2}}} \dfrac{1}{(u a)^{\frac{d_s}{2}}} p\left(y^*(u a)^{\frac{d_s}{2}}\right) \right. \\
    \left. \dfrac{1}{(u+v b)^{\frac{d_s}{2}}} p\left(y^*(u+v b)^{\frac{d_s}{2}}\right) - \dfrac{p(0)}{(u-a u+bv)^{\frac{d_s}{2}}} \dfrac{1}{(u a)^{\frac{d_s}{2}}} p\left(y^*(u a)^{\frac{d_s}{2}}\right) \dfrac{1}{(u+v )^{\frac{d_s}{2}}} p\left(y^*(u+v )^{\frac{d_s}{2}}\right) \right].
\end{multline}
Performing the last spatial integration yields:
\begin{multline}
    -p(0)^2 \int_0^1 \dd u \, u \int_0^{1-u} \dd v\, v \int_0^1 \dd a \int_0^1 \dd b \left[\dfrac{1}{(u+au)^{\frac{d_s}{2}}(u-ua+v+v b)^{\frac{d_s}{2}}} - \dfrac{1}{(u+au)^{\frac{d_s}{2}}(u-ua+v+v b)^{\frac{d_s}{2}}} \right. \\
    \left. + \dfrac{1}{(u-au+v)^{\frac{d_s}{2}}(u+ua+v b)^{\frac{d_s}{2}}} - \dfrac{1}{(u-au+vb)^{\frac{d_s}{2}}(u+v+ua)^{\frac{d_s}{2}}}\right].
\end{multline}
The first $2$ terms cancel out, and we arrive at:
\begin{multline}
    p(0)^2 \int_0^1 \dd u \, u \int_0^{1-u} \dd v\, v \int_0^1 \dd a \int_0^1 \dd b \left[\dfrac{1}{(u-au+vb)^{\frac{d_s}{2}}(u+ua+v )^{\frac{d_s}{2}}} - \dfrac{1}{(u-au+v)^{\frac{d_s}{2}}(u+vb+ua)^{\frac{d_s}{2}}}\right].
\end{multline}

For the last $2-1-2$ term, we have: 
\begin{multline}
    \int_0^1 \dfrac{\dd u}{u^{d_s-1}} \int_0^{1-u} \dfrac{\dd v}{v^{d_s-1}} \iint \dd x^* \dd y^* \int_0^1 \dfrac{\dd a}{[a(1-a)]^{\frac{d_s}{2}}} p\left(y^*(ua)^{\frac{d_s}{2}}\right) p\left((x^*-y^*)(u-ua)^{\frac{d_s}{2}}\right) \int_0^1 \dd b \\
    \left[ p\left((y^*-x^*)v^{\frac{d_s}{2}}\right) \dfrac{1}{(u+b v)^{\frac{d_s}{2}}} p\left(y^*(u+b v)^{\frac{d_s}{2}}\right) - \dfrac{1}{b^{\frac{d_s}{2}}} p\left((y^*-x^*)(bv)^{\frac{d_s}{2}}\right) \dfrac{1}{(u+ v)^{\frac{d_s}{2}}} p\left(y^*(u+bv)^{\frac{d_s}{2}}\right)\right].
\end{multline}
after integration over $z^*$. Integrating over $x^*$ yields:
\begin{multline}
     p(0) \int_0^1 \dfrac{\dd u}{u^{\frac{d_s}{2}-1}} \int_0^{1-u} \dd v \, v \int \dd y^* \int_0^1 \dfrac{\dd a}{a^{\frac{d_s}{2}}} \int_0^1 \dd b p\left(y^*(ua)^{\frac{d_s}{2}}\right)\\
    \left[\dfrac{1}{(u-ua+v)^{\frac{d_s}{2}}}\dfrac{1}{(u+bv)^{\frac{d_s}{2}}} p\left(y^*(u+bv)^{\frac{d_s}{2}}\right) - \dfrac{1}{(u-ua+bv)^{\frac{d_s}{2}}}\dfrac{1}{(u+v)^{\frac{d_s}{2}}} p\left(y^*(u+v)^{\frac{d_s}{2}}\right)  \right]
\end{multline}
which we integrate over $y^*$ and arrive at:
\begin{multline}
    p(0)^2\int_0^1 \dd u \, u \int_0^{1-u} \dd v\, v \int_0^1 \dd a \int_0^1 \dd b \left[\dfrac{1}{(u-au+v)^{\frac{d_s}{2}}(u+ua+v b)^{\frac{d_s}{2}}} - \dfrac{1}{(u-au+bv)^{\frac{d_s}{2}}(u+v+ua)^{\frac{d_s}{2}}}\right].
\end{multline}
As this term is exactly the opposite of the one we obtained previously in the $1-2-2$ product, the only non-zero contributions to the variance come from the quadruple integrals over $\Theta$. This considerable simplification is possible in the symmetric case and should be relaxed for asymmetric universality classes. 

Collecting previous results and restoring the multiplicative $4 t^{4-d_s}/g_0^4$, we arrive at the following expression for the variance:
\begin{tcolorbox}[
colback=white,
colframe=black,
arc=2mm]
\begin{varwidth}{\textwidth}
\begin{multline}
    \langle [F_1(t) - F_2(t)]^2\rangle \underset{t \rightarrow +\infty}{\sim} \dfrac{4}{g_0^4} t^{4-d_s} \int_0^1 \dd u \int_0^{1-u} \dd v \int_0^u \dd a \left\{ \int_0^a \dd b \left[\Theta(a-b,u-a,v) - \Theta(a-b,u-a,u+v+b) \right. \right.\\
    \left. \left. + \Theta(a-b,u+b,v) - \Theta(a-b,u+b,u+v-a) \right] + \int_0^u \dd b \left[\Theta(u-a,a+b,v) - \Theta(u-a,a+b,v+u-b)\right]\right\}.
    \label{eq 84}
\end{multline}
\end{varwidth}
\end{tcolorbox}

\subsubsection{Exact formula for $\Var[X_n]$ in the intermediate regime}

Inserting the leading asymptotics for the mean range $\langle \mathcal{N}_1(t) \rangle \sim \frac{t}{g_0}$ and Eq.~\eqref{eq 84} into Eq.~\eqref{eq 36 prime}, we get for the variance of the fraction at fixed time:
\begin{multline}
    \Var[X_t] \underset{t \rightarrow +\infty}{\sim} \dfrac{1}{4g_0^2} t^{2-d_s} \int_0^1 \dd u \int_0^{1-u} \dd v \int_0^u \dd a \left\{ \int_0^a \dd b \left[\Theta(a-b,u-a,v) - \Theta(a-b,u-a,u+v+b) \right. \right.\\
    \left. \left. + \Theta(a-b,u+b,v) - \Theta(a-b,u+b,u+v-a) \right] + \int_0^u \dd b \left[\Theta(u-a,a+b,v) - \Theta(u-a,a+b,v+u-b)\right]\right\}.
\end{multline}
We finally transfer this to the case of the random stopping time $t_n$. Because of self-averaging, $t_n$ is sharply peaked around its mean value $\frac{g_0 n}{2}$. We therefore obtain the final result for the variance of $X_n$:
\begin{tcolorbox}[
colback=white,
colframe=black,
arc=2mm]
\begin{varwidth}{\textwidth}
\begin{multline}
    \Var[X_n] \underset{n \rightarrow +\infty}{\sim} \dfrac{2^{d_s}}{16 g_0^{d_s}} n^{2-d_s} \int_0^1 \dd u \int_0^{1-u} \dd v \int_0^u \dd a \left\{ \int_0^a \dd b \left[\Theta(a-b,u-a,v) - \Theta(a-b,u-a,u+v+b) \right. \right.\\
    \left. \left. + \Theta(a-b,u+b,v) - \Theta(a-b,u+b,u+v-a) \right] + \int_0^u \dd b \left[\Theta(u-a,a+b,v) - \Theta(u-a,a+b,v+u-b)\right]\right\}.
    \label{eq 86}
\end{multline}
\end{varwidth}
\end{tcolorbox}
In this expression, the integral is given by the scaling limit of the walk (since $\Theta$ is expressed in term of the scaling function of the propagator) but the prefactor of the variance still depends on the microscopic details of the walk through $g_0$.

\subsubsection{Explicit characterization for Markovian universality classes}

One would like compute the integral appearing in Eq.~\eqref{eq 86}, or at least express it in a nicer way allowing for numerical evaluation. Because the universality classes appearing as scaling limits of Markovian random walks ($d$-dimensional Lévy walks) are naturally defined through Fourier transform, we first express the integral in Fourier space. The convolution identity for the propagator allows to define the Lévy exponent $\psi(k)$ such that:
\begin{equation}
    \int \dd x \dfrac{1}{u^{\frac{d_s}{2}}} p\left(x u^{-\frac{d_s}{2d}}\right) e^{i k \cdot x} = e^{-u \psi(k)}.
    \label{eq 109}
\end{equation}
Because we consider scaling limits of random walks, $p$ must be the propagator of a \emph{stable} Lévy process. In particular, the left-hand side of equation Eq.~\eqref{eq 109} is a function of $ku^{\frac{d_s}{2d}}$. This fixes the Lévy exponent up to a proportionality constant, which is the generalized diffusion coefficient:
\begin{equation}
    \psi(k) = D |k|^{\frac{2d}{d_s}},
\end{equation}
as expected for Lévy walks. We now express $\Theta$ using Fourier transforms:
\begin{equation} \begin{split}
    \Theta(u,v,w) &= \int \dd x \dfrac{1}{(2\pi)^d} \int \dd k e^{-u D |k|^{\frac{2d}{d_s}}} e^{-i k\cdot x} \dfrac{1}{(2\pi)^d} \int \dd k' e^{-v D |k'|^{\frac{2d}{d_s}}} e^{-i k'\cdot x} \dfrac{1}{(2\pi)^d} \int \dd k'' e^{-w D |k''|^{\frac{2d}{d_s}}} e^{-i k''\cdot x} \\
    &= \dfrac{1}{(2\pi)^{2d}} \iiint \dd k \dd k' \dd k'' e^{-u D |k|^{\frac{2d}{d_s}}-v D |k'|^{\frac{2d}{d_s}}-w D |k''|^{\frac{2d}{d_s}}} \delta(k+k'+k'') \\
    &= \dfrac{1}{(2\pi)^{2d}} \iint \dd k \dd k' e^{-u D |k|^{\frac{2d}{d_s}}-v D |k'|^{\frac{2d}{d_s}}-w D |k+k'|^{\frac{2d}{d_s}}}.
\end{split} \end{equation}
This form is especially interesting for further integration over $u,v,w$ since it decorrelates these variables allowing the analytical computations of the integrals. Indeed, plugging this representation into the formula for the variance and doing the $4$ temporal integrals, one obtains a quite lengthy double integral expression for the variance:
\begin{tcolorbox}[
colback=white,
colframe=black,
arc=2mm]
\begin{varwidth}{\textwidth}
\begin{equation} \begin{split}
     \Var[X_n] &\underset{n \rightarrow +\infty}{\sim} \dfrac{2^{d_s}}{16 (2\pi)^{2 d} g_0^{d_s}} n^{2-d_s} \iint \dd k \dd k' \mathcal{I}\left(D|k|^{\frac{2d}{d_s}},D|k'|^{\frac{2d}{d_s}},D|k+k'|^{\frac{2d}{d_s}}\right) \\
     &\underset{n \rightarrow +\infty}{\sim} \dfrac{2^{d_s}}{16 (2\pi)^{2 d} D^{d_s} g_0^{d_s}} \iint \dd k \dd k' \mathcal{I}\left(|k|^{\frac{2d}{d_s}},|k'|^{\frac{2d}{d_s}},|k+k'|^{\frac{2d}{d_s}}\right)
\end{split} \end{equation}
\end{varwidth}
\end{tcolorbox}
The integrand $\mathcal{I}$ is easily obtained with for example Mathematica and can be used for numerical evaluation. It is given by:
\begin{multline}
    \mathcal{I}(x,y,z) = \frac{1}{2} \left(-\frac{2 e^{-x}}{x^2 y (x-z)}-\frac{1}{x^2 (x+y) (x+z)}-\frac{1}{x^2 y z}-\frac{2 e^{-2 y}
   z}{y^2 (y-x) \left(2 y^2-3 y z+z^2\right)}+\frac{2 e^{-y}}{x y^2 (y-z)} \right. \\
   \left. +\frac{1}{y^2 (x+y)
   (y+z)}-\frac{1}{x y^2 z}+\frac{2 e^{-z} \left(\frac{2 (x-2 y)}{(x-y) (z-2 y)}+\frac{2 y}{(y-x)
   (x+y-z)}-\frac{x+y}{x z}+\frac{e^{-x} y}{x (x+z)}+\frac{2 y}{z^2}-\frac{e^{-y}}{y+z}\right)}{y (x-z)
   (y-z)}-\frac{3}{x y z^2} \right. \\
   \left. +\frac{e^{-2 z}}{z^2 (x-z) (z-y)}-\frac{2 z e^{-x-y}}{x (x-y) (x+y) (x-z)
   (x+y-z)}+\frac{2}{x y z}\right)+\frac{z e^{-x-y}}{y \left(y^2-x^2\right) (y-z) (x+y-z)}.
\end{multline}
We can simplify the expression for the variance further if we restrict ourselves to the one-dimensional case. Decomposing the integral to explicit the absolute values, we obtain $3$ contributions:
\begin{multline}
    \iint \dd k \dd k' \mathcal{I}\left(|k|^{\frac{2}{d_s}},|k'|^{\frac{2}{d_s}},|k+k'|^{\frac{2}{d_s}}\right) = 2 \int_0^{+\infty} \dd k \int_0^{+\infty} \dd k' \mathcal{I}\left(k^{\frac{2}{d_s}},k'^{\frac{2}{d_s}},(k+k')^{\frac{2}{d_s}}\right)\\
    + 2 \int_0^{+\infty} \dd k \int_k^{+\infty} \dd k' \mathcal{I}\left(k^{\frac{2}{d_s}},k'^{\frac{2}{d_s}},(k'-k)^{\frac{2}{d_s}}\right) + 2 \int_0^{+\infty} \dd k \int_0^k \dd k' \mathcal{I}\left(k^{\frac{2}{d_s}},k'^{\frac{2}{d_s}},(k-k')^{\frac{2}{d_s}}\right) \\
    = 2 \int_0^{+\infty} \dd k \int_0^{+\infty} \dd k' \left\{ \mathcal{I}\left(k^{\frac{2}{d_s}},k'^{\frac{2}{d_s}},(k+k')^{\frac{2}{d_s}}\right) + \mathcal{I}\left(k^{\frac{2}{d_s}},(k+k')^{\frac{2}{d_s}},k'^{\frac{2}{d_s}}\right) + \mathcal{I}\left((k+k')^{\frac{2}{d_s}},k'^{\frac{2}{d_s}},k^{\frac{2}{d_s}}\right) \right\}
\end{multline}
To exploit symmetries, we explicit the factor of $2$ as a sum and swap $k$ and $k'$ in one of the integrals:
\begin{equation}
    \iint \dd k \dd k' \mathcal{I}\left(|k|^{\frac{2}{d_s}},|k'|^{\frac{2}{d_s}},|k+k'|^{\frac{2}{d_s}}\right) = \int_0^{+\infty} \dd k \int_0^{+\infty} \dd k'  \mathcal{I}_{\text{sym}} \left(k^{\frac{2}{d_s}},k'^{\frac{2}{d_s}},(k+k')^{\frac{2}{d_s}}\right).
\end{equation}
The integrand $\mathcal{I}_{\text{sym}}$ has a slightly simpler expression compared to $\mathcal{I}$:
\begin{eqnarray}
\mathcal{I}_{\text{sym}}(x,y,z) &=& \frac{6x y z-5(xy+yz+ zx)}{x^2 y^2 z^2}+\frac{16 x^{-2}e^{-x}}{(x-2 y) (x-2 z)}
+\frac{16 y^{-2} e^{-y}}{(y-2 x) (y-2z)} + \frac{16 z^{-2} e^{-z}}{(z-2 x)(z-2 y)}\\
&+&\frac{e^{-2 x} (y z/x^2-4)}{(y-2 x)(y-x)(z-2 x)(z-x)}+\frac{e^{-2 y} (xz/y^2-4)}{(x-2 y) (x-y)(z-2 y) (z-y)}\\
&+&\frac{e^{-2 z} (x y/z^2-4)}{(x-2 z) (x-z) (y-2 z) (y-z)}\,.
\end{eqnarray}
We now rescale $k'$ with $k$ introducing the ratio $r=\frac{k'}{k}$ and we set $u=k^{\frac{2}{d_s}}$. This gives:
\begin{equation}
    \iint \dd k \dd k' \mathcal{I}\left(|k|^{\frac{2}{d_s}},|k'|^{\frac{2}{d_s}},|k+k'|^{\frac{2}{d_s}}\right) = \dfrac{d_s}{2} \int_0^{+\infty} \int_0^{+\infty} \dd u \, u^{d_s-1}  \mathcal{I}_{\text{sym}}\left(u,u r^{\frac{2}{d_s}},u(1+r)^{\frac{2}{d_s}}\right).
\end{equation}
Integrating over $u$, we obtain:
\begin{multline}
     \iint \dd k \dd k' \mathcal{I}\left(|k|^{\frac{2}{d_s}},|k'|^{\frac{2}{d_s}},|k+k'|^{\frac{2}{d_s}}\right) = \dfrac{8 \pi d_s \csc(\pi d_s)}{\Gamma(5-d_s)} \int_0^{+\infty} \dd r \left\{\frac{r^{\frac{4}{d_s}-2}}{\left(r^{\frac{2}{d_s}}-2\right) \left(r^{\frac{2}{d_s}}-2 (r+1)^{\frac{2}{d_s}}\right)} \right. \\
   +\frac{2^{-d_s} \left((r
   (r+1))^{\frac{2}{d_s}}-4\right)}{\left(r^{\frac{2}{d_s}}-2\right) \left(r^{\frac{2}{d_s}}-1\right)
   \left((r+1)^{\frac{2}{d_s}}-2\right)
   \left((r+1)^{\frac{2}{d_s}}-1\right)}+\frac{(r+1)^{\frac{4}{d_s}-2}}{\left((r+1)^{\frac{2}{d_s}}-2\right)
   \left((r+1)^{\frac{2}{d_s}}-2 r^{\frac{2}{d_s}}\right)} \\
   +\frac{1}{\left(2 r^{\frac{2}{d_s}}-1\right)
   \left(2 (r+1)^{\frac{2}{d_s}}-1\right)}+\frac{2^{-d_s} \left((r+1)^{\frac{2}{d_s}}-4
   r^{\frac{4}{d_s}}\right) r^{\frac{4}{d_s}-2}}{\left(r^{\frac{2}{d_s}}-1\right) \left(2
   r^{\frac{2}{d_s}}-1\right) \left(r^{\frac{2}{d_s}}-(r+1)^{\frac{2}{d_s}}\right) \left(2
   r^{\frac{2}{d_s}}-(r+1)^{\frac{2}{d_s}}\right)}\\
   \left.+ \frac{2^{-d_s} (r+1)^{\frac{4}{d_s}-2} \left(r^{\frac{2}{d_s}}-4
   (r+1)^{\frac{4}{d_s}}\right)}{\left(r^{\frac{2}{d_s}}-2 (r+1)^{\frac{2}{d_s}}\right)
   \left(r^{\frac{2}{d_s}}-(r+1)^{\frac{2}{d_s}}\right) \left((r+1)^{\frac{2}{d_s}}-1\right) \left(2
   (r+1)^{\frac{2}{d_s}}-1\right)} \right\}.
   \label{eq 119}
\end{multline}
The above integral can be straightforwardly evaluated numerically. It diverges for $d_s \rightarrow 3$, where the variance is no longer dominated by scaling terms and where the scaling is not $t$ but $t\log{t}$. To recover the variance prefactor for a given walk, one then simply needs to compute the generalized diffusion coefficient $D$ (related to the large time length scale of the walk) and the microscopic constant $g_0$.

\subsection{Marginal regime $d_s=2$}

\subsubsection{Logarithmic asymptotic expansion}

In the marginal case, the asymptotic expansion of the variance is in inverse powers of logarithms. The expression Eq.~\eqref{eq 43} is still valid but individual expansions of the quantities involved in it will slightly differ. As before, we need the behaviors of $G(x,\xi)$ and $G_0(\xi)$ in the limit $\xi \rightarrow 1$. For $G$, nothing changes compared to the transient case, but now since $d_s=2$, we have:
\begin{equation}
    G(x,\xi) \sim \mathcal{G}\left(x (1-\xi)^{\frac{1}{d}}\right).
\end{equation}
For $G_0$ however, there is now a logarithmic divergence as $\xi \rightarrow 1$, with the explicit asymptotic expansion:
\begin{equation}
    G_0(\xi) \underset{\xi \rightarrow 1}{\sim} - p(0) \log(1-\xi) + c + o(1),
\end{equation}
where $c$ depends on the microscopic details of the walk:
\begin{equation}
    c = 1+ \sum_{n=1}^{+\infty} \left(P_n(0|0) -  \dfrac{p(0)}{n}\right).
\end{equation}
Keeping the same notations for rescaled spatial variables but now denoting $\epsilon = \dfrac{1}{-p(0)\log(1-\xi)}$, we get:
\begin{equation}
    F(x,\xi) \sim \mathcal{G}(x^*) \left[\epsilon -c\epsilon^2+ c^2 \epsilon^3\right],
\end{equation}
\begin{equation}
    S(x,\xi) \sim \dfrac{1}{1-\xi} \left\{ 1-\mathcal{G}(x^*)\left[\epsilon -c\epsilon^2\right]\right\},
\end{equation}
\begin{eqnarray}
    F^{\{y\}}(x,\xi) &\sim& \epsilon\, \mathcal{G}(x^*) - \epsilon^2\left[\mathcal{G}(y^*)\mathcal{G}(x^*-y^*) 
    + c\mathcal{G}(x^*)\right] \nonumber \\
    &+&\epsilon^3\left[\mathcal{G}(x^*)\mathcal{G}(y^*-x^*)\mathcal{G}(x^*-y^*) + 2c \mathcal{G}(y^*) \mathcal{G}(x^*-y^*)+c^2\mathcal{G}(x^*)\right],
\end{eqnarray}
\begin{multline}
    P^{\{x,y\}}(z,\xi) \sim \mathcal{G}(z^*)  - \epsilon \left[\mathcal{G}(x^*)\mathcal{G}(z^*-x^*) +\mathcal{G}(y^*)\mathcal{G}(z^*-y^*)\right] \\
    + \epsilon^2\left[\mathcal{G}(y^*)\mathcal{G}(x^*-y^*)\mathcal{G}(z^*-x^*) + c\mathcal{G}(x^*)\mathcal{G}(z^*-x^*) + \mathcal{G}(x^*)\mathcal{G}(y^*-x^*)\mathcal{G}(z^*-y^*) + c\mathcal{G}(y^*)\mathcal{G}(z^*-y^*) \right],
\end{multline}
where we kept the $3$ leading terms in each case. Compared to the case $2<d_s<3$, the expansion seems slightly more complicated because of the extra $c$ terms. But we expect these terms to fully cancel out in the prefactor of the variance, as this prefactor should only depend on the scaling limit of the walk. As going from these Laplace-transformed expansions back to real space is quite technical, we detail the inversion for $F(x,\xi)$. From the definition of $\mathcal{G}$ in the case $d_s=2$, we first see that $\mathcal{L}^{-1}\{\mathcal{G}(x^*)\} = \frac{1}{t}p\left(x t^{-\frac{1}{d}}\right)$, where $p$ is the scaling function for the propagator. We therefore express $F(x,t)$ as a convolution:
\begin{equation}
    F(x,t) = -\int_0^t \dd \tau \, \frac{p\!\left(x \tau^{-\frac{1}{d}}\right)}{p(0)\,\tau}\, 
    \mathcal{L}^{-1}\left\{\dfrac{1}{\log s} + \dfrac{\kappa}{(\log s)^2} + \dfrac{\kappa^2}{(\log s)^3}\right\}(t-\tau).
    \label{eq 104}
\end{equation}
Hereinafter, we often use a shorthand notation $\kappa = c/p(0)$. We will also use the known \cite{henyey_number_1982, wijland_statistical_1997} trick to compute the asymptotics of inverse Laplace transforms of inverse logarithms:
\begin{equation} 
\begin{split}
    \mathcal{L}^{-1}\left\{\dfrac{1}{s^n (\log s)^k}\right\} &= \mathcal{L}^{-1}\left\{\dfrac{(-1)^k}{\Gamma(k)} \int_0^{+\infty} \dd u \, u^{k-1} \, s^{u-n} \right\} = \dfrac{(-1)^k t^{n-1}}{\Gamma(k)}\int_0^{\infty} \dd u \dfrac{u^{k-1}}{t^u \Gamma(n-u)}\\
    & \underset{v= u \log t}{=} \dfrac{(-1)^k t^{n-1}}{\Gamma(k) (\log t)^k} \int_0^{+\infty} \dd v \dfrac{v^{k-1} e^{-v}}{\Gamma\left(n-\frac{v}{\log t}\right)} \\
    &\sim \dfrac{(-1)^k t^{n-1}}{\Gamma(k) (\log t)^k} \int_0^{\infty} \dd v \, v^{k-1} \, e^{-v} \sum_{l\geq 0}\dfrac{1}{l!}\dfrac{d^l}{d\, x^l}\dfrac{1}{\Gamma(x)}\bigg\vert_{x=n} \left(-\dfrac{v}{\log t}\right)^l \\
 &\sim \dfrac{(-1)^k t^{n-1}}{\Gamma(k) (\log t)^k} \sum_{l\geq 0}\dfrac{(-1)^l(k+l-1)!}{l! (\log t)^l}\dfrac{d^l}{d\, x^l}\dfrac{1}{\Gamma(x)}\bigg\vert_{x=n} \\
 &= \dfrac{t^{n-1}}{ (\log t)^k}\sum_{l\geq 0} \dfrac{(-1)^{k+l}}{(\log t)^l} \binom{k+l-1}{l} \dfrac{d^l}{d\, x^l}\dfrac{1}{\Gamma(x)}\bigg\vert_{x=n}.
    \label{eq 134}
\end{split} 
\end{equation}
To find the asymptotic of the inverse Laplace transform appearing in Eq.~\eqref{eq 104}, we use Eq.~\eqref{eq 134} with $n=0$ and $k=1,2,3$ together with the asymptotic expansion of the gamma function:
\begin{equation}
\Gamma(1+x) = 1 - \gamma x + \frac{\pi^2 + 6\gamma^2}{12}\,x^2 + O(x^3),
\end{equation}
where $\gamma=0.577\,215\ldots$ is the Euler constant. In our case with $n=0$, this gives:
\begin{eqnarray}
    \mathcal{L}^{-1}\left\{\dfrac{1}{\log s} + \dfrac{\kappa}{(\log s)^2} + \dfrac{\kappa^2}{(\log s)^3}\right\} 
    &\sim& \dfrac{1}{t (\log t)^2}- \dfrac{2\gamma}{t (\log t)^3} - \dfrac{\pi^2 -6 \gamma^2}{2t (\log t)^4} -\dfrac{2\kappa}{t (\log t)^3} + \dfrac{6\gamma\kappa}{t (\log t)^4}   + \dfrac{3 \kappa^2}{t (\log t)^4}  \nonumber \\
    &=& \dfrac{1}{t (\log t)^2} - \dfrac{2\gamma  +2c\kappa}{t (\log t)^3} + \dfrac{6\kappa^2 + 12 \gamma \kappa- \pi^2 + 6 \gamma^2}{2 t (\log t)^4},
\end{eqnarray}
leading to:
\begin{equation}
    F(x,t) \sim \int_0^t \dd \tau \, \dfrac{p\!\left(x \tau^{-\frac{1}{d}}\right) }{p(0)\tau(t-\tau)} \left[- \dfrac{1}{ \log(t-\tau)^2} + \dfrac{2\gamma  +2c\kappa}{\log(t-\tau)^3} - \dfrac{6\kappa^2 + 12 \gamma \kappa- \pi^2 + 6 \gamma^2}{2  \log(t-\tau)^4} \right].
    \label{eq 107}
\end{equation}
for $F(x,t)$. We have to regularize this integral, since the inverse logarithms diverge when $t-\tau =1$. In the scaling limit, this will cause a divergence when $u=\frac{\tau}{t}=1$. The trick is to rewrite Eq.~\eqref{eq 107} in a seemingly more complicated way bu singling out the mass at $t=\tau$:
\begin{multline}
    \int_0^t \dd \tau \, \left[\frac{1}{\tau} p\left(x \tau^{-\frac{1}{d}}\right) - \dfrac{1}{t}p\left(x t^{-\frac{1}{d}}\right) \right] \dfrac{1}{p(0)(t-\tau)} 
    \left[- \dfrac{1}{ \log(t-\tau)^2} + \dfrac{2\gamma  +2c\kappa}{\log(t-\tau)^3} - \dfrac{6\kappa^2 + 12 \gamma \kappa- \pi^2 + 6 \gamma^2}{2  \log(t-\tau)^4} \right] \\
    + \dfrac{1}{t}p\left(x t^{-\frac{1}{d}}\right) \int_0^t \dd \tau \, \dfrac{1}{p(0)(t-\tau)} 
    \left[- \dfrac{1}{ \log(t-\tau)^2} + \dfrac{2\gamma  +2c\kappa}{\log(t-\tau)^3} - \dfrac{6\kappa^2 + 12 \gamma \kappa- \pi^2 + 6 \gamma^2}{2  \log(t-\tau)^4} \right],
\end{multline}
and recognize the second as a primitive with explicit Laplace transform. This term becomes:
\begin{equation}
 -\dfrac{1}{p(0)t}p\left(x t^{-\frac{1}{d}}\right) \mathcal{L}^{-1}\left\{ \dfrac{1}{s \log s} + \dfrac{\kappa}{s (\log s)^2} + \dfrac{\kappa^2}{s(\log s)^3}\right\}(t).
\end{equation}
Now we set $\tau = u t$ and expand the inverse powers of logs in series of $\frac{1}{\log{t}}$. We also use Eq.~\eqref{eq 134} with $n=1$. This gives:
\begin{multline}
    p(0)F(x,t) \sim \int_0^1 \dd u \left[\dfrac{1}{u}p\left(x t^{-\frac{1}{d}} u^{-\frac{1}{d}}\right) - p\left(x t^{-\frac{1}{d}}\right)\right] \dfrac{1}{t(1-u)} \left[-\dfrac{1}{(\log t)^2} + \dfrac{2\log(1-u)}{(\log t)^3}-\dfrac{3\log(1-u)^2}{(\log t)^4} \right. \\
    \left. + 2(\gamma +\kappa)\left(\dfrac{1}{(\log t)^3}-\dfrac{3\log(1-u)}{(\log t)^4}\right) - \dfrac{6\kappa^2 + 12 \gamma \kappa- \pi^2 + 6 \gamma^2}{2 (\log t)^4}\right] \\
    + \dfrac{1}{t}p\left(x t^{-\frac{1}{d}}\right) \left[\dfrac{1}{\log t}-\dfrac{\gamma}{(\log t)^2} + \dfrac{6\gamma^2-\pi^2}{6(\log t)^3} - \dfrac{\kappa}{(\log t)^2}\left(1-\dfrac{2\gamma}{\log t}\right) + \dfrac{\kappa^2}{(\log t)^3 }\right],
\end{multline}
and we see that the leading order is $\mathcal{O}\left(\frac{1}{t\log{t}}\right)$ and we can stop at the third order in inverse logarithms. The resulting expansion reads:
\begin{eqnarray}
p(0)F(x,t) &\sim& \dfrac{1}{t \log t}p\left(x t^{-\frac{1}{d}}\right) + \dfrac{1}{t (\log t)^2}\int_0^1 \dfrac{\dd u}{1-u} \left[p\left(x t^{-\frac{1}{d}}\right) - \dfrac{p\left(x t^{-\frac{1}{d}} u^{-\frac{1}{d}}\right)}{u}\right] - \dfrac{\gamma + \kappa}{ t (\log t)^2} p\left(x t^{-\frac{1}{d}}\right) \nonumber \\
   & -& \dfrac{2}{t (\log t)^3} \int_0^1 \dfrac{\dd u}{1-u} \left[p\left(x t^{-\frac{1}{d}}\right) - \dfrac{p\left(x t^{-\frac{1}{d}} u^{-\frac{1}{d}}\right)}{u}\right] 
   \left[\log(1-u) + \gamma + \kappa\right] \nonumber \\
    &+& \dfrac{6\kappa^2 + 12 \gamma \kappa- \pi^2 + 6 \gamma^2}{6 t (\log t)^3}p\left(x t^{-\frac{1}{d}}\right) 
    + \mathcal{O}\left(\frac{1}{t (\log t)^4}\right).
\end{eqnarray}
The expansion of the survival is easier and takes a similar form:
\begin{eqnarray}
    S(x,t) &\sim & 1 - \dfrac{1}{p(0) \log t}\int_0^1 \dd u \dfrac{1}{u}p\left(x t^{-\frac{1}{d}} u^{-\frac{1}{d}}\right) \nonumber \\
    &+& \dfrac{1}{p(0) (\log t)^2} \int_0^1 \dd u \dfrac{1}{u} p\left(x t^{-\frac{1}{d}} u^{-\frac{1}{d}}\right) \left[\log(1-u) + \gamma + \kappa\right]
    +O[(\log t)^{-3}].
\end{eqnarray}
The expansions can be matched using integration by parts.

For the two-target FPTs, the expansion has a similar form but convolution is now made with convolutions of propagators instead of a single $p$. We obtain:
\begin{multline}
    F^{\{y\}}(x,t) \sim F(x,t) +\dfrac{2}{p(0)^2 t (\log{t})^3} \int_0^1 \dfrac{\dd u}{1-u}\left[\dfrac{1}{u}g_3\left(y t^{-\frac{1}{d}} u^{-\frac{1}{d}},(x-y) t^{-\frac{1}{d}} u^{-\frac{1}{d}}\right) - g_3\left(y t^{-\frac{1}{d}} ,(x-y) t^{-\frac{1}{d}} \right)\right]\\
     - \left( \dfrac{1}{p(0)^2 t (\log{t})^2} - \dfrac{2 \gamma}{p(0)^2 t (\log{t})^3}\right) g_3\left(y t^{-\frac{1}{d}} ,(x-y) t^{-\frac{1}{d}} \right) \\
     + \dfrac{1}{p(0)^3 t (\log{t})^3} \left[2 c g_3\left(y t^{-\frac{1}{d}} ,(x-y) t^{-\frac{1}{d}} \right) + g_4\left(x t^{-\frac{1}{d}} , (y-x) t^{-\frac{1}{d}}, (x-y) t^{-\frac{1}{d}} \right)\right]\\
    \sim \dfrac{1}{p(0) t \log t}p\left(x t^{-\frac{1}{d}}\right) + \dfrac{1}{p(0)t (\log t)^2}\int_0^1 \dfrac{\dd u}{1-u} \left[p\left(x t^{-\frac{1}{d}}\right) - \dfrac{p\left(x t^{-\frac{1}{d}} u^{-\frac{1}{d}}\right)}{u}\right] - \dfrac{\gamma + \kappa}{p(0) t (\log{t})^2} p\left(x t^{-\frac{1}{d}}\right) \\
    - \dfrac{1}{p(0)^2 t (\log{t})^2} g_3\left(y t^{-\frac{1}{d}} ,(x-y) t^{-\frac{1}{d}} \right)
     - \dfrac{2}{p(0) t (\log{t})^3} \int_0^1 \dfrac{\dd u}{1-u} \left[p\left(x t^{-\frac{1}{d}}\right) - \dfrac{p\!\left(x t^{-\frac{1}{d}} u^{-\frac{1}{d}}\right)}{u}\right] \left[\log(1-u) + \gamma  + \kappa\right] \\
    + \dfrac{2 \gamma}{p(0)^2 t (\log t)^3} g_3\left(y t^{-\frac{1}{d}} ,(x-y) t^{-\frac{1}{d}} \right) 
    + \dfrac{6\kappa^2 + 12 \gamma \kappa- \pi^2 + 6 \gamma^2}{6 p(0) t (\log t)^3}p\left(x t^{-\frac{1}{d}}\right)\\
    + \dfrac{2}{p(0)^2 t (\log{t})^3} \int_0^1 \dfrac{\dd u}{1-u}\left[\dfrac{1}{u}\,g_3\!\left(y t^{-\frac{1}{d}} u^{-\frac{1}{d}},(x-y) t^{-\frac{1}{d}} u^{-\frac{1}{d}}\right) - g_3\left(y t^{-\frac{1}{d}} ,(x-y) t^{-\frac{1}{d}} \right)\right] \\
    + \dfrac{1}{p(0)^3 t (\log{t})^3} \left[2 c g_3\left(y t^{-\frac{1}{d}},(x-y) t^{-\frac{1}{d}} \right) + g_4\left(x t^{-\frac{1}{d}} , (y-x) t^{-\frac{1}{d}}, (x-y) t^{-\frac{1}{d}} \right)\right].
\end{multline}
As before, the functions $g_3$ and $g_4$ are the convolutions of the propagator:
\begin{equation} \begin{split}
    & g_3(x,y) = \int_0^1 \dfrac{\dd u}{u(1-u)} p\left(x u^{-\frac{1}{d}}\right) p\left(y (1-u)^{-\frac{1}{d}}\right), \\
    & g_4(x,y,z) = \int_0^1 \dd u \int_0^u \dfrac{\dd v}{v(u-v)(1-v)} p\left(x v^{-\frac{1}{d}}\right) p\left(y (u-v)^{-\frac{1}{d}}\right) p\left(z (1-u)^{-\frac{1}{d}}\right).
\end{split} \end{equation}
In this expansion, there are $3$ terms of order $2$ and $4$ terms of order $3$, compared to one term by order in the transient case. But we expect very few of these terms to actually play a role. Finally for $P^{\{x,y\}}$, we have:
\begin{multline}
    P^{\{x,y\}}(z,t) \sim \dfrac{1}{t}p\left(z t^{-\frac{1}{d}}\right) -\dfrac{1}{p(0) t \log{t}}\left[ g_3\left(x t^{-\frac{1}{d}} ,(z-x) t^{-\frac{1}{d}} \right) + g_3\left(y t^{-\frac{1}{d}} ,(z-y) t^{-\frac{1}{d}} \right)\right] \\
    - \dfrac{1}{p(0)t (\log{t})^2} \left\{\int_0^1 \dfrac{\dd u}{1-u}\left[ g_3\left(x t^{-\frac{1}{d}} ,(z-x) t^{-\frac{1}{d}} \right) - \dfrac{1}{u}g_3\left(x t^{-\frac{1}{d}} u^{-\frac{1}{d}},(z-x) t^{-\frac{1}{d}} u^{-\frac{1}{d}}\right) \right] \right. \\
    \left. + \int_0^1 \dfrac{\dd u}{1-u}\left[ g_3\left(y t^{-\frac{1}{d}} ,(z-y) t^{-\frac{1}{d}} \right) - \dfrac{1}{u}g_3\left(y t^{-\frac{1}{d}} u^{-\frac{1}{d}},(z-y) t^{-\frac{1}{d}} u^{-\frac{1}{d}}\right) \right] \right\} \\
    + \dfrac{\gamma}{p(0) t (\log{t})^2} \left[ g_3\left(x t^{-\frac{1}{d}} ,(z-x) t^{-\frac{1}{d}} \right) + g_3\left(y t^{-\frac{1}{d}} ,(z-y) t^{-\frac{1}{d}} \right)\right] + \dfrac{1}{p(0)^2 t (\log{t})^2}\left[g_4\left(y t^{-\frac{1}{d}} , (x-y) t^{-\frac{1}{d}}, (z-x) t^{-\frac{1}{d}} \right) \right. \\
    \left. + g_4\left(x t^{-\frac{1}{d}} , (y-x) t^{-\frac{1}{d}}, (z-y) t^{-\frac{1}{d}} \right) + c g_3\left(x t^{-\frac{1}{d}} ,(z-x) t^{-\frac{1}{d}} \right) + c g_3\left(y t^{-\frac{1}{d}} ,(z-y) t^{-\frac{1}{d}} \right)\right].
\end{multline}
It proves convenient to introduce the scaling functions:
\begin{equation}
\frac{p^{(2)}(x)}{p(0)}= \int_0^1 \dfrac{\dd u}{1-u} \left[p\left(x \right) - \dfrac{p\left(x u^{-\frac{1}{d}}\right)}{u}\right] - (\gamma + \kappa) p\left(x \right),
\end{equation}
and:
\begin{eqnarray}
\frac{p^{(3)}(x)}{p(0)^2} &=& \dfrac{6 \gamma^2 -\pi^2  + 12\gamma \kappa +6\kappa^2}{6 }p\left(x \right) \nonumber \\
& - & 2\int_0^1 \dfrac{\dd u}{1-u} \left[p\left(x\right) - \dfrac{p\left(x u^{-\frac{1}{d}}\right)}{u}\right] \left[\log(1-u) + \gamma +\kappa\right].
\end{eqnarray}
Then:
\begin{equation}
    F(x,t) \sim \dfrac{1}{p(0)t \log{t}}p\left(x t^{-\frac{1}{d}}\right) + \dfrac{1}{p(0)^2 t (\log{t})^2}p^{(2)}\left(x t^{-\frac{1}{d}}\right) + \dfrac{1}{p(0)^3 t (\log{t})^3}p^{(3)}\left(x t^{-\frac{1}{d}}\right). 
\end{equation}
We also introduce $g_2$ and its higher order generalization:
\begin{equation} \begin{split}
    & g_2(x) = \int_0^1 \dd u \dfrac{1}{u}p\left(x u^{-\frac{1}{d}}\right),\\
    & \frac{g_2^{(2)}(x)}{p(0)} = \int_0^1 \dd u \dfrac{1}{u}p\left(x u^{-\frac{1}{d}}\right) \left[\log(1-u) +\gamma +\kappa\right],
\end{split} \end{equation}
and we have:
\begin{equation}
    S(x,t) \sim 1 - \dfrac{1}{p(0) \log{t}} g_2\left(x t^{-\frac{1}{d}}\right) + \dfrac{1}{p(0)^2 (\log{t})^2} g_2^{(2)} \left(x t^{-\frac{1}{d}}\right).
\end{equation}
Introducing one more scaling function:
\begin{equation}
    \frac{g_3^{(2)}(x,y)}{p(0)}= \int_0^1 \dfrac{\dd u}{1-u}\left[ \dfrac{1}{u}g_3\left(x u^{-\frac{1}{d}},y u^{-\frac{1}{d}}\right) - g_3\left(x,y\right) \right] 
    + (\gamma +\kappa) g_3(x,y),
\end{equation}
we can rewrite previous expansions in a more compact form:
\begin{multline}
    F^{\{y\}}(x,t) \sim \dfrac{1}{p(0)t \log{t}}p\left(x t^{-\frac{1}{d}}\right) + \dfrac{1}{p(0)^2 t (\log{t})^2}\left[p^{(2)}\left(x t^{-\frac{1}{d}}\right) - g_3\left(y t^{-\frac{1}{d}} ,(x-y) t^{-\frac{1}{d}} \right) \right] \\
    + \dfrac{1}{p(0)^3 t (\log{t})^3}\left[p^{(3)}\left(x t^{-\frac{1}{d}}\right) + 2 g_3^{(2)}\left(y t^{-\frac{1}{d}} ,(x-y) t^{-\frac{1}{d}}\right) + g_4\left(x t^{-\frac{1}{d}} , (y-x) t^{-\frac{1}{d}}, (x-y) t^{-\frac{1}{d}} \right) \right],
\end{multline}
\begin{multline}
    P^{\{x,y\}}(z,t) \sim \dfrac{1}{t}p\left(z t^{-\frac{1}{d}}\right) - \dfrac{1}{p(0)t \log{t}}\left[g_3\left(x t^{-\frac{1}{d}} ,(z-x) t^{-\frac{1}{d}} \right) + g_3\left(y t^{-\frac{1}{d}} ,(z-y) t^{-\frac{1}{d}} \right)\right] \\
    + \dfrac{1}{p(0)^2 t (\log{t})^2}\left[g_4\left(y t^{-\frac{1}{d}} , (x-y) t^{-\frac{1}{d}}, (z-x) t^{-\frac{1}{d}} \right) + g_4\left(x t^{-\frac{1}{d}} , (y-x) t^{-\frac{1}{d}}, (z-y) t^{-\frac{1}{d}} \right) \right. \\
    \left. + g_3^{(2)}\left(x t^{-\frac{1}{d}} ,(z-x) t^{-\frac{1}{d}}\right) + g_3^{(2)}\left(y t^{-\frac{1}{d}} ,(z-y) t^{-\frac{1}{d}}\right)\right].
\end{multline}
As for the transient case, the first and second order cancel out in the expression of the variance. It is clear for the first order since the dominant terms have a exact same expression as for the transient case. For the second order, we have two new contributions. The first one:
\begin{multline}
    \sum_{x,y,z} \sum_{0\leq i<j \leq t} \dfrac{1}{p(0)^2 i (\log{i})^2}p^{(2)}\left(x i^{-\frac{1}{d}}\right) \dfrac{1}{i}p\left(z i^{-\frac{1}{d}}\right) \left[ \dfrac{1}{p(0)j \log{j}}p\left((y-x) j^{-\frac{1}{d}}\right) - \dfrac{1}{p(0)j \log{j}}p\left((y-z) j^{-\frac{1}{d}}\right)\right],
\end{multline}
vanishes as we can redefine $y=y+x$ in the first term and $y=y+z$ in the second. The other new contribution:
\begin{multline}
    \sum_{x,y,z} \sum_{0\leq i<j \leq t} \dfrac{1}{p(0)i \log{i}}p\left(x i^{-\frac{1}{d}}\right) \dfrac{1}{i}p\left(z i^{-\frac{1}{d}}\right) \left[ \dfrac{1}{p(0)^2 j (\log{j})^2}p^{(2)}\left((y-x) j^{-\frac{1}{d}}\right) - \dfrac{1}{p(0)^2 j (\log{j})^2}p^{(2)}\left((y-z) j^{-\frac{1}{d}}\right)\right].
\end{multline}
also vanishes as it is antisymmetric with respect to the exchange $x\leftrightarrow z$.

We therefore go to third order. Here we have a priori $10$ terms to compute. But some of them will also vanish by symmetry. Replacing the sums by integrals and using $\log{i} \sim \log{j} \sim \log{t}$ at first order, we get for the $3-1-1-1$ term:
\begin{multline}
    \int_0^1 \dfrac{\dd u}{u^2} \int_0^{1-u} \dfrac{\dd v}{v} \iiint \dd x \, \dd y \,  \dd z \left[p^{(3)}\left(x u^{-\frac{1}{d}}\right) + 2 g_3^{(2)}\left(y u^{-\frac{1}{d}} ,(x-y) u^{-\frac{1}{d}}\right)\right. \\
    \left. + g_4\left(x u^{-\frac{1}{d}} , (y-x) u^{-\frac{1}{d}}, (x-y) u^{-\frac{1}{d}} \right) \right] p\left(z u^{-\frac{1}{d}}\right) \left[p\left((y-x) v^{-\frac{1}{d}}\right) - p\left((y-z) v^{-\frac{1}{d}}\right)\right].
\end{multline}
In this equation and the following formulas up to Eq.~\eqref{eq 142} we suppress the multiplicative factor $\dfrac{4 t^2 }{p(0)^4 (\log t)^4}$.

We already know that the $g_4$ term will contribute towards the prefactor. The $p^{(3)}$ term is zero after the integration over $y$. For now, we stop here, such that the first term is written as:
\begin{multline}
   \int_0^1 \dfrac{\dd u}{u^2} \int_0^{1-u} \dfrac{\dd v}{v} \iiint \dd x \, \dd y \,  \dd z \left[ 2 g_3^{(2)}\left(y u^{-\frac{1}{d}} ,(x-y) u^{-\frac{1}{d}}\right) + g_4\left(x u^{-\frac{1}{d}} , (y-x) u^{-\frac{1}{d}}, (x-y) u^{-\frac{1}{d}} \right) \right]\\
    p\left(z u^{-\frac{1}{d}}\right) \left[p\left((y-x) v^{-\frac{1}{d}}\right) - p\left((y-z) v^{-\frac{1}{d}}\right)\right].
    \label{eq 128}
\end{multline}
For the $1-3-1-1$ term, we obtain:
\begin{multline}
   \int_0^1 \dfrac{\dd u}{u^2} \int_0^{1-u} \dfrac{\dd v}{v} \iiint \dd x \, \dd y \,  \dd z \left[g_4\left(y u^{-\frac{1}{d}} , (x-y) u^{-\frac{1}{d}}, (z-x) u^{-\frac{1}{d}} \right) + g_4\left(x u^{-\frac{1}{d}} , (y-x) u^{-\frac{1}{d}}, (z-y) u^{-\frac{1}{d}} \right) \right. \\
    \left. + g_3^{(2)}\left(x u^{-\frac{1}{d}} ,(z-x) u^{-\frac{1}{d}}\right) + g_3^{(2)}\left(y u^{-\frac{1}{d}} ,(z-y) u^{-\frac{1}{d}}\right)\right] p\left(x u^{-\frac{1}{d}}\right) \left[p\left((y-x) v^{-\frac{1}{d}}\right) - p\left((y-z) v^{-\frac{1}{d}}\right)\right].
\end{multline}
We immediately see that the terms involving $g_3^{(2)}\left(y u^{-\frac{1}{d}} ,(z-y) u^{-\frac{1}{d}}\right)$ will cancel out with half of the $g_3^{(2)}$ in Eq.~\eqref{eq 128}, as these are the same products with $x \rightarrow z$. The terms with $g_3^{(2)}\left(x u^{-\frac{1}{d}} ,(z-x) u^{-\frac{1}{d}}\right)$ also vanish after integration over $y$. This is also the case for the product involving $g_4\left(x u^{-\frac{1}{d}} , (y-x) u^{-\frac{1}{d}}, (z-y) u^{-\frac{1}{d}} \right)$ as it is antisymmetric with respect to the exchange $y-x \rightarrow y-z$ (as for the transient case). We therefore get combining the first two contributions to the variance:
\begin{multline}
   \int_0^1 \dfrac{\dd u}{u^2} \int_0^{1-u} \dfrac{\dd v}{v} \iiint \dd x \, \dd y \,  \dd z \left[ g_3^{(2)}\left(y u^{-\frac{1}{d}} ,(x-y) u^{-\frac{1}{d}}\right) + g_4\left(x u^{-\frac{1}{d}} , (y-x) u^{-\frac{1}{d}}, (x-y) u^{-\frac{1}{d}} \right) \right]\\
    p\left(z u^{-\frac{1}{d}}\right) \left[p\left((y-x) v^{-\frac{1}{d}}\right) - p\left((y-z) v^{-\frac{1}{d}}\right)\right] + \int_0^1 \dfrac{\dd u}{u^2} \int_0^{1-u} \dfrac{\dd v}{v} \\
    \iiint \dd x \, \dd y \,  \dd z \, g_4\left(y u^{-\frac{1}{d}} , (x-y) u^{-\frac{1}{d}}, (z-x) u^{-\frac{1}{d}} \right) p\left(x u^{-\frac{1}{d}}\right) \left[p\left((y-x) v^{-\frac{1}{d}}\right) - p\left((y-z) v^{-\frac{1}{d}}\right)\right].
\end{multline}
Now for the $1-1-3-1$ term:
\begin{multline}
   \int_0^1 \dfrac{\dd u}{u^2} \int_0^{1-u} \dfrac{\dd v}{v} \iiint \dd x \, \dd y \,  \dd z \, p\left(x u^{-\frac{1}{d}}\right) p\left(z u^{-\frac{1}{d}}\right) \\
    \left[p\left((y-x) v^{-\frac{1}{d}}\right) g_2^{(2)} \left((y-z) v^{-\frac{1}{d}}\right) - p\left((y-z) v^{-\frac{1}{d}}\right) g_2^{(2)} \left((y-x) v^{-\frac{1}{d}}\right)\right].
\end{multline}
This term is antisymmetric in $x\leftrightarrow z$ so it equals zero. This is also true for the $1-1-1-3$ term:
\begin{multline}
    \int_0^1 \dfrac{\dd u}{u^2} \int_0^{1-u} \dfrac{\dd v}{v} \iiint \dd x \, \dd y \,  \dd z \, p\left(x u^{-\frac{1}{d}}\right) p\left(z u^{-\frac{1}{d}}\right) \left[p^{(3)}\left((y-x) v^{-\frac{1}{d}}\right)  - p^{(3)}\left((y-z) v^{-\frac{1}{d}}\right) \right].
\end{multline}
We now turn our attention to terms involving second order corrections, starting with the $2-2-1-1$ term:
\begin{multline}
    - \int_0^1 \dfrac{\dd u}{u^2} \int_0^{1-u} \dfrac{\dd v}{v} \iiint \dd x \, \dd y \,  \dd z \, \left[p^{(2)}\left(x u^{-\frac{1}{d}}\right) - g_3\left(y u^{-\frac{1}{d}} ,(x-y) u^{-\frac{1}{d}} \right) \right] \\
    \left[g_3\left(x u^{-\frac{1}{d}} ,(z-x) u^{-\frac{1}{d}} \right) + g_3\left(y u^{-\frac{1}{d}} ,(z-y) u^{-\frac{1}{d}} \right)\right]\left[p\left((y-x) v^{-\frac{1}{d}}\right) - p\left((y-z) v^{-\frac{1}{d}}\right)\right].
\end{multline}
The product where we pick $g_3$ first and then $g_3\left(y u^{-\frac{1}{d}} ,(z-y) u^{-\frac{1}{d}} \right)$ vanishes as it is antisymmetric in $x \leftrightarrow z$. The one with $p^{(2)}$ first and then $g_3\left(x u^{-\frac{1}{d}} ,(z-x) u^{-\frac{1}{d}} \right)$ is zero after integration over $y$. We therefore get $2$ non zero terms:
\begin{multline}
     - \int_0^1 \dfrac{\dd u}{u^2} \int_0^{1-u} \dfrac{\dd v}{v} \iiint \dd x \, \dd y \,  \dd z \, \left[p^{(2)}\left(x u^{-\frac{1}{d}}\right)g_3\left(y u^{-\frac{1}{d}} ,(z-y) u^{-\frac{1}{d}} \right) \right. \\
     \left. - g_3\left(y u^{-\frac{1}{d}} ,(x-y) u^{-\frac{1}{d}} \right) g_3\left(x u^{-\frac{1}{d}} ,(z-x) u^{-\frac{1}{d}} \right) \right] \left[p\left((y-x) v^{-\frac{1}{d}}\right) - p\left((y-z) v^{-\frac{1}{d}}\right)\right].
\end{multline}
We group the $2-1-2-1$ term:
\begin{multline}
     \int_0^1 \dfrac{\dd u}{u^2} \int_0^{1-u} \dfrac{\dd v}{v} \iiint \dd x \, \dd y \,  \dd z \, \left[p^{(2)}\left(x u^{-\frac{1}{d}}\right) - g_3\left(y u^{-\frac{1}{d}} ,(x-y) u^{-\frac{1}{d}} \right) \right] \\
    p\left(z u^{-\frac{1}{d}}\right) \left[-  g_2\left((y-z) v^{-\frac{1}{d}}\right) p\left((y-x) v^{-\frac{1}{d}}\right) + g_2\left((y-x) v^{-\frac{1}{d}}\right) p\left((y-z) v^{-\frac{1}{d}}\right)\right]
\end{multline}
and the $1-2-2-1$ term:
\begin{multline}
    \int_0^1 \dfrac{\dd u}{u^2} \int_0^{1-u} \dfrac{\dd v}{v} \iiint \dd x \, \dd y \,  \dd z \, p\!\left(x u^{-\frac{1}{d}}\right) \left[ - g_3\left(x u^{-\frac{1}{d}} ,(z-x) u^{-\frac{1}{d}} \right) - g_3\left(y u^{-\frac{1}{d}} ,(z-y) u^{-\frac{1}{d}} \right) \right] \\
    \left[-  g_2\left((y-z) v^{-\frac{1}{d}}\right) p\left((y-x) v^{-\frac{1}{d}}\right) + g_2\left((y-x) v^{-\frac{1}{d}}\right) p\left((y-z) v^{-\frac{1}{d}}\right)\right].
\end{multline}
Indeed, as before, the products involving $g_3\left(x u^{-\frac{1}{d}} ,(z-x) u^{-\frac{1}{d}} \right)$ vanish as swapping $z-y$ and $y-x$ leaves $z-x$ unchanged, and the two $g_3$ terms left (one in each term) group to produce a contribution antisymmetric in $x\leftrightarrow z$. We are left with the new term:
\begin{multline}
    \int_0^1 \dfrac{\dd u}{u^2} \int_0^{1-u} \dfrac{\dd v}{v} \iiint \dd x \, \dd y \,  \dd z \, p^{(2)}\left(x u^{-\frac{1}{d}}\right) p\left(z u^{-\frac{1}{d}}\right) \\
    \left[-  g_2\left((y-z) v^{-\frac{1}{d}}\right) p\left((y-x) v^{-\frac{1}{d}}\right) + g_2\left((y-x) v^{-\frac{1}{d}}\right) p\left((y-z) v^{-\frac{1}{d}}\right)\right],
\end{multline}
which vanishes after expliciting $g_2$ and integrating over $y$. We also group the terms $2-1-1-2$ and $1-2-1-2$:
\begin{multline}
     \int_0^1 \dfrac{\dd u}{u^2} \int_0^{1-u} \dfrac{\dd v}{v} \iiint \dd x \, \dd y \,  \dd z \, \left[p^{(2)}\left(x u^{-\frac{1}{d}}\right)  - g_3\left(y u^{-\frac{1}{d}} ,(x-y) u^{-\frac{1}{d}} \right) \right] p\left(z u^{-\frac{1}{d}}\right) \\
    \left[p^{(2)}\left((y-x) v^{-\frac{1}{d}}\right) - p^{(2)}\left((y-z) v^{-\frac{1}{d}}\right)\right] - \dfrac{4 t^2 }{p(0)^4 (\log t)^4} \int_0^1 \dfrac{\dd u}{u^2} \int_0^{1-u} \dfrac{\dd v}{v} \iiint \dd x \, \dd y \,  \dd z \, p\left(x u^{-\frac{1}{d}}\right) \\
    \left[g_3\left(x u^{-\frac{1}{d}} ,(z-x) u^{-\frac{1}{d}} \right) + g_3\left(y u^{-\frac{1}{d}} ,(z-y) u^{-\frac{1}{d}} \right)\right] \left[p^{(2)}\left((y-x) v^{-\frac{1}{d}}\right) - p^{(2)}\left((y-z) v^{-\frac{1}{d}}\right)\right]
\end{multline}
Integrating over $y$ we find that the $p^{(2)}$ term in the first integral and the $g_3\left(x u^{-\frac{1}{d}} ,(z-x) u^{-\frac{1}{d}} \right)$ in the second one vanish. We are left with two $g_3$ terms that together form an antisymmetric combination in $x \leftrightarrow z$. This whole contribution is therefore zero. The last term ($1-1-2-2$) is also antisymmetric in $x\leftrightarrow z$:
\begin{multline}
    \int_0^1 \dfrac{\dd u}{u^2} \int_0^{1-u} \dfrac{\dd v}{v} \iiint \dd x \, \dd y \,  \dd z \,  p\left(x u^{-\frac{1}{d}}\right)  p\left(z u^{-\frac{1}{d}}\right) \\
    \left[ - g_2\left((y-z) v^{-\frac{1}{d}}\right) p^{(2)}\left((y-x) v^{-\frac{1}{d}}\right) + g_2\left((y-x) v^{-\frac{1}{d}}\right) p^{(2)}\left((y-z) v^{-\frac{1}{d}}\right) \right].
\end{multline}
A priori, we therefore have two new contributions to the variance prefactor compared to the transient case:
\begin{multline}
    \int_0^1 \dfrac{\dd u}{u^2} \int_0^{1-u} \dfrac{\dd v}{v} \iiint \dd x \, \dd y \,  \dd z \, g_3^{(2)}\left(y u^{-\frac{1}{d}} ,(x-y) u^{-\frac{1}{d}}\right) p\left(z u^{-\frac{1}{d}}\right) \left[p\left((y-x) v^{-\frac{1}{d}}\right) - p\left((y-z) v^{-\frac{1}{d}}\right)\right],
\end{multline}
and
\begin{equation}
    -\int_0^1 \dfrac{\dd u}{u^2} \int_0^{1-u} \dfrac{\dd v}{v} \iiint \dd x \, \dd y \,  \dd z \, p^{(2)}\left(x u^{-\frac{1}{d}}\right)g_3\left(y u^{-\frac{1}{d}} ,(z-y) u^{-\frac{1}{d}} \right) \left[p\left((y-x) v^{-\frac{1}{d}}\right) - p\left((y-z) v^{-\frac{1}{d}}\right)\right].
    \label{eq 142}
\end{equation}
These contributions can be explicitly computed (they involve only spatial integrals of the propagators squared) and they cancel each other. The expression for the prefactor in the marginal case is therefore exactly the same as for the transient case, with the correspondence $g_0 \leftrightarrow p(0) \log{t}$. 

\subsubsection{Exact formula for $\Var[X_n]$ in the marginal regime}

Exploiting this correspondence and restoring the $ \dfrac{4 t^2 }{p(0)^4 (\log t)^4}$ factor, we have:
\begin{equation} \begin{split}
     \langle [F_1(t) - F_2(t)]^2\rangle \underset{t \rightarrow +\infty}{\sim} \dfrac{4}{p(0)^4 (2\pi)^{2d} D^2} \dfrac{t^2}{(\log{t})^4} \iint \dd^d k \dd^d k' \mathcal{I}\left(|k|^d,|k'|^d,|k+k'|^d\right).
\end{split} \end{equation}
Combining with the asymptotics $\langle \mathcal{N}_1(t) \rangle \sim \frac{t}{p(0) \log{t}}$, we arrive at:
\begin{equation} \begin{split}
     \Var[X_t] \underset{t \rightarrow +\infty}{\sim} \dfrac{1}{4 p(0)^2 (2\pi)^{2d} D^2} \dfrac{1}{(\log{t})^2} \iint \dd^d k \dd^d k' \mathcal{I}\left(|k|^d,|k'|^d,|k+k'|^d\right).
\end{split} \end{equation}
Here one does not even care about the prefactors in the scaling of $t_n$ with $n$, as $\log(t_n) \sim \log{n}$ at leading order. Thus the asymptotics of the variance of the fraction in the marginal case reads:
\begin{equation}
\tcbhighmath[
colback=white,
colframe=black,
arc=2mm]{
    \Var[X_n] \underset{n \rightarrow +\infty}{\sim} \dfrac{1}{4 p(0)^2 (2\pi)^{2d} D^2} \dfrac{1}{(\log{n})^2} \iint \dd^d k \dd^d k' \mathcal{I}\left(|k|^d,|k'|^d,|k+k'|^d\right).
    }
    \label{eq 170}
\end{equation}
As claimed, in the marginal case the variance prefactor is totally independent on the microscopic details of the walk and does not even depend on the intrinsic scale of the dynamics (as $p(0) \sim \frac{1}{D}$ in the marginal case). 

\subsubsection{Explicit examples}

We now apply this general formula to the $2$ most relevant cases, namely the Cauchy walk in $1$ dimension and the simple random walk in $2$ dimensions.

For the $1$-dimensional Cauchy walk,we need to compute the integral in Eq.~\eqref{eq 119} in the $d_s \rightarrow 2$ limit. We obtain:
\begin{equation} \begin{split}
    & \iint \dd k \dd k' \mathcal{I}\left(|k|,|k'|,|k+k'|\right) \\
    = & \int_0^{+\infty} \dd r \left\{ \frac{4 \log (r+1)}{r}-\frac{4 \left(25 \log (r)-34 \log (r+1)+32 \arccoth(2 r+1)+18 \log (2)\right)}{9(r-1)^2}  \right. \\
         & +\frac{2 \left(143 \log (r)-128 \log (r+1)+256 \arccoth(2 r+1)-15 \log (2)\right)}{15(r-2)}\\   
         & -\frac{2 \left(803 \log (r)-560 \log (r+1)+1120 \arccoth(2 r+1)\right)}{81 (r-1)} \\
   & -\frac{16 \left(\log (r)-\log (r+1)+2 \arccoth(2 r+1)\right)}{9 (r-1)^3} \\
   &+\frac{2 \left(-277 \log (r)+277\log (r+1)+256 \arccoth(2 r+1)+405 \log (2)\right)}{405 (r+2)}\\
   &-\frac{4 \left(-16 \log (r)+\log (r+1)-2 \arccoth(2 r+1)-15 \log (2)\right)}{15 (2 r-1)}\\
   & \left. -\frac{4 \left(\log (r)+404 \log (r+1)+2
   \arccoth(2 r+1)+405 \log (2)\right)}{405 (2 r+1)} \right\},
\end{split} \end{equation}
which actually reduces all the way to $8 \log{2}$. We can therefore write the asymptotics of the fraction variance very nicely in this case. Using $p(0) = \frac{1}{\pi D}$, we get:
\begin{equation}
\tcbhighmath[
colback=white,
colframe=black,
arc=2mm]{
    \Var[X_n] \sim \dfrac{\log{2}}{2(\log{n})^2}.
    }
\end{equation}

Now in the case of the $2$ dimensional simple random walk, we perform the computation in real space because the propagator has a simple Gaussian form:
\begin{equation}
    p(x,u)=\frac{1}{4\pi D u}
    \exp\left(-\frac{|x|^2}{4Du}\right),
\end{equation}
From this, we get the expression of $\Theta$:
\begin{equation}
    \Theta(u,v,w)=\frac{1}{16\pi^2D^2}\frac{1}{uv+uw+vw}.
\end{equation}
We therefore write the asymptotic decay of the variance in real space from Eq.~\eqref{eq 86} with the correspondence $g_0 \rightarrow p(0) \log{n}$:
\begin{multline}
    \Var[X_n] \underset{n \rightarrow +\infty}{\sim} \dfrac{4}{16 p(0)^2 (\log{n})^2} \dfrac{1}{16 \pi^2 D^2}\int_0^1 \dd u \int_0^{1-u} \dd v \int_0^u \dd a \left\{ \int_0^a \dd b \left[\dfrac{1}{(u-a)(a-b) + (u-b)v} \right. \right.\\
    - \dfrac{1}{(u+b)(a-b) + (u+a)(u+v-a)}  \left. + \dfrac{1}{(a-b)(u+b) + v(u+a)} - \dfrac{1}{(u-a)(a-b) + (u-b)(u+v+b)} \right]\\
    \left. + \int_0^u \dd b \left[\dfrac{1}{(u-a)(a+b) + v(u+b)} - \dfrac{1}{(u-a)(a+b)+(u-b+v)(u+b)}\right]\right\},
\end{multline}
with the prefactor simplifying all the way up to $1/4$.

We now identify these rational combinations as the same integrand on different integration regions. Setting $x,y = u-a, a-b$ in the first and fourth terms, $x,y = u+b,a-b$ in the second and third ones, and $x,y = u-a,a+b$ in the last two, we can express the prefactor as:
\begin{multline}
    \dfrac{1}{4} \int_0^1 \dd u \int_0^{1-u} \dd v  \left\{ \int_0^u \dd x \int_0^{u-x} \dd y \left[ \dfrac{1}{x y + v(x+y)} -\dfrac{1}{x y + (x+y)(2u-x-y+v)} \right] \right. \\
    + \int_u^{2u} \dd x \int_0^{2u-x} \dd y \left[ \dfrac{1}{x y + v(x+y)} -\dfrac{1}{x y + (x+y)(2u-x-y+v)} \right] \\
    + \left. \int_0^u \dd x \int_{u-x}^{2u-x} \dd y \left[ \dfrac{1}{x y + v(x+y)} -\dfrac{1}{x y + (x+y)(2u-x-y+v)} \right] \right\},
\end{multline}
where we recover thrice the same integrand. The integration regions add up nicely so that we get:
\begin{equation}
    \dfrac{1}{4} \int_0^1 \dd u \int_0^{1-u} \dd v \int_0^{2 u} \dd x \int_0^{2u-x} \dd y \left[ \dfrac{1}{x y + v(x+y)} -\dfrac{1}{x y + (x+y)(2u-x-y+v)} \right].
\end{equation}
We now rescale all variables by u by setting $v= u r$, $x=u s$, $y = u t$. This gives:
\begin{equation}
    \dfrac{1}{4} \int_0^1 \dd u \int_0^{1/u -1} u \, \dd r \int_0^2 u \, \dd s \int_0^{2-s} u \, \dd t \left[ \dfrac{1}{u^2(s t + r(s+t))} -\dfrac{1}{u^2(s t + (s+t)(2-s-t+r))} \right].
\end{equation}
Swapping the integration order between $u$ and $r$ and integrating over $u$, one obtains:
\begin{equation}
    \dfrac{1}{8} \int_0^{+\infty} \dfrac{\dd r}{(1+r)^2} \int_0^2 \dd s \int_0^{2-s} \dd t \left[ \dfrac{1}{s t + r(s+t)} -\dfrac{1}{s t + (s+t)(2-s-t+r)} \right].
\end{equation}
We redefine t = s+t to lighten the expression slightly:
\begin{equation}
    \dfrac{1}{8} \int_0^{+\infty} \dfrac{\dd r}{(1+r)^2} \int_0^2 \dd s \int_s^2 \dd t \left[ \dfrac{1}{s (t-s) + r t} -\dfrac{1}{s (t-s) + t(2-t+r)} \right],
\end{equation}
set $s= t z$ and swap the integration order:
\begin{equation}
    \dfrac{1}{8} \int_0^{+\infty} \dfrac{\dd r}{(1+r)^2} \int_0^2 \dd t \int_0^1 \dd z \left[ \dfrac{1}{t z(1-z) + r } -\dfrac{1}{t z(1-z) + 2-t+r} \right].
\end{equation}
We can now integrate over $t$:
\begin{equation}
    \dfrac{1}{8} \int_0^{+\infty} \dfrac{\dd r}{(1+r)^2} \int_0^1 \dd z \left[ \frac{\log \left(\frac{r-2 (z-1) z}{r+2}\right)}{z^2-z+1}+\frac{\log \left(\frac{r}{r-2 (z-1)
   z}\right)}{(z-1) z} \right],
\end{equation}
then over r to obtain:
\begin{equation}
    \dfrac{1}{8} \int_0^1 \dd z \frac{2 [(1-z) z \log (4)-\log (4 (1-z) z)]}{((z-1) z+1) (2 (z-1) z+1)}.
\end{equation}
Setting $q = 2z-1$, one gets:
\begin{equation}
    \dfrac{1}{8} \int_{-1}^1 \dd q \frac{2(1-q^2) \log (4)-8\log (1-q^2)}{((q^2-1)+4) ((q^2-1)+2)},
\end{equation}
which can be simplified using the parity of the integrand to yield:
\begin{equation}
    \int_0^1 \dd q \frac{(1-q^2) \log (2)-2\log (1-q^2)}{(q^2+3)(q^2+1)}.
\end{equation}
At this point, Mathematica outputs an expression containing Trigamma functions, that can be expressed nicely as particular values of the Clausen function $\Cl_2(\theta) = \sum_{n=1}^{+\infty} \frac{\sin(n\theta)}{n^2}$. Using:
\begin{equation}
    \begin{cases}
    \psi_1(1/6) = 2\pi^2 +6\sqrt{3} [\Cl_2(\pi/3) +\Cl_2(2\pi/3)],\\
    \psi_1(1/3) = 2\pi^2/3 +3\sqrt{3} \Cl_2(2\pi/3),\\
    \psi_1(2/3) = 2\pi^2/3 -3\sqrt{3} \Cl_2(2\pi/3),\\
    \psi_1(5/6) = 2\pi^2 -6\sqrt{3} [\Cl_2(\pi/3) +\Cl_2(2\pi/3)],
    \end{cases}
\end{equation}
we get:
\begin{equation}
    \int_0^1 \dd q \frac{(1-q^2) \log (2)-2\log (1-q^2)}{(q^2+3)(q^2+1)} = \Cl_2(\pi/2) - \dfrac{\sqrt{3}}{4} \Cl_2(2\pi/3) - \dfrac{\sqrt{3}}{6} \Cl_2(\pi/3).
\end{equation}
Using $\Cl_2(2\pi/3) = \frac{2}{3} \Cl_2(\pi/3)$, we obtain the final expression for the prefactor and thus the complete decay of the variance for $2$-dimensional Brownian motion:
\begin{equation}
\tcbhighmath[
colback=white,
colframe=black,
arc=2mm]{
    \Var[X_n] \underset{n \rightarrow +\infty}{\sim} \dfrac{\Cl_2(\pi/2) - 3^{-1/2} \Cl_2(\pi/3)}{(\log{n})^2}.
    }
\end{equation}

\subsection{Strongly transient regime $d_s \geq 3$}

To complete the quantitative description of the variance prefactor in the self-averaging case, we study the regime $d_s \geq 3$, where overlap between walker trajectories is totally negligible even when describing leading order fluctuations of the fraction. We therefore start back from Eq.~\eqref{eq 36} but now $F_1(t)$ and $F_2(t)$ are independent at leading order and can be replaced by the individual ranges $\mathcal{N}_1(t)$ and $\mathcal{N}_2(t)$. We therefore get:
\begin{equation}
    \Var[X_t] \sim \dfrac{\Var[\mathcal{N}_1(t)]}{8 \langle \mathcal{N}_1(t) \rangle^2},
\end{equation}
The asymptotics of the range mean and variance can then be obtained using the approach of \cite{mariz_statistics_2001, kohler_variance_1990} and we can extract these results directly to obtain the fraction variance. In all cases for transient walks, we know that $\langle \mathcal{N}_1(t) \rangle \sim \frac{t}{g_0}$ but for the variance there are different expressions depending on the value of $d_s$. More precisely,
\begin{multline}
    \Var[\mathcal{N}_1(t)] \sim
    \begin{cases}
    \dfrac{3 \beta(d) C^3}{d g_0^4} t \log{t} & , d_s=3 \\[20pt]
    \displaystyle \left\{1 - \dfrac{2}{g_0} +2 \sum_{x\neq 0} G(x,1) \left[\dfrac{g_0 - G(-x,1)}{g_0^2 - G(x,1)G(-x,1)} - \dfrac{1}{g_0} + \dfrac{G(-x,1)}{g_0^2} \right]\right\} \dfrac{t}{g_0} & , 3<d_s\leq 4 \\[20pt]
    \displaystyle \left\{ 2 \sum_{x\neq 0} G(x,1) \left[\dfrac{g_0 - G(-x,1)}{g_0^2 - G(x,1)G(-x,1)} - \dfrac{1}{g_0}\right]- 1 +2 \dfrac{c}{g_0^2}\right\} \dfrac{t}{g_0} & , d_s > 4.
    \end{cases}
\end{multline}
Here $\beta(d)$ is the area of the $d-1$-dimensional hypersphere with the convention $\beta(1)=2$, $C$ is the prefactor of the divergence of the scaling function $\mathcal{G}$ at the origin ($\mathcal{G}(x) \sim C |x|^{-\frac{d}{3}}$ for $d_s=3$), and the constant $c$ appearing in the prefactor for $d_s>4$ is another series:
\begin{equation}
    c = \sum_{n=1}^{+\infty} n P_n(0|0),
\end{equation}
which converges for $d_s>4$. This translates into the corresponding expression for the variance of $X_n$:
\begin{tcolorbox}[
colback=white,
colframe=black,
arc=2mm]
\begin{varwidth}{\textwidth}
\begin{multline}
    \Var[X_n] \sim
    \begin{cases}
    \dfrac{3 \beta(d) C^3}{8 d g_0^3}\dfrac{\log{n}}{n} &, d_s = 3 \\[20 pt]
    \displaystyle \left\{1 - \dfrac{2}{g_0} +2 \sum_{x\neq 0} G(x,1) \left[\dfrac{g_0 - G(-x,1)}{g_0^2 - G(x,1)G(-x,1)} - \dfrac{1}{g_0} + \dfrac{G(-x,1)}{g_0^2} \right]\right\} \dfrac{1}{8 n} & , 3<d_s \leq 4 \\[20 pt]
    \displaystyle \left\{ 2 \sum_{x\neq 0} G(x,1) \left[\dfrac{g_0 - G(-x,1)}{g_0^2 - G(x,1)G(-x,1)} - \dfrac{1}{g_0}\right]- 1 +2 \dfrac{c}{g_0^2}\right\} \dfrac{1}{8 n} & , d_s > 4.
    \end{cases}
\end{multline}
\end{varwidth}
\end{tcolorbox}
As seen in these expressions, the variance prefactor depends for strongly transient walks in a quite complicated way on the microscopic details, and involves a lattice sum over the Green function. We explicit the important case of $3$-dimensional simple random walk below.

In this case where $d_s = 3$ and $d=3$, $C=\frac{3}{2\pi}$ and $g_0$ can be computed \cite{hughes_random_2023}:
\begin{equation}
    g_0 = \dfrac{\sqrt{6}}{32 \pi^3}\Gamma\left(\dfrac{1}{24}\right)\Gamma\left(\dfrac{5}{24}\right)\Gamma\left(\dfrac{7}{24}\right)\Gamma\left(\dfrac{11}{24}\right).
\end{equation}
This gives:
\begin{equation}
\tcbhighmath[
colback=white,
colframe=black,
arc=2mm]{
    \Var[X_n] \sim \dfrac{3 \sqrt{6} \; 2^9 \pi^7}{\Gamma^3\left(\dfrac{1}{24}\right)\Gamma^3\left(\dfrac{5}{24}\right)\Gamma^3\left(\dfrac{7}{24}\right)\Gamma^3\left(\dfrac{11}{24}\right)} \dfrac{\log{n}}{n} \approx 0.049 \dfrac{\log{n}}{n}.
    }
    \label{eq 153}
\end{equation}

\section{Asymptotic value of the variance for $1$-dimensional nearest-neighbour diffusion}

In the previous section, we obtained precise asymptotics for $\Var[X_n]$ including prefactors for all marginal and transient walks. For recurrent walks, a quantitative description is arguably even more important since in this case $\Var[X_\infty]$ quantifies persistent fluctuations away from the mean in the large time limit. But the tools we used in the previous section will fail in this regime. Indeed, because the random time $t_n$ is not self-averaging, it is impossible to transfer fixed time variance estimations to the problem considered in this work, and one needs to describe very precisely the exploration process of the $2$ walkers. Such a description seems out of reach for general walks but for the $1$-dimensional simple random walk, we perform here the exact calculation of $\Var[X_\infty]$.

\subsection{Two-walkers splitting probabilities and basic objects}

In this subsection, we will use two-walker splitting probabilities and we therefore introduce notations.  Let $R^+_{k,k+n}(z|x_1,x_2)$ denote the probability that starting from a situation with one walker located at $x_1$, one walker located at $x_2$, and two boundaries at $k$ and $k+n$, the first exit is by the second walker through the top boundary (the one located at $k+n$) and the first walker is in $z$ at the exit time:
\begin{equation}
    R^+_{k,k+n}(z|x_1,x_2) = \mathbb{P}(\tau^2_{k+n}<\tau^1_k , \tau^2_{k+n}<\tau^2_k , \tau^2_{k+n}<\tau^1_{k+n} , X^1_{\tau^2_{k+n}} = z | X^1_0 = x_1 , X^2_0 = x_2),
\end{equation}
with $\tau^i_k$ denoting the first hitting time of site $k$ by walker $i$. We do not need $R^+$ (probably very complicated) exact expression, only that it possesses a scaling form:
\begin{equation}
    R^+_{k,k+n}(z|x_1,x_2) \sim \dfrac{1}{n}\rho^+_{\frac{x_1-k}{n},\frac{x_2-k}{n}} \left(\dfrac{z-k}{n}\right),
\end{equation}
in the limit where the interval width $n$ goes to infinity. $k$ can be totally removed as a parameter because of translational invariance. The scaling function $\rho^+$ is obtained by solving Laplace equation inside a square with boundary conditions $\rho^+_{0,x_2}(z) = 0$, $\rho^+_{x_1,0}(z) = 0$, $\rho^+_{1,x_2}(z) = 0$, $\rho^+_{x_1,1}(z) = \delta(z-x_1)$ and has the expression:
\begin{equation}
    \rho^+_{x,y}(z) = 2 \sum_{n=1}^{+\infty} \dfrac{\sin(n\pi z)\sin(n\pi x)\sinh(n\pi y)}{\sinh(n\pi)}.
\end{equation}
We also introduce $\pi^+$ which integrate over the position of the first walker:
\begin{equation}
    \pi^+_{x,y} = \int_0^1 \dd z \rho^+_{x,y}(z).
\end{equation}

\subsection{Enumeration of discovery scenarios}

To compute the variance of $X_\infty$, we compute the correlations between discoverer labels introduced in the main text. Denoting by $I_k = \mathds{1}_{\text{walker } 1 \text{ discovered the } k \text{-th site}}$, we show that:
\begin{equation}
    \langle I_k I_{k+k'} \rangle - \langle I_k \rangle \langle I_{k+k'} \rangle \sim c\left(\dfrac{k'}{k}\right),
\end{equation}
in the limit where $k$ and $k'$ both go to infinity with fixed ratio. This is an example of aging correlations that cause for recurrent walks non-ergodicity and infinite-time fluctuations of the variance.

We therefore start from a situation where the $2$ walkers have explored $n$ sites collectively. Because of translational invariance, we take the visited interval to be $[0,n]$, and we denote by walker $1$ the walker that just discovered $n$, and by walker $2$ the other walker, sitting at position $n-z$ ($z$ is the gap between the walkers).

We will count all the possible ways the $2$ walkers can discover the next $k$ sites with walker $1$ discovering the $k+k'$-th one. This way we count all possible trajectories where $I_k = I_{k+k'} = 1$. Denoting by $k''$ the number of sites among $k'$ discovered to the right of $k$, we have $5$ possibilities:
\begin{enumerate}
    \item walker $2$ touches $k''-k'$ and then walker $1$ touches $k+k''$
    \item walker $1$ touches $k''-k'$ and then walker $1$ touches $k+k''$
    \item walker $1$ touches $k+k''$ and then walker $1$ touches $k''-k'$
    \item walker $2$ touches $k+k''$ and then walker $1$ touches $k''-k'$
    \item if $k''=k'$, there is also the possibility for walker $1$ to touch $k+k'$ before any of the two walkers touches $-1$
\end{enumerate}
Since walker $1$ starts at the right edge of the interval, there will always be some exploration to the right in the scaling limit, so the analogous of the fifth option were all sites are discovered to the left is not relevant in this limit.

We express the probability of the first option carefully using the splitting probabilities in the square. We will integrate over the position of walker $1$ when walker $2$ touches $k''-k'$. This gives:
\begin{equation}
    \frac{1}{2} \int_{k''-k'}^{k+k''} \dd z' \dfrac{1}{k+k'}\rho^+_{\frac{k''}{k+k'},\frac{k''+z}{k+k'}}\left(\dfrac{k+k''-z'}{k+k'}\right)\pi^+_{\frac{1}{k+k'+1},\frac{1+z'-k''+k'}{k+k'+1}}.
\end{equation}
The initial factor $\frac{1}{2}$ quantifies the probability that $I_k = 1$, the factor $\rho^+$ corresponds to the probability that walker $2$ touches the bottom boundary before walker 1 exits the interval, and that walker $1$ ends in a given position, and the factor $\pi^+$ then describes the probability for walker $1$ to touch the top boundary before walker $2$ exits the interval. Making the change of variables $u = \dfrac{k+k''-z'}{k+k'}$, we obtain:
\begin{equation}
    \frac{1}{2} \int_{0}^{1} \dd u \, \rho^+_{\frac{k''}{k+k'},\frac{k''+z}{k+k'}}\left(u\right)\pi^+_{\frac{1}{k+k'+1},\frac{1+(k+k')(1-u)}{k+k'+1}}.
\end{equation}
We now take the large $k$ (scaling limit) behavior of this probability, since we are interested in the asymptotic value of the variance. Introducing the rescaled parameters $x=\frac{k''}{k}$, $r=\frac{k'}{k}$, $\zeta=\frac{z}{k}$ and taking $n\rightarrow +\infty$, we obtain:
\begin{equation} \begin{split}
    \frac{1}{2} \int_{0}^{1} \dd u \, \rho^+_{\frac{x}{1+r},\frac{x+\zeta}{1+r}}\left(u\right)\pi^+_{\frac{1}{n(1+r)},1-u}\\
    \underset{k \rightarrow +\infty}{\sim} \dfrac{1}{2k(1+r)} \int_{0}^{1} \dd u \, \rho^+_{\frac{x}{1+r},\frac{x+\zeta}{1+r}}\left(u\right)\partial_1 \pi^+_{0,1-u}.
\end{split} \end{equation}
After integration over $x$, this probability has a well-defined finite limit as $k\rightarrow +\infty$. Indeed, since $x=\frac{k''}{k}$,
\begin{equation} \begin{split}
    \sum_{k''=0}^{k'} \dfrac{1}{2k(1+r)} \int_{0}^{1} \dd u \, \rho^+_{\frac{x}{1+r},\frac{x+\zeta}{1+r}}\left(u\right)\partial_1 \pi^+_{0,1-u}\\
    \underset{n \rightarrow +\infty}{\rightarrow} \dfrac{1}{2(1+r)} \int_0^r \dd x \int_{0}^{1} \dd u \, \rho^+_{\frac{x}{1+r},\frac{x+\zeta}{1+r}}\left(u\right)\partial_1 \pi^+_{0,1-u}
\end{split} \end{equation}
Terms $2$ to $4$ are treated in a very similar manner, the only changes being swaps/reversals in the arguments of $\rho^+$ and $\pi^+$. For example, trajectories where walker $1$ is also the one touching the bottom boundary have probability:
\begin{equation}
    \frac{1}{2}\int_{k''-k'}^{k+k''} \dd z' \dfrac{1}{k+k'}\rho^+_{\frac{k''+z}{k+k'},\frac{k''}{k+k'}}\left(\dfrac{k+k''-z'}{k+k'}\right)\pi^+_{\frac{1+z'-k''+k'}{k+k'+1},\frac{1}{k+k'+1}},
\end{equation}
leading to another contribution in the scaling limit:
\begin{equation}
    \dfrac{1}{2(1+r)} \int_0^r \dd x \int_{0}^{1} \dd u \, \rho^+_{\frac{x+\zeta}{1+r},\frac{x}{1+r}}\left(u\right)\partial_2 \pi^+_{1-u,0}.
\end{equation}
When the $k+k'$-th site discovered is on the right, we simply reverse the arguments $x\rightarrow 1-x$. This gives:
\begin{equation}
    \frac{1}{2}\int_{k''-k'}^{k+k''} \dd z' \dfrac{1}{k+k'}\rho^+_{\frac{k+k' -k''-z}{k+k'},\frac{k+k'-k''}{k+k'}}\left(\dfrac{z'+k'-k''}{k+k'}\right)\pi^+_{\frac{k+k''-z'+1}{k+k'+1},\frac{1}{k+k'+1}},
\end{equation}
for term $3$, and:
\begin{equation}
    \frac{1}{2}\int_{k''-k'}^{k+k''} \dd z' \dfrac{1}{k+k'}\rho^+_{\frac{k+k' -k''}{k+k'},\frac{k+k'-k''-z}{k+k'}}\left(\dfrac{z'+k'-k''}{k+k'}\right)\pi^+_{\frac{1}{k+k'+1},\frac{k+k''-z'+1}{k+k'+1}},
\end{equation}
for term $4$. The associated scaling limits are:
\begin{equation}
    \dfrac{1}{2(1+r)} \int_0^r \dd x \int_{0}^{1} \dd u \, \rho^+_{1-\frac{x+\zeta}{1+r},1-\frac{x}{1+r}}\left(u\right)\partial_2 \pi^+_{1-u,0},
\end{equation}
and:
\begin{equation}
    \dfrac{1}{2(1+r)} \int_0^r \dd x \int_{0}^{1} \dd u \, \rho^+_{1-\frac{x}{1+r},1-\frac{x+\zeta}{1+r}}\left(u\right)\partial_1 \pi^+_{0,1-u}.
\end{equation}
Term $5$ has a slightly different (and simpler) structure. Here, we simply need walker $1$ to exit trough the top boundary before any of the walkers crosses $0$. This gives, in the scaling limit, a term:
\begin{equation}
    \frac{1}{2}\pi^+_{\frac{1-\zeta}{1+r},\frac{1}{1+r}}.
\end{equation}
Now that we quantified the probabilities of all these processes, we can express the probability that $I_k = I_{k+k'} = 1$ given a certain value of the original gap $z = k\zeta$. The only thing left to do is to inject the asymptotic distribution of the gap. This distribution can be obtained from two-walker splitting probabilities and is expressed nicely as $\rho(\zeta) = 2 \zeta$. This gives:
\begin{equation}
\lim_{k\to\infty} \mathbb{P}\left(I_k = I_{k(1+r)} = 1\right) = f(r)
\end{equation}
with
\begin{tcolorbox}[
colback=white,
colframe=black,
arc=2mm]
\begin{varwidth}{\textwidth}
\begin{multline}
\label{fr:def}
 f(r) = \dfrac{1}{2(1+r)} \int_0^1 \dd \zeta \, 2 \zeta \int_0^r \dd x  \int_0^1 \dd u \left\{ \left[ \rho^+_{\frac{x+\zeta}{1+r},\frac{x}{1+r}}\left(u\right) +  \rho^+_{1-\frac{x+\zeta}{1+r},1-\frac{x}{1+r}}\left(u\right)\right] \partial_2 \pi^+_{1-u,0}\right.\\
    \left. + \left[ \rho^+_{\frac{x}{1+r},\frac{x+\zeta}{1+r}}\left(u\right) +  \rho^+_{1-\frac{x}{1+r},1-\frac{x+\zeta}{1+r}}\left(u\right)\right] \partial_1 \pi^+_{0,1-u}\right\} + \frac{1}{2}\int_0^1 \dd \zeta \, 2 \zeta \, \pi^+_{\frac{1-\zeta}{1+r},\frac{1}{1+r}}.
\end{multline}
\end{varwidth}
\end{tcolorbox}

\subsection{Scaling function $c(r)$}

Defining $c(r)=  f(r)- \frac{1}{4}$ the scaling function for correlations, we can inject the explicit forms for $\rho^+$ and $\pi^+$ to obtain:
\begin{eqnarray}
c(r)
&=&
\frac{8}{1+r}
\int_0^1 d\zeta \, \zeta
\int_0^r dx
\int_0^1 du
\Biggl\{
\sum_{n\ge 1}
\frac{\sin(n\pi u)}{\sinh(n\pi)} 
\sum_{n'\ge 0}
\frac{
\sin\!\bigl((2n'+1)\pi(1-u)\bigr)
}{
\sinh\!\bigl((2n'+1)\pi\bigr)
}\nonumber\\
&&\times
\Biggl[
\sin\!\left(
n\pi \frac{x+\zeta}{1+r}
\right)
\sinh\!\left(
n\pi \frac{x}{1+r}
\right)
+
\sin\!\left(
n\pi - n\pi \frac{x+\zeta}{1+r}
\right)
\sinh\!\left(
n\pi - n\pi \frac{x}{1+r}
\right)
\Biggr]
\nonumber\\
&&+ \sum_{n\ge 1}
\frac{\sin(n\pi u)}{\sinh(n\pi)}
\Biggl[
\sin\!\left(
n\pi \frac{x}{1+r}
\right)
\sinh\!\left(
n\pi \frac{x+\zeta}{1+r}
\right)
+
\sin\!\left(
n\pi - n\pi \frac{x}{1+r}
\right)
\nonumber\\
&&\qquad\qquad\qquad\qquad\times
\sinh\!\left(
n\pi - n\pi \frac{x+\zeta}{1+r}
\right)
\Biggr]
\sum_{n'\ge 0}
\frac{
\sinh\!\bigl((2n'+1)\pi(1-u)\bigr)
}{
\sinh\!\bigl((2n'+1)\pi\bigr)
}
\Biggr\}
\nonumber\\
&&
-\frac14
+4
\int_0^1 d\zeta \, \zeta \,\sum_{n\ge 0}
\frac{
\sin\!\left(
(2n+1)\pi\frac{1-\zeta}{1+r}
\right)
\sinh\!\left(
(2n+1)\pi\frac{1}{1+r}
\right)
}{
(2n+1)\pi\,
\sinh\!\bigl((2n+1)\pi\bigr)
}.
\end{eqnarray}
Integrating first over $u$ and using the identities:
\begin{equation} \begin{split}
& \int_0^1 \dd u \sin(n \pi u) \sin((2n'+1)\pi(1-u)) = \frac{(-1)^{n+1}}{2}\delta_{n,2n'+1}, \\
& \int_0^1 \dd u \sin(n \pi u) \sinh((2n'+1)\pi(1-u)) = \frac{n \sinh((2n'+1)\pi)}{\pi(n^2 + (2n'+1)^2},
\end{split} \end{equation}
one gets:
\begin{eqnarray}
c(r)
&=&
\frac{1}{1+r}
\int_0^1 d\zeta \, \zeta
\int_0^r dx
\Biggl\{
2\sum_{n\geq 1}
4\sum_{n'\geq 0}
\frac{(-1)^{n+1}\delta_{n,\,2n'+1}}
     {2\sinh(n\pi)\sinh((2n'+1)\pi)}
\nonumber\\
&&\times
\Biggl[
\sin\!\left(
n\pi\frac{x+\zeta}{1+r}
\right)
\sinh\!\left(
n\pi\frac{x}{1+r}
\right)
+
\sin\!\left(
n\pi-n\pi\frac{x+\zeta}{1+r}
\right)
\sinh\!\left(
n\pi-n\pi\frac{x}{1+r}
\right)
\Biggr]
\nonumber\\
&&+
2\sum_{n\geq 1}
4\sum_{n'\geq 0}
\frac{
n\,\sinh((2n'+1)\pi)
}{
\pi\bigl(n^2+(2n'+1)^2\bigr)
\sinh((2n'+1)\pi)
\sinh(n\pi)
}
\nonumber\\
&&\times
\Biggl[
\sin\!\left(
n\pi\frac{x}{1+r}
\right)
\sinh\!\left(
n\pi\frac{x+\zeta}{1+r}
\right)
+
\sin\!\left(
n\pi-n\pi\frac{x}{1+r}
\right)
\nonumber\\
&&\qquad\qquad\qquad\qquad\times
\sinh\!\left(
n\pi-n\pi\frac{x+\zeta}{1+r}
\right)
\Biggr]
\Biggr\}
\nonumber\\
&&
-\frac{1}{4}
+4
\int_0^1 d\zeta \, \zeta \,
\sum_{n\geq 0}
\frac{
\sin\!\left(
(2n+1)\pi\frac{1-\zeta}{1+r}
\right)
\sinh\!\left(
(2n+1)\pi\frac{1}{1+r}
\right)
}{
(2n+1)\pi\,
\sinh((2n+1)\pi)
}.
\end{eqnarray}
The sum over $n'$ is straightforward in the first term. In the second term, it is performed using the identity:
\begin{equation}
    \sum_{n'=0}^{+\infty} \dfrac{n }{\pi (n^2 +(2n'+1)^2)} = \dfrac{1}{4} \tanh\left(\dfrac{n \pi}{2}\right).
\end{equation}
This results in:
\begin{eqnarray}
c(r)
&=&
\frac{1}{1+r}
\int_0^1 d\zeta \, \zeta
\int_0^r dx
\Biggl\{
4\sum_{n\geq 0}
\frac{1}{\sinh^2\!\bigl((2n+1)\pi\bigr)}
\nonumber\\
&&\times
\Biggl[
\sin\!\left(
(2n+1)\pi\frac{x+\zeta}{1+r}
\right)
\sinh\!\left(
(2n+1)\pi\frac{x}{1+r}
\right)
+
\sin\!\left(
(2n+1)\pi
-(2n+1)\pi\frac{x+\zeta}{1+r}
\right)
\nonumber\\
&&\qquad\qquad\qquad\qquad\times
\sinh\!\left(
(2n+1)\pi
-(2n+1)\pi\frac{x}{1+r}
\right)
\Biggr]
\nonumber\\
&&+
2\sum_{n\geq 1}
\frac{\tanh\!\left(\frac{n\pi}{2}\right)}
     {\sinh(n\pi)}
\Biggl[
\sin\!\left(
n\pi\frac{x}{1+r}
\right)
\sinh\!\left(
n\pi\frac{x+\zeta}{1+r}
\right)
+
\sin\!\left(
n\pi
-n\pi\frac{x}{1+r}
\right)
\nonumber\\
&&\qquad\qquad\qquad\qquad\times
\sinh\!\left(
n\pi
-n\pi\frac{x+\zeta}{1+r}
\right)
\Biggr]
\Biggr\}
\nonumber\\
&&
-\frac{1}{4}
+ 4
\int_0^1 d\zeta \, \zeta \,
\sum_{n\geq 0}
\frac{
\sin\!\left(
(2n+1)\pi\frac{1-\zeta}{1+r}
\right)
\sinh\!\left(
(2n+1)\pi\frac{1}{1+r}
\right)
}{
(2n+1)\pi\,\sinh\!\bigl((2n+1)\pi\bigr)
}
\end{eqnarray}
We now integrate over $\zeta$ and $x$. The relevant blocks are:
\begin{multline}
    \int_0^1 \dd \zeta 2 \zeta \int_0^r \dd x \left[\sin\left((2n+1) \pi \dfrac{x+\zeta}{1+r}\right) \sinh\left((2n+1)\pi\dfrac{x}{1+r}\right) + \sin\left((2n+1) \pi - (2n+1)\pi \dfrac{x+\zeta}{1+r}\right) \right. \\
    \left. \sinh\left((2n+1) \pi -(2n+1) \pi \dfrac{x}{1+r}\right)\right] =  \frac{(r+1)^3}{(2n+1)^3 \pi^3} \left(-\sin \left(\frac{\pi  (2 n+1)}{r+1}\right)+\frac{\pi  (2 n+1) \cos \left(\frac{\pi  (2
   n+1)}{r+1}\right)}{r+1}+\sinh \left(\frac{\pi  (2 n+1)}{r+1}\right) \right. \\
   +\sinh \left(\frac{\pi  (2 n+1)
   r}{r+1}\right)+\frac{\pi  (2 n+1) \sinh (\pi  (2 n+1)) \sin \left(\frac{\pi  (2 n+1)
   r}{r+1}\right)}{r+1}+\cos \left(\frac{\pi  (2 n+1)}{r+1}\right) \left(-\sinh \left(\frac{\pi  (2
   n+1)}{r+1}\right)\right) \\
   -\sinh (\pi  (2 n+1)) \cos \left(\frac{\pi  (2 n+1) r}{r+1}\right)+\cos
   \left(\frac{\pi  (2 n+1) r}{r+1}\right) \sinh \left(\frac{\pi  (2 n+1) r}{r+1}\right)+\frac{\pi  (2 n+1)
   \cosh (\pi  (2 n+1)) \cos \left(\frac{\pi  (2 n+1) r}{r+1}\right)}{r+1} \\
   +\left(\sin \left(\frac{\pi  (2
   n+1)}{r+1}\right)-\frac{\pi  (2 n+1)}{r+1}\right) \cosh \left(\frac{\pi  (2 n+1)}{r+1}\right)+\cosh (\pi 
   (2 n+1)) \sin \left(\frac{\pi  (2 n+1) r}{r+1}\right)\\
   \left. -\frac{\left((r+1) \sin \left(\frac{\pi  (2 n+1)
   r}{r+1}\right)+\pi  (-2 n-1)\right) \cosh \left(\frac{\pi  (2 n+1) r}{r+1}\right)}{r+1}-\sinh (\pi  (2
   n+1))\right),
\end{multline}
\begin{multline}
    \int_0^1 \dd \zeta 2 \zeta \int_0^r \dd x \left[\sin\left(n \pi \dfrac{x}{1+r}\right) \sinh\left(n\pi\dfrac{x+\zeta}{1+r}\right) + \sin\left(n \pi - n\pi \dfrac{x}{1+r}\right) \sinh\left(n \pi -n \pi \dfrac{x+\zeta}{1+r}\right)\right] \\
    = \frac{(r+1)^3}{n^3 \pi^3} \left(\sin \left(\frac{\pi  n}{r+1}\right)-\frac{\pi  n \cos \left(\frac{\pi 
   n}{r+1}\right)}{r+1}-\sinh \left(\frac{\pi  n}{r+1}\right)+(-1)^n \sinh \left(\frac{\pi  n
   r}{r+1}\right)+\frac{\pi  n \sinh (\pi  n) \sin \left(\frac{\pi  n r}{r+1}\right)}{r+1} \right. \\
   +\cos
   \left(\frac{\pi  n}{r+1}\right) \sinh \left(\frac{\pi  n}{r+1}\right)+\sinh (\pi  n) \cos \left(\frac{\pi
    n r}{r+1}\right)-\cos \left(\frac{\pi  n r}{r+1}\right) \sinh \left(\frac{\pi  n
   r}{r+1}\right)-\frac{\pi  n \cosh (\pi  n) \cos \left(\frac{\pi  n r}{r+1}\right)}{r+1} +\left(\frac{\pi 
   n}{r+1} \right. \\
   \left. \left. -\sin \left(\frac{\pi  n}{r+1}\right)\right) \cosh \left(\frac{\pi  n}{r+1}\right)-\cosh (\pi  n)
   \sin \left(\frac{\pi  n r}{r+1}\right)+\frac{\left((r+1) \sin \left(\frac{\pi  n r}{r+1}\right)+\pi 
   (-1)^n n\right) \cosh \left(\frac{\pi  n r}{r+1}\right)}{r+1}-(-1)^n \sinh (\pi  n)\right),
\end{multline}
and:
\begin{equation}
    \int_0^1 \dd \zeta 2 \zeta \sin\left((2n+1)\pi\dfrac{1-\zeta}{1+r}\right) = \dfrac{2(1+r)}{(2n+1)^2 \pi^2}\left((2n+1)\pi - (1+r) \sin\left(\dfrac{(2n+1)\pi}{1+r}\right) \right).
\end{equation}
We therefore obtain the expression for $c(r)$:
\begin{eqnarray}
c(r)
&=&
2(1+r)^2
\sum_{n\geq 0}
\frac{1}
     {(2n+1)^3\pi^3\sinh^2\!\bigl((2n+1)\pi\bigr)}
\Biggl(
-\sin\!\left(\frac{\pi(2n+1)}{r+1}\right)
+\frac{\pi(2n+1)\cos\!\left(\frac{\pi(2n+1)}{r+1}\right)}{r+1}
\nonumber\\
&&\qquad
+\sinh\!\left(\frac{\pi(2n+1)}{r+1}\right)
+\sinh\!\left(\frac{\pi(2n+1)r}{r+1}\right)
+\frac{\pi(2n+1)\sinh\!\bigl(\pi(2n+1)\bigr)
       \sin\!\left(\frac{\pi(2n+1)r}{r+1}\right)}
      {r+1}
\nonumber\\
&&\qquad
-\cos\!\left(\frac{\pi(2n+1)}{r+1}\right)
 \sinh\!\left(\frac{\pi(2n+1)}{r+1}\right)
-\sinh\!\bigl(\pi(2n+1)\bigr)
 \cos\!\left(\frac{\pi(2n+1)r}{r+1}\right)
\nonumber\\
&&\qquad
+\cos\!\left(\frac{\pi(2n+1)r}{r+1}\right)
 \sinh\!\left(\frac{\pi(2n+1)r}{r+1}\right)
+\frac{\pi(2n+1)\cosh\!\bigl(\pi(2n+1)\bigr)
       \cos\!\left(\frac{\pi(2n+1)r}{r+1}\right)}
      {r+1}
\nonumber\\
&&\qquad
+\left(
\sin\!\left(\frac{\pi(2n+1)}{r+1}\right)
-\frac{\pi(2n+1)}{r+1}
\right)
\cosh\!\left(\frac{\pi(2n+1)}{r+1}\right)
+\cosh\!\bigl(\pi(2n+1)\bigr)
 \sin\!\left(\frac{\pi(2n+1)r}{r+1}\right)
\nonumber\\
&&\qquad
-\frac{
\Bigl(
(r+1)\sin\!\left(\frac{\pi(2n+1)r}{r+1}\right)
-\pi(2n+1)
\Bigr)
\cosh\!\left(\frac{\pi(2n+1)r}{r+1}\right)
}
{r+1}
-\sinh\!\bigl(\pi(2n+1)\bigr)
\Biggr)
\nonumber\\
&&
+(1+r)^2
\sum_{n\geq 1}
\frac{\tanh\!\left(\frac{n\pi}{2}\right)}
     {n^3\pi^3\sinh(n\pi)}
\Biggl(
\sin\!\left(\frac{\pi n}{r+1}\right)
-\frac{\pi n\cos\!\left(\frac{\pi n}{r+1}\right)}{r+1}
-\sinh\!\left(\frac{\pi n}{r+1}\right)
+(-1)^n\sinh\!\left(\frac{\pi nr}{r+1}\right)
\nonumber\\
&&\qquad
+\frac{\pi n\sinh(\pi n)
       \sin\!\left(\frac{\pi nr}{r+1}\right)}
      {r+1}
+\cos\!\left(\frac{\pi n}{r+1}\right)
 \sinh\!\left(\frac{\pi n}{r+1}\right)
+\sinh(\pi n)
 \cos\!\left(\frac{\pi nr}{r+1}\right)
\nonumber\\
&&\qquad
-\cos\!\left(\frac{\pi nr}{r+1}\right)
 \sinh\!\left(\frac{\pi nr}{r+1}\right)
-\frac{\pi n\cosh(\pi n)
       \cos\!\left(\frac{\pi nr}{r+1}\right)}
      {r+1}
+\left(
\frac{\pi n}{r+1}
-\sin\!\left(\frac{\pi n}{r+1}\right)
\right)
\cosh\!\left(\frac{\pi n}{r+1}\right)
\nonumber\\
&&\qquad
-\cosh(\pi n)
 \sin\!\left(\frac{\pi nr}{r+1}\right)
+\frac{
\Bigl(
(r+1)\sin\!\left(\frac{\pi nr}{r+1}\right)
+\pi(-1)^n n
\Bigr)
\cosh\!\left(\frac{\pi nr}{r+1}\right)
}
{r+1}
+(-1)^{n+1}\sinh(\pi n)
\Biggr)
\nonumber\\
&&
-\frac14
+4(1+r)
\sum_{n\geq 0}
\frac{
\sinh\!\left(\frac{(2n+1)\pi}{1+r}\right)
}
{
(2n+1)^3\pi^3\sinh\!\bigl((2n+1)\pi\bigr)
}
\Biggl(
(2n+1)\pi
-(1+r)
\sin\!\left(
\frac{(2n+1)\pi}{1+r}
\right)
\Biggr).
\label{eq 179}
\end{eqnarray}

\subsection{Exact value for the variance}

We now obtain the value of the variance from the two-point correlation function. As the discovery-share is defined as:
\begin{equation}
    X_n = \dfrac{1}{n} \sum_{k=1}^n I_k,
\end{equation}
we have:
\begin{equation}
    \Var[X_n] = \dfrac{1}{n^2} \sum_{k,k'=1}^n \langle I_k I_{k'} \rangle - \langle I_k \rangle \langle I_{k'} \rangle .
\end{equation}
Taking the scaling limit then gives:
\begin{equation}
   \Var[X_n] \underset{n \rightarrow +\infty}{\sim} \dfrac{2}{n^2} \sum_{1\leq k < k' \leq n} c\left(\dfrac{k'-k}{k}\right) \underset{n \rightarrow +\infty}{\rightarrow} 2\int_0^1 \dd u \int_0^u \dd v\, c\left(\dfrac{u-v}{v}\right).
\end{equation}
After some elementary manipulations, we obtain:
\begin{equation}
    \Var[X_\infty] = \int_0^{+\infty} \dd r \dfrac{c(r)}{(1+r)^2}.
\end{equation}
Computing this integral and defining:
\begin{equation}
    \begin{cases}
    f_1(x) = 4 \int_0^x \dfrac{\dd u}{u} [1 - \cos(u)]\sinh(u),  \\
    f_2(x) = 4 \int_0^x \dfrac{\dd u}{u} \sin(u)\sinh(u), \\
    f_3(x) = 4 \int_0^x \dfrac{\dd u}{u} \sin(u) [\sinh(x) - \sinh(x-u)],
    \end{cases}
\end{equation}
we can express the variance as:
\begin{tcolorbox}[
colback=white,
colframe=black,
arc=2mm]
\begin{varwidth}{\textwidth}
\begin{eqnarray}
    \Var[X_\infty] & = & \sum_{n\geq 0}\dfrac{1}{\pi^2(2n+1)^2}\left[\dfrac{f_1((2n+1)\pi)}{\sinh((2n+1)\pi)} - \dfrac{f_2((2n+1)\pi)}{1+\cosh((2n+1)\pi)}\right]- \frac{1}{4} \nonumber \\
    &-& \sum_{n\geq 1}\dfrac{1}{2 n^2 \pi^2 (1+\cosh(n \pi))}\left[f_2(n\pi) + (-1)^{n} f_3(n\pi)\right].
    \label{eq 230}
\end{eqnarray}
\end{varwidth}
\end{tcolorbox}
Numerically, $\Var[X_\infty] = 0.03433$, although numerical evaluations of these sums can be quite intricate.

\section{One-sided geometry, an example of fixed discovery order}\label{sec:one-sided}

This section develops in more detail an alternative to one-dimensional nearest neighbor diffusion, where the two walkers still perform simple random walks along the whole line, but food is only present on the positive part of the axis, such that discoveries are only counted on positive sites. This specification induces a deterministic discovery order where the $n$-th site discovered is exactly at position $n$. This is in stark contrast to the natural setting described in the previous section where discoveries were made on both sides of the origin, and it turns out this is a considerable simplification in the context of the discovery problem, in that we can obtain the full distribution of the limiting random variable $X_\infty$, and this distribution is much simpler than the limiting law in the two-sided case.

We now describe the rewriting used to quantify fluctuations of $X_n$ in the large $n$ limit (and actually for all values of $n$ in this case).

\subsection{The time-lag process}

We start by rewriting the random variables $X_n$ in term of first-passage times by each of the walkers at positive integers. Because the discovery order is fixed, $I_k$ equals $1$ if and only if walker $1$ reached site $k$ before walker $2$. We therefore have:
\begin{equation}
    X_n = \dfrac{1}{n} \sum_{k=1}^n I_k = \dfrac{1}{n} \sum_{k=1}^n \mathds{1}_{T_1(k) < T_2(k)},
\end{equation}
where $T_i(x)$ denotes the first-passage time by walker $i$ at site $x$. 

We now define the time-lag process $\{Z_k\}_{k\geq 0}$ according to:
\begin{equation}
    Z_k = T_2(k) - T_1(k),
\end{equation}
such that:
\begin{equation}
    X_n = \dfrac{1}{n} \sum_{k=1}^n \mathds{1}_{Z_k \geq 0}.
\end{equation}
We therefore mapped the fraction of discovered sites to the fraction of "time" spent above $0$ by an auxiliary random process. This definition of the time-lag process is not new and was already used in \cite{randon-furling_markovian_2015} to obtain the long-time decay of the ordering probability of maxima of independent Brownian walkers. As was done there, we now prove that this process is a jump process, namely that the increments $Z_{k+1}- Z_k$ are independent and identically distributed. Such a property is quite natural, as each of the increments $T_i(k+1) - T_i(k)$ describe the time needed for one walker to go up by one unit, so they are identically distributed by translational invariance and independent because of the Markovian nature of the dynamics. Furthermore, because the two walkers are independent, this structure transfers to the time-lag process $Z_k$ as it is the difference of two independent jump processes.

To fully characterize $Z$, we only need to specify the increment statistics, namely the law of $Z_1$. The distribution of the first-passage time at site $1$ by a simple random walk starting at $0$ is a classic \cite{hughes_random_2023} quantity most commonly expressed using generating functions:
\begin{equation}
    \mathbb{E}\left[\xi^{T(1)}\right] = \dfrac{1-\sqrt{1-\xi^2}}{\xi}.
\end{equation}
Plugging in $\xi = e^{i k}$ gives the characteristic function:
\begin{equation}
    \mathbb{E}\left[e^{i k T(1)}\right] = e^{-i k} - e^{-i k}\sqrt{1-e^{2 i k}}
\end{equation}
As the first-passage times for walker $1$ and $2$ are independent and identically distributed for symmetrical walkers, we immediately obtain the characteristic function of $Z_1$:
\begin{equation}
    \mathbb{E}\left[e^{i k Z_1}\right] = \mathbb{E}\left[e^{i k T(1)}\right] \mathbb{E}\left[e^{-i k T(1)}\right] = \left(1-\sqrt{1-e^{2 i k}}\right) \left(1-\sqrt{1-e^{-2 i k}}\right).
\end{equation}

\subsection{scaling limit of the time-lag process and limit theorems for the occupation time}

To obtain the asymptotic distribution of the discovery-share $X_n$, one thus needs to express the asymptotic distribution of the time spent above $0$ by a specific jump processes. This is a central problem in random walk theory, with fundamental results dating back to \cite{levy_sur_1940}. In particular, for all symmetric jump processes, the fraction of time spent above zero is asymptotically distributed according to the celebrated arcsine law. If $\{X_k\}_{\{k \geq 0 \} }$ is a symmetric jump process, then:
\begin{equation}
    \mathbb{P}\left(\dfrac{1}{n} \sum_{k=1}^n \mathds{1}_{X_k \geq 0} \leq x \right) \underset{n \rightarrow +\infty}{\rightarrow} \dfrac{2}{\pi}\arcsin\left(\sqrt{x}\right),
\end{equation}
with the associated limiting probability density:
\begin{equation}
    \rho(x) = \dfrac{1}{\pi \sqrt{x(1-x)}}.
\end{equation}
For the discovery-share, we therefore prove the following. If the two Brownian walkers are identical and discoveries are only counted on one side of the origin, then the discovery-share is asymptotically arcsine distributed:
\begin{equation}
\tcbhighmath[
colback=white,
colframe=black,
arc=2mm]{
    \dfrac{1}{n} \sum_{k=1}^n I_k \overset{\mathcal{L}}{\longrightarrow} \Beta\left(\frac{1}{2},\frac{1}{2}\right).
    }
\end{equation}
The associated limiting density diverges at $x=0$ and $1$, such that trajectories where one walker discovers nearly all the sites are the most probable. This is easily understood in the one-sided case, since discovering sites gives an clear advantage for further discoveries along the same direction. In the two-sided case, because there are two different directions for exploration, equal share become more typical. One can compare analytical predictions for the variance, as we have for the arcsine law:
\begin{equation}
    \Var[X_\infty] = \dfrac{1}{8} = 0.125,
\end{equation}
a value much larger than $\Var[X_\infty] \approx 0.03433$ in the two-sided case. When discoveries are only possible on one side of the starting point, highly unequal situations where one walker discovered nearly all the sites and $X_n$ deviates a lot from $\frac{1}{2}$ become favored compared to equal share situations.

This subsection's results can be generalized to the physically relevant case where the walkers are not identical but have different hopping rates. In the scaling limit, this translates into two different diffusion coefficients $D_1$ and $D_2$. The time-lag process is not symmetric anymore in this regime. The characteristic function of the increment $Z_1$ is now expressed as:
\begin{equation}
    \mathbb{E}\left[e^{i k Z_1}\right] = e^{i k \frac{D_2-D_1}{D_1 D_2}} \left(1-\sqrt{1-e^{2 i \frac{k}{D_2}}}\right) \left(1-\sqrt{1-e^{-2 i \frac{k}{D_1}}}\right).
\end{equation}
The asymptotic occupation time statistics for $Z$ are fixed by its scaling limit. To identify this scaling limit, we expand the characteristic function at small $k$:
\begin{equation}
    \mathbb{E}\left[e^{i k Z_1}\right] \underset{k \rightarrow 0}{=} 1 - \dfrac{\sqrt{D_1}+\sqrt{D_2}}{\sqrt{D_1 D_2}} \left(1 - i \dfrac{\sqrt{D_1}-\sqrt{D_2}}{\sqrt{D_1}+\sqrt{D_2}} \sgn{k} \right) \sqrt{|k|} + \mathcal{O}(k),
\end{equation}
where we wrote the short-$k$ correction in the canonical form for stable processes.

This means that the time-lag process $Z$ asymptotically behaves (in the sense of the generalized central limit theorem) as a stable process with parameters $\alpha = \frac{1}{2}$ and $\beta = \dfrac{\sqrt{D_1}-\sqrt{D_2}}{\sqrt{D_1}+\sqrt{D_2}}$. In the case $D_1 = D_2$, this process is symmetric and its occupation time is asymptotically arcsine distributed, but in the general case we have the following generalization \cite{spitzer_combinatorial_1956}:
\begin{equation}
\tcbhighmath[
colback=white,
colframe=black,
arc=2mm]{
    \dfrac{1}{n} \sum_{k=1}^n I_k \overset{\mathcal{L}}{\longrightarrow} \Beta\left(c,1-c\right),
    }
\end{equation}
where:
\begin{equation}
    c = \frac{1}{2}+\frac{2}{\pi}\arctan\left(\dfrac{\sqrt{D_1}-\sqrt{D_2}}{\sqrt{D_1}+\sqrt{D_2}}\right) = \dfrac{2}{\pi}\arctan\left(\sqrt{\dfrac{D_1}{D_2}}\right).
\end{equation}
In the heterogeneous case, the mean share is not $\frac{1}{2}$ anymore but is related to the difference in search efficiency by the walkers:
\begin{equation}
    \langle X_n \rangle \underset{n \rightarrow +\infty}{\rightarrow} \dfrac{2}{\pi}\arctan\left(\sqrt{\dfrac{D_1}{D_2}}\right).
\end{equation}
The limiting density of $X_n$ still exhibits divergences at $x=0$ and $1$, such that trajectories where one walker discovers nearly all the sites (even the slower one) are more probable than trajectories where $X_n$ is close to its mean.

The rewriting in term of first-passage times used in this subsection to obtain limiting distributions of the discovery-share crucially uses the deterministic order of discoveries in the one-sided geometry. In the two-sided case, the presence of left and right discoveries means the share is not controlled only by first-passing times at integers, but also by the random order in which sites themselves are discovered by the walkers. Whereas the joint law of all first-passing times at positive sites by a simple one-dimensional random walk is described by a Markovian jump process, the joint law of first-passing times at all sites lacks such a structure.

\section{Structure of correlations and explicit scaling functions}

In this section, we prove the results shown in Eq.~\eqref{eq:cov_all} in the main text and show examples of the scaling functions $c(r)$ for some relevant universality classes. As always, we will separate between the recurrent case $d_s < 2$ and the marginal/ transient case $d_s \geq 2$. It should be noted that results summarized in Eq.~\eqref{eq:cov_all} and developed in this section are valid in the regime where both $i$, $j$ and the time difference $i-j$ are large and of comparable order. In particular, correlations between $I_i$ and $I_j$ for $i$, $j$ large but with $\mathcal{O}(1)$ difference is not contained in our approach. This distinction is especially relevant in the strongly transient case, where such short-range correlations dominate the variance at leading order and fix both the decay and the prefactor.

\subsection{Transient case}

As the structure of correlations was already used to compute the prefactor of the variance in the intermediate case, we start with the transient case, more precisely from Eq.~\eqref{eq 43}. In this exact expression, the correlation between discovery at times $i$ and $j$ is summed over $i$ and $j$ to obtain the variance. Identifying the spatial integral with the covariance expansion:
\begin{equation}
    \Var[X_t] \sim 2 \sum_{i=0}^t \sum_{j=1}^{t-i} \dfrac{1}{t^2}\mathrm{Cov}(I_i,I_{i+j})
\end{equation}
and renormalizing by the mean range squared, we can therefore extract the exact formula (writing the time difference $j-i$ as $j$):
\begin{equation}
    \mathrm{Cov}(I_i, I_{i+j}) = \dfrac{g_0^2}{8} \sum_{x,y} F^{\{y\}}_i(x) F_j(y-x) \sum_{z \neq x,y} P^{\{x,y\}}_i(z)S^{\{y-z\}}_j \\
    -  \dfrac{g_0^2}{8} \sum_{x,y} F^{\{y\}}_i(x) S_j^{\{y-x\}} \sum_{z \neq x,y} P^{\{x,y\}}_i(z)F_j(y-z).
    \label{eq 246}
\end{equation}
Our goal is thus to obtain the scaling behavior of this sum in the regime where $i$ and $j$ are large and of comparable order. The computation of the prefactor of the variance in the intermediate regime involved a continuous version of this sum, where it was dominated by terms with $x,y,z$ large. In this case, we obtained an explicit expression for the triple spatial integrals involving the overlap function $\Theta(u,v,w)$, after two cancellations at first and second order.
Therefore, as soon as the spatial sum expressing the covariance is dominated by scaling terms and can be replaced by an integral, we already computed the scaling behavior of the covariance. For now, we compute the scaling function in this regime.

Replacing the spatial sums by integrals and the first-passage quantities appearing in Eq.~\eqref{eq 246} by their scaling expansions gives with the same notations as in section \ref{sec:quant}:
\begin{multline}
    \mathrm{Cov}(I_i,I_{i+j}) \sim \dfrac{g_0^2}{8} \iiint \dd x \dd y \dd z \left[\dfrac{1}{ g_0 } p\left(x\right) - \dfrac{1}{ g_0^2 i^{\frac{d_s}{2}-1}} g_3\left(y ,(x-y)\right)  + \dfrac{1}{ g_0^3 i^{d_s-2}} g_4\left(x ,(y-x) , (x-y) \right) \right] \\
    \left[p\left(z \right) - \dfrac{1}{ g_0 i^{\frac{d_s}{2}-1}}  \left\{ g_3\left(x,(z-x)\right) + g_3\left(y ,(z-y) \right) \right\} + \dfrac{1}{g_0^2  i^{d_s-2}} \left\{ g_4\left(x,(y-x) , (z-y) \right) + g_4\left(y ,(x-y) , (z-x)\right) \right\} \right] \\
    \left[ \dfrac{i^{\frac{d_s}{2}}}{g_0 j^{\frac{d_s}{2}}} p\left((y-x) i^{\frac{d_s}{2d}} j^{-\frac{d_s}{2d}}\right) \left\{1 - \dfrac{1}{g_0 j^{\frac{d_s}{2}-1}} g_2\left((y-z) i^{\frac{d_s}{2d}} j^{-\frac{d_s}{2d}}\right) \right\} \right. \\
    \left. - \dfrac{i^{\frac{d_s}{2}}}{g_0 j^{\frac{d_s}{2}}} p\left((y-z) i^{\frac{d_s}{2d}} j^{-\frac{d_s}{2d}}\right) \left\{1 - \dfrac{1}{g_0 j^{\frac{d_s}{2}-1}} g_2\left((y-x) i^{\frac{d_s}{2d}} j^{-\frac{d_s}{2d}}\right) \right\} \right].
\end{multline}
Introducing $r= \frac{j}{i}$ and expanding the third order terms since we already showed that these are the first ones to give a non-zero contribution, one gets:
\begin{multline}
    \mathrm{Cov}(I_i,I_{i+j}) \sim \dfrac{g_0^2}{8} \iiint \dd x \dd y \dd z \left\{ \dfrac{1}{ g_0^3 i^{d_s-2}} g_4\left(x ,(y-x) , (x-y) \right) p(z)  \left[\dfrac{1}{g_0 r^{\frac{d_s}{2}}} p\left((y-x) r^{-\frac{d_s}{2d}}\right) - \dfrac{1}{g_0 r^{\frac{d_s}{2}}} p\left((y-z) r^{-\frac{d_s}{2d}}\right) \right] \right. \\
    + \dfrac{1}{ g_0 } p\left(x\right) \dfrac{1}{g_0^2  i^{d_s-2}} \left[ g_4\left(x,(y-x) , (z-y) \right) + g_4\left(y ,(x-y) , (z-x)\right) \right] \left[\dfrac{1}{g_0 r^{\frac{d_s}{2}}} p\left((y-x) r^{-\frac{d_s}{2d}}\right) - \dfrac{1}{g_0 r^{\frac{d_s}{2}}} p\left((y-z) r^{-\frac{d_s}{2d}}\right) \right] \\
    + \dfrac{1}{ g_0^2 i^{\frac{d_s}{2}-1}} g_3\left(y ,(x-y)\right) \dfrac{1}{ g_0 i^{\frac{d_s}{2}-1}}  \left[ g_3\left(x,(z-x)\right) + g_3\left(y ,(z-y) \right) \right] \left[\dfrac{1}{g_0 r^{\frac{d_s}{2}}} p\left((y-x) r^{-\frac{d_s}{2d}}\right) - \dfrac{1}{g_0 r^{\frac{d_s}{2}}} p\left((y-z) r^{-\frac{d_s}{2d}}\right) \right] \\
    - \left[\dfrac{1}{ g_0^2 i^{\frac{d_s}{2}-1}} g_3\left(y ,(x-y)\right) p(z) + \dfrac{1}{ g_0 } p\left(x\right) \dfrac{1}{ g_0 i^{\frac{d_s}{2}-1}}  \left[ g_3\left(x,(z-x)\right) + g_3\left(y ,(z-y) \right) \right] \right] \\
    \left. \left[- \dfrac{1}{g_0 r^{\frac{d_s}{2}}} p\left((y-x) r^{-\frac{d_s}{2d}}\right) \dfrac{1}{g_0 j^{\frac{d_s}{2}-1}} g_2\left((y-z) r^{-\frac{d_s}{2d}}\right) + \dfrac{1}{g_0 r^{\frac{d_s}{2}}} p\left((y-z) r^{-\frac{d_s}{2d}}\right) \dfrac{1}{g_0 j^{\frac{d_s}{2}-1}} g_2\left((y-x) r^{-\frac{d_s}{2d}}\right) \right] \right\}.
\end{multline}
To match the form used in the main text, we factor out a factor $j^{2-d_s}$ such that the integral is now only a function of $r$, as claimed:
\begin{multline}
    \mathrm{Cov}(I_i,I_{i+j}) \sim \dfrac{1}{8 g_0^2 j^{d_s-2}} \iiint \dd x \dd y \dd z \left\{ r^{d_s-2} g_4\left(x ,(y-x) , (x-y) \right) p(z)  \left[\dfrac{1}{r^{\frac{d_s}{2}}} p\left((y-x) r^{-\frac{d_s}{2d}}\right) - \dfrac{1}{r^{\frac{d_s}{2}}} p\left((y-z) r^{-\frac{d_s}{2d}}\right) \right] \right. \\
    + p\left(x\right) r^{d_s-2} \left[ g_4\left(x,(y-x) , (z-y) \right) + g_4\left(y ,(x-y) , (z-x)\right) \right] \left[\dfrac{1}{r^{\frac{d_s}{2}}} p\left((y-x) r^{-\frac{d_s}{2d}}\right) - \dfrac{1}{ r^{\frac{d_s}{2}}} p\left((y-z) r^{-\frac{d_s}{2d}}\right) \right] \\
    +  r^{d_s-2} g_3\left(y ,(x-y)\right) \left[ g_3\left(x,(z-x)\right) + g_3\left(y ,(z-y) \right) \right] \left[\dfrac{1}{r^{\frac{d_s}{2}}} p\left((y-x) r^{-\frac{d_s}{2d}}\right) - \dfrac{1}{ r^{\frac{d_s}{2}}} p\left((y-z) r^{-\frac{d_s}{2d}}\right) \right] \\
    - r^{\frac{d_s}{2}-1}\left[g_3\left(y ,(x-y)\right) p(z) +  p\left(x\right) \left[ g_3\left(x,(z-x)\right) + g_3\left(y ,(z-y) \right) \right] \right] \\
    \left. \left[- \dfrac{1}{ r^{\frac{d_s}{2}}} p\left((y-x) r^{-\frac{d_s}{2d}}\right) g_2\left((y-z) r^{-\frac{d_s}{2d}}\right) + \dfrac{1}{ r^{\frac{d_s}{2}}} p\left((y-z) r^{-\frac{d_s}{2d}}\right)  g_2\left((y-x) r^{-\frac{d_s}{2d}}\right) \right] \right\}.
\end{multline}
The only step left is to relate the time difference $j$ to the difference in discovered sites $l$. This is easily done in the self-averaging case using $j \sim \frac{g_0 l}{2}$. The integral part of the covariance is left unchanged, since $i$ and $j$ are rescaled in the same way and thus $r$ is not modified. This proves the form shown in the main text for the covariance:
\begin{equation}
\tcbhighmath[
colback=white,
colframe=black,
arc=2mm]{
    \mathrm{Cov}(I_k,I_{k+l}) \sim \dfrac{1}{l^{d_s-2}} c\left(\dfrac{l}{k}\right),
    }
    \label{eq: scaling behavior covariance inter}
\end{equation}
with the scaling function given by:
\begin{multline}
    c(r) = \dfrac{2^{d_s}}{32 g_0^{d_s}} \iiint \dd x \dd y \dd z \left\{ r^{d_s-2} g_4\left(x ,(y-x) , (x-y) \right) p(z)  \left[\dfrac{1}{r^{\frac{d_s}{2}}} p\left((y-x) r^{-\frac{d_s}{2d}}\right) - \dfrac{1}{r^{\frac{d_s}{2}}} p\left((y-z) r^{-\frac{d_s}{2d}}\right) \right] \right. \\
    + p\left(x\right) r^{d_s-2} \left[ g_4\left(x,(y-x) , (z-y) \right) + g_4\left(y ,(x-y) , (z-x)\right) \right] \left[\dfrac{1}{r^{\frac{d_s}{2}}} p\left((y-x) r^{-\frac{d_s}{2d}}\right) - \dfrac{1}{ r^{\frac{d_s}{2}}} p\left((y-z) r^{-\frac{d_s}{2d}}\right) \right] \\
    +  r^{d_s-2} g_3\left(y ,(x-y)\right) \left[ g_3\left(x,(z-x)\right) + g_3\left(y ,(z-y) \right) \right] \left[\dfrac{1}{r^{\frac{d_s}{2}}} p\left((y-x) r^{-\frac{d_s}{2d}}\right) - \dfrac{1}{ r^{\frac{d_s}{2}}} p\left((y-z) r^{-\frac{d_s}{2d}}\right) \right] \\
    - r^{\frac{d_s}{2}-1}\left[g_3\left(y ,(x-y)\right) p(z) +  p\left(x\right) \left[ g_3\left(x,(z-x)\right) + g_3\left(y ,(z-y) \right) \right] \right] \\
    \left. \left[- \dfrac{1}{ r^{\frac{d_s}{2}}} p\left((y-x) r^{-\frac{d_s}{2d}}\right) g_2\left((y-z) r^{-\frac{d_s}{2d}}\right) + \dfrac{1}{ r^{\frac{d_s}{2}}} p\left((y-z) r^{-\frac{d_s}{2d}}\right)  g_2\left((y-x) r^{-\frac{d_s}{2d}}\right) \right] \right\}.
\end{multline}
Of course, the expression of $c(r)$ can be simplified, exactly as was done in the computation of the variance prefactor. Performing exactly the same steps (cancelling out terms because of symmetries and rewriting everything in term of $\Theta$) gives:
\begin{tcolorbox}[
colback=white,
colframe=black,
arc=2mm]
\begin{varwidth}{\textwidth}
\begin{multline}
    c(r) = \dfrac{2^{d_s} r^{d_s-2}}{32 g_0^{d_s}} \int_0^1 \dd a \left\{ \int_0^a \dd b \left[ \Theta(a-b,1-a,r) - \Theta(a-b,1-a,1+r+b) \right. \right. \\
    \left. \left. + \Theta(a-b,1+b,r) - \Theta(a-b,1+b,1+r-a) \right] + \int_0^1 \dd b \left[ \Theta(1-a,a+b,r) - \Theta(1-a,a+b,1+r-b) \right] \right \}.
    \label{eq: covariance scaling function transient}
\end{multline}
\end{varwidth}
\end{tcolorbox}
For the specific case of one-dimensional Levy walks, we simplify the integral over $\Theta$ exactly as for the integral involved in the variance prefactor. Expressing $\Theta$ in Fourier space and evaluating over $b$ and $a$ gives:
\begin{equation}
    c(r) = \dfrac{2^{d_s} r^{d_s-2}}{32 (2 \pi)^2 g_0^{d_s} D^{d_s}} \int_0^{+\infty} \dd k \int_0^{+\infty} \dd k' I\left(k^{\frac{2}{d_s}},k'^{\frac{2}{d_s}},(k+k')^{\frac{2}{d_s}},r\right),
\end{equation}
with the integrand:
\begin{multline}
    I(x,y,z,r) = \frac{2 e^{-(r+2) (x+y+z)}}{x y
   z (x-y) (x-z) (y-z)} \\
   \left(e^{2 z} \left(e^{2 y} (y-z) \left(e^{2 x} (x-y) (x-z)-y z\right)+e^{2 x} x z (x-z)\right)-x y
   e^{2 (x+y)} (x-y)\right) \left(x e^{r (y+z)}+y e^{r (x+z)}+z e^{r (x+y)}\right).
\end{multline}
Setting $k' = k u$, we can perform the integration over $k$ resulting in:
\begin{equation}
    c(r) = \dfrac{2^{d_s} r^{d_s-2}}{32 (2 \pi)^2 g_0^{d_s} D^{d_s}} \int_0^{+\infty} \dd u K\left(u^{\frac{2}{d_s}},(1+u)^{\frac{2}{d_s}},r\right),
\end{equation}
with:
\begin{multline}
    K(x,y,r) = \frac{d_s \Gamma \left(d_s-2\right)}{(x-1) x (y-1) y (x-y)} \left(x^2 y r^{2-d_s}-x^2 r^{2-d_s}-x y^2 r^{2-d_s}+x r^{2-d_s}+y^2
   r^{2-d_s}-y r^{2-d_s} \right. \\
   +r^2 x^5 y (r x)^{-d_s}-r^2 x^5 (r x)^{-d_s}-r^2 x^4 y^2 (r x)^{-d_s}+r^2 x^4 (r
   x)^{-d_s}+r^2 x^3 y^2 (r x)^{-d_s}-r^2 x^3 y (r x)^{-d_s}+r^2 x^2 y^4 (r y)^{-d_s} \\
   -r^2 x^2 y^3 (r
   y)^{-d_s}-r^2 x y^5 (r y)^{-d_s}+r^2 x y^3 (r y)^{-d_s}+r^2 y^5 (r y)^{-d_s}-r^2 y^4 (r y)^{-d_s}-(r+2)^2
   x^3 y^2 ((r+2) x)^{-d_s} \\
   +x^3 (r x+2 y)^{2-d_s}-x^3 y (r x+2)^{2-d_s}+(r+2)^2 x^3 y ((r+2)
   x)^{-d_s} +(r+2)^2 x^2 y^3 ((r+2) y)^{-d_s}-x^2 y^2 (r y+2)^{2-d_s}\\
   +x^2 y^2 (r x+2)^{2-d_s}+x^2 (r+2
   y)^{2-d_s}-x^2 (r x+2 y)^{2-d_s}-x^2 y (r+2)^{2-d_s}+x y^3 (r y+2)^{2-d_s}-(r+2)^2 x y^3 ((r+2)
   y)^{-d_s}\\
   -(y-1) y^2 (r y+2 x)^{2-d_s}-y^2 (r+2 x)^{2-d_s}+x y^2 (r+2)^{2-d_s}-x (r+2 y)^{2-d_s}+y (r+2
   x)^{2-d_s}\left. \right).
\end{multline}
Such an expression, although quite lengthy, allows for easy numerical evaluation of the scaling function for $1$-dimensional Levy walks in the intermediate regime. Coming back to the assumption we made at the start of this subsection of the spatial sum being dominated by scaling terms, it turns out the integral over $u$ needed to compute $c(r)$ is divergent when $d_s \geq 4$. This does not signal that a regularization is needed, but is really the indicator that the scaling behavior:
\begin{equation}
    \mathrm{Cov}(I_k,I_{k+l}) \sim \dfrac{1}{l^{d_s-2}} c\left(\dfrac{l}{k}\right)
    \label{eq 213}
\end{equation}
with $c$ given by Eq.~\eqref{eq: covariance scaling function transient} does not hold anymore when $d_s \geq 4$. The large scale structure of correlations takes a very different form above this threshold, and is probably asymptotically stationary. However this regime is not contained in our scaling approach and is not relevant for describing fluctuations of the discovery-share.\\
The scaling behavior Eq.~\eqref{eq 213}, however, is well and truly valid for transient walks with $2<d_s<4$, even above $d_s > 3$. However, it dominates the variance only when $2 \leq d_s < 3$, whereas for $d_s \geq 3$ the variance is dominated by terms where $l \ll k$ that produce a $\mathcal{O}(n)$ contribution. 

We now compute the asymptotics of $c(r)$ in the special case of $1$-dimensional Lévy walks. For $r\rightarrow 0$, we have:
\begin{equation}
    K\left(u^{\frac{2}{d_s}},(1+u)^{\frac{2}{d_s}},r\right) \sim \dfrac{d_s \Gamma(d_s-2)}{r^{d_s-2} (xy)^{1+d_s}} (x^{d_s} y^3 + x^3 y^{d_s} + (xy)^{d_s}),
\end{equation}
and after replacing $x$ and $y$ and integrating over $u$, one obtains:
\begin{equation}
    c(r) \underset{r \sim 0}{\sim} \dfrac{2^{d_s} r^{d_s-2}}{32 (2 \pi)^2 g_0^{d_s} D^{d_s}}\frac{-\pi ^{3/2} 16^{\frac{1}{d_s}} r^{2-d_s} \tan \left(\frac{\pi }{d_s}\right) \sec \left(\frac{2 \pi
   }{d_s}\right) \Gamma \left(d_s-2\right)}{\Gamma \left(\frac{3}{2}-\frac{2}{d_s}\right) \Gamma
   \left(1+\frac{2}{d_s}\right)}.
\end{equation}
Thus $c(0)$ is finite with value:
\begin{equation}
    c(0) = -\dfrac{2^{d_s}16^{\frac{1}{d_s}} \tan \left(\frac{\pi }{d_s}\right) \sec \left(\frac{2 \pi
   }{d_s}\right) \Gamma \left(d_s-2\right)}{128 \sqrt{\pi} g_0^{d_s} D^{d_s} \Gamma \left(\frac{3}{2}-\frac{2}{d_s}\right) \Gamma
   \left(1+\frac{2}{d_s}\right)}.
\end{equation}
In the limit $r \sim +\infty$, we have:
\begin{equation} \begin{split}
    K\left(u^{\frac{2}{d_s}},(1+u)^{\frac{2}{d_s}},r\right) & \underset{v=r u^{\frac{2}{d_s}}}{=} \dfrac{d_s}{2 r^{\frac{d_s}{2}}} \int_0^{+\infty} \dd v \, v^{\frac{d_s}{2}-1} K\left(\frac{v}{r},\left(1+\left(\frac{v}{r}\right)^{\frac{d_s}{2}}\right)^{\frac{2}{d_s}},r\right) \\
    & \sim \dfrac{d_s^2 \Gamma(d_s-2)}{r^{1+\frac{d_s}{2}}}\int_0^{+\infty} \dd v \frac{(v+2) \left(d_s+v\right) v^{d_s}+v (v+2)^{d_s} \left(d_s-v-2\right)}{(v+2)^{d_s}v^{\frac{d_s}{2}}},
\end{split} \end{equation}
such that after integrating over $v$ and including the prefactor of $c(r)$:
\begin{equation} \begin{split}
    c(r) & \underset{r \sim +\infty}{\sim}  \dfrac{2^{d_s} r^{d_s-2}}{32 (2 \pi)^2 g_0^{d_s} D^{d_s}} \dfrac{2^{5-\frac{d_s}{2}} d_s \Gamma\left(\frac{d_s}{2}+1\right)^2}{(d_s-6)(d_s-4) r^{1+\frac{d_s}{2}}} \\
   & \sim \dfrac{2^{\frac{d_s}{2}} d_s \Gamma\left(1+\frac{d_s}{2}\right)^2}{(2\pi)^2 g_0^{d_s} D^{d_s} (d_s-6)(d_s-4) r^{3-\frac{d_s}{2}}}.
\end{split} \end{equation}
Using these asymptotics and including the prefactor in front of the scaling function, we can extract the leading order for the decorrelation of $I_k$ and $I_{k+l}$ at very large $l$:
\begin{equation}
    \mathrm{Cov}(I_k,I_{k+l}) \underset{l \gg k}{\sim} \dfrac{2^{\frac{d_s}{2}} d_s \Gamma\left(1+\frac{d_s}{2}\right)^2 k^{3-\frac{d_s}{2}}}{(2\pi)^2 g_0^{d_s} D^{d_s} (6-d_s)(4-d_s) l^{1+\frac{d_s}{2}}}.
\end{equation}

These asymptotics could be generalized for higher-dimensional universality classes, although the integral involved in the general expression Eq.~\eqref{eq: covariance scaling function transient} becomes harder to evaluate for these classes in the general case.

\subsection{Marginal case}

The scaling behavior Eq.~\eqref{eq: scaling behavior covariance inter} also generalizes immediately to marginal walks, with the coefficient $g_0^{d_s}$ in $c$ factoring into a $(\log{l})^2$ in front of the scaling function and a $p(0)^2$ in $c$. This gives for marginal walks:
\begin{equation}
\tcbhighmath[
colback=white,
colframe=black,
arc=2mm]{
    \mathrm{Cov}(I_k,I_{k+l}) \sim \dfrac{1}{(\log{l})^2} c\left(\dfrac{l}{k}\right),
    }
\end{equation}
with $c$ only slightly modified:
\begin{tcolorbox}[
colback=white,
colframe=black,
arc=2mm]
\begin{varwidth}{\textwidth}
\begin{multline}
    c(r) = \dfrac{1}{8 p(0)^2} \int_0^1 \dd a \left\{ \int_0^a \dd b \left[ \Theta(a-b,1-a,r) - \Theta(a-b,1-a,1+r+b) \right. \right. \\
    \left. \left. + \Theta(a-b,1+b,r) - \Theta(a-b,1+b,1+r-a) \right] + \int_0^1 \dd b \left[ \Theta(1-a,a+b,r) - \Theta(1-a,a+b,1+r-b) \right] \right \}.
    \label{eq: covariance scaling function marginal}
\end{multline}
\end{varwidth}
\end{tcolorbox}
The marginal case covers two very important universality classes, namely $1$-dimensional Cauchy walks and $2$-dimensional diffusive walks that converge towards Brownian motion. We thus compute $c(r)$ for these $2$ walks. Here, we will directly compute $\Theta$ as propagators in real space are easily expressed for these universality classes.

For $1$-dimensional Cauchy walks, $p(x) = \frac{D}{\pi(D^2 + x^2)}$, such that:
\begin{equation}
    \Theta(u,v,w) = \dfrac{u+v+w}{D^2 \pi^2 (u+v)(u+w)(v+w)}.
\end{equation}
The two integrals over $a$ and $b$ can be performed explicitly using Mathematica. Because $p(0) = \frac{1}{\pi D}$, there is no dependence on the generalized diffusion coefficient and we obtain:
\begin{equation}
\tcbhighmath[
colback=white,
colframe=black,
arc=2mm]{
    c(r) = \dfrac{1}{4}\Li_2\left(\dfrac{r}{2(1+r)}\right) + \dfrac{1}{8}\log\left(\dfrac{r}{2(1+r)}\right)^2 - \dfrac{\pi^2}{48},
    }
\end{equation}
where $\Li_2$ is the dilogarithm. This function decreases monotonically with $r$ and possesses the asymptotics:
\begin{multline}
\begin{cases}
c(r) \underset{r \rightarrow 0}{\sim} \dfrac{(\log{r})^2}{8},\\[10pt]
c(r) \underset{r \rightarrow +\infty}{\sim} \dfrac{1}{4 r^2}.
\end{cases}
\end{multline}

For $2$-dimensional Brownian motion, one gets using $p(x) = \frac{e^{-\frac{|x|^2}{4D}}}{4\pi D}$:
\begin{equation}
\Theta(u,v,w) = \dfrac{1}{16 D^2 \pi^2 (u v + u w + v w)}.
\end{equation}
Subsequent integration over $a$ and $b$ is significantly harder than in the Cauchy case. The scaling function $c(r)$ involves dilogarithms of more complicated arguments. If we define the following functions:
\begin{equation}
\begin{split}
	K(r) & = \log\left(\dfrac{r}{1+r}\right)^2 -\log\left(1+r+\sqrt{2r+1}\right)		\log\left(\dfrac{r}{1+r}\right)\\
	& + \Li_2\left(\dfrac{1}{1+r}\right) - \Li_2\left(\dfrac{1+r}{1+r+\sqrt{1+2r}}\right)+\Li_2\left(1-\dfrac{\sqrt{1+2r}}{1+r}\right),
	\end{split}
\end{equation}
\begin{equation}
	C(r) = \Cl_2\left(2r+\dfrac{2\pi}{3}\right) - \Cl_2\left(2r-\dfrac{2\pi}{3}\right),
\end{equation}
and:
\begin{equation}
	H(x,y) = (\log{y})^2 - \log{x}\log{y} - \Li_2(x y) + \Li_2\left(\dfrac{y}{x}\right),
\end{equation}
then we can express $c(r)$ as:
\begin{tcolorbox}[
colback=white,
colframe=black,
arc=2mm]
\begin{varwidth}{\textwidth}
\begin{multline}
	c(r) = \dfrac{1}{8}\left[K(r) + H\left(\dfrac{r}{1+r},\dfrac{\sqrt{2r+1}-1}{\sqrt{2r+1}+1}\right) - H\left(\dfrac{r}{1+r},\dfrac{\sqrt{4r+1}-1}{\sqrt{4r+1}+1}\right)\right] -\dfrac{1}{8\sqrt{3}}\left[\dfrac{2}{3}\Cl_2\left(\dfrac{\pi}{3}\right)  \right. \\
	\left. + C\left(\arctan\left(\sqrt{\dfrac{3(1+r)}{r+5}}\right)\right)- 2 C\left(\arctan\left(\sqrt{\dfrac{3}{2r+1}}\right)\right) - C\left(\arctan\left(\sqrt{\dfrac{3(1+r)}{r+5}}\right)+\arctan\left(\sqrt{\dfrac{3}{4r+5}}\right)\right) \right. \\
	\left.  - C\left(\arctan\left(\sqrt{\dfrac{3(1+r)}{r+5}}\right)-\arctan\left(\sqrt{\dfrac{3}{4r+5}}\right)\right) + C\left(\arctan\left(\sqrt{\dfrac{3}{4r+5}}\right)\right)\right].
\end{multline}
\end{varwidth}
\end{tcolorbox}
This function has the same asymptotics as the scaling function in the Cauchy case, and is in fact quite close to it for all values of $r$. This means we can write the decay of discovery order covariances in the limit $l \gg k$ for these two universality classes as:
\begin{equation}
    \mathrm{Cov}(I_k,I_{k+l}) \underset{l \gg k}{\sim} \dfrac{k^2}{4 l^2 (\log{l})^2}.
\end{equation}

\subsection{Recurrent case}

As explained in the section on the one-dimensional simple random walk, the recurrent case is considerably harder to describe quantitatively, because of the breakdown of self-averaging and the impossibility of transferring calculations at fixed time to the random time $t_n$. Therefore no general formula for scaling functions is available. However, in this subsection we prove the aging scaling form of correlations shown in the main text, and we explicit the important example of $1$-dimensional Brownian motion already computed in the previous section. To prove the scaling form of correlations, we follow the same approach as in the first section, decomposing over time at which discoveries were made and using known scaling of classic random walk observables.
\begin{equation} \begin{split}
    \mathbb{P}(I_k = 1, I_{k+l} = 1) & = \sum_{x,y\neq x} \mathbb{P}(1 \text{ discovers } x \text{ as the } k \text{-th site},  1 \text{ discovers } y \text{ as the } k+l \text{-th site})\\
    & = \sum_{x,y \neq x} \sum_{t,t'} \mathbb{P}(1 \text{ discovers } x \text{ as the } k \text{-th site}, t_k = t, 1 \text{ discovers } y \text{ as the } k+l \text{-th site}, t_{k+l} = t+t') \\
    & = \sum_{x,y \neq x} \sum_{t,t'} \mathbb{P}(X_t=x, \tau^2_x > t,  t_k = t, X_{t+t'}=y, \tau^2_y > t+t', t_{k+l} = t+t'),
\end{split} \end{equation}
where $\tau^i_x$ denotes the first-passage time by walker $i$ at site $x$ and $X_t$ is the position of the walker at time $t$. Inserting the known scalings for the first-passage time distribution, the survival, and the first range time $t_k$ for recurrent walks and using $X_t \sim t^{\frac{d_s}{2d}}$ finally gives:
\begin{equation}
    \mathbb{P}(I_k = 1, I_{k+l} = 1) \sim \sum_{x,y \neq x} \sum_{t,t'} \dfrac{1}{t^{\frac{d_s}{2}}}\dfrac{1}{k^{\frac{2}{d_s}}}\dfrac{1}{(t+t')^{\frac{d_s}{2}}}\dfrac{1}{(k+l)^{\frac{2}{d_s}}} F\left(\dfrac{x}{t^{\frac{d_s}{2d}}}, \dfrac{t}{k^{\frac{2}{d_s}}}, \dfrac{y}{(t+t')^{\frac{d_s}{2d}}}, \dfrac{t+t'}{(k+l)^{\frac{2}{d_s}}} \right).
\end{equation}
In this expression, the scaling function $F$ is totally out of reach for general walks, since the values of $t_k$ and $t_{k+l}$ depend on the entire trajectories of the walkers and therefore induce correlations between the $4$ first-passage time involved in the probability. In particular, $F$ does not factorize into two terms each describing the trajectory of one walker and the paths of the walkers before and after $t$ can not be decoupled, despite the Markovian nature of the dynamics. But describing precisely $F$ is not needed to prove the scaling form of correlations. Replacing all the sums by integrals, one obtains:
\begin{equation} \begin{split}
    & \mathbb{P}(I_k = 1, I_{k+l} = 1) \sim \iint \dd^d x \dd^d y \int_0^{+\infty} \dd t \int_0^{+\infty} \dd t' \dfrac{1}{t^{\frac{d_s}{2}}}\dfrac{1}{k^{\frac{2}{d_s}}}\dfrac{1}{(t+t')^{\frac{d_s}{2}}}\dfrac{1}{(k+l)^{\frac{2}{d_s}}} F\left(\dfrac{x}{t^{\frac{d_s}{2d}}}, \dfrac{t}{k^{\frac{2}{d_s}}}, \dfrac{y}{(t+t')^{\frac{d_s}{2d}}}, \dfrac{t+t'}{(k+l)^{\frac{2}{d_s}}} \right) \\
    & \underset{x^* = \frac{x}{k^{1/d}}, y^* = \frac{y}{k^{1/d}}}{\sim} \iint k^2 \dd^d x^* \dd^d y^* \int_0^{+\infty} \dd t \int_0^{+\infty} \dd t' \dfrac{1}{t^{\frac{d_s}{2}}}\dfrac{1}{k^{\frac{2}{d_s}}}\dfrac{1}{(t+t')^{\frac{d_s}{2}}}\dfrac{1}{(k+l)^{\frac{2}{d_s}}} F\left(\dfrac{k^{\frac{1}{d}} x^*}{t^{\frac{d_s}{2d}}}, \dfrac{t}{k^{\frac{2}{d_s}}}, \dfrac{k^{\frac{1}{d}} y^*}{(t+t')^{\frac{d_s}{2d}}}, \dfrac{t+t'}{(k+l)^{\frac{2}{d_s}}} \right) \\
    & \underset{\tau = \frac{t}{k^{2/d_s}}, \tau' = \frac{t'}{k^{2/d_s}}}{\sim} \iint k^2 \dd^d x^* \dd^d y^* \int_0^{+\infty} k^{\frac{4}{d_s}} \dd \tau \\
    & \int_0^{+\infty} \dd \tau' \dfrac{1}{k \tau^{\frac{d_s}{2}}}\dfrac{1}{k^{\frac{2}{d_s}}}\dfrac{1}{k(\tau+\tau')^{\frac{d_s}{2}}}\dfrac{1}{(k+l)^{\frac{2}{d_s}}} F\left(\dfrac{k^{\frac{1}{d}} x^*}{\tau^{\frac{d_s}{2d}} k^{\frac{1}{d}}}, \tau, \dfrac{k^{\frac{1}{d}} y^*}{k^{\frac{1}{d}} (\tau+\tau')^{\frac{d_s}{2d}}}, \dfrac{(\tau+\tau')k^{\frac{2}{d_s}}}{(k+l)^{\frac{2}{d_s}}} \right).
\end{split} \end{equation}
Setting $r = l/k$ and substracting the mean squared, we obtain the claimed behavior $\mathrm{Cov}(I_k,I_{k+l}) \sim c(l/k)$ with:
\begin{equation}
c(r) = \iint \dd^d x^* \dd^d y^* \int_0^{+\infty} \dd \tau \int_0^{+\infty} \dd \tau' \dfrac{1}{ \tau^{\frac{d_s}{2}}}\dfrac{1}{(\tau+\tau')^{\frac{d_s}{2}}}\dfrac{1}{(1+r)^{\frac{2}{d_s}}} F\left(\dfrac{x^*}{\tau^{\frac{d_s}{2d}}}, \tau, \dfrac{ y^*}{(\tau+\tau')^{\frac{d_s}{2d}}}, \dfrac{(\tau+\tau')}{(1+r)^{\frac{2}{d_s}}} \right) -\dfrac{1}{4},
\end{equation}
the scaling function being in general impossible to compute. In the special case of $1$-dimensional Brownian motion, we obtained the expression of $c(r)$ shown in Eq.~\eqref{eq 179}. One can check the asymptotic regimes $r\rightarrow 0$ and $r\rightarrow +\infty$. For $r=0$, the first two sums vanish and the expression reduces to:
\begin{equation}
    c(0) = - \dfrac{1}{4} + 4 \sum_{n=0}^{+\infty} \dfrac{1}{(2n+1)^2 \pi^2 } = \dfrac{1}{4},
\end{equation}
as it should since labels of infinitely close discoveries are perfectly correlated and therefore $\langle I_k I_{k+1} \rangle \rightarrow \frac{1}{2}$ as $k \rightarrow +\infty$. We also prove that $c(r) \sim r^{-2}$ in the regime $r \rightarrow +\infty$. Expanding the summands at order $\mathcal{O}(r^{-2})$, one obtains:
\begin{multline}
    c(r) \sim 2 \sum_{n=0}^{+\infty} \dfrac{1}{\sinh((2n+1)\pi)}\left(\dfrac{1}{(2n+1)\pi} - \dfrac{(2n+1)\pi}{12 r^2}\right) + \sum_{n=1}^{+\infty} \tanh\left(\dfrac{n \pi}{2}\right) \left(\dfrac{(-1)^{n+1}}{n \pi} + \dfrac{(-1)^n n \pi}{4 r^2} \right)\\
    - \dfrac{1}{4} + 4 \sum_{n=0}^{+\infty} \dfrac{(2n+1)\pi}{6 r^2 \sinh((2n+1)\pi)}.
\end{multline}
Although this is not clear from the this expression, the zero-th order term vanishes as needed for decorrelation at infinity. The sums can be computed as particular values of elliptic theta functions. We get:
\begin{equation} \begin{split}
    & 2\sum_{n=0}^{+\infty} \dfrac{1}{(2n+1)\pi \sinh((2n+1)\pi)} + \sum_{n=1}^{+\infty} \tanh\left(\dfrac{n \pi}{2}\right) \dfrac{(-1)^{n+1}}{n \pi} = \dfrac{1}{\pi} \log\left[2 \prod_{n=1}^{+\infty} \dfrac{(1+e^{-2\pi n})^2}{(1-e^{-(2n-1)\pi})^2}\right] \\
    = & \dfrac{1}{\pi} \log\left[ e^{\frac{\pi}{4}}\dfrac{\theta_2(e^{-\pi})}{\theta_4(e^{-\pi})}\right] = \dfrac{1}{\pi}\log(e^{\frac{\pi}{4}}) = \dfrac{1}{4},
\end{split} \end{equation}
where we used the modular symmetry $\theta_2(q) = \sqrt{\frac{\pi}{-\log(q)}}\theta_4\left(e^{\frac{\pi^2}{\log(q)}}\right)$ for the very special value $q=e^{-\pi}$.
We therefore get the expected cancellation with the constant term and the leading behavior at infinity is:
\begin{equation}
    c(r) \sim \left( \dfrac{1}{4}\sum_{n=1}^{+\infty} (-1)^n n \pi \tanh\left(\dfrac{n \pi}{2}\right) + \dfrac{4}{3} \sum_{n=0}^{+\infty} \dfrac{(2n+1)\pi}{\sinh((2n+1)\pi)} \right) \dfrac{1}{r^2}.
\end{equation}
The first sum is formally divergent but admits the regularization:
\begin{equation}
    \sum_{n=1}^{+\infty} (-1)^n n \pi \tanh\left(\dfrac{n \pi}{2}\right) = \sum_{n=1}^{+\infty} (-1)^n n \pi - \sum_{n=1}^{+\infty} (-1)^n n \pi \left(1- \tanh\left(\dfrac{n \pi}{2}\right) \right) = - \dfrac{\pi}{4} + \sum_{n=1}^{+\infty} (-1)^{n+1} n \pi \left(1- \tanh\left(\dfrac{n \pi}{2}\right) \right),
\end{equation}
where we used $\eta(-1) = \frac{1}{4}$. The decay of $c(r)$ thus rewrites into:
\begin{equation}
    c(r) \sim \left( \dfrac{4}{3} \sum_{n=0}^{+\infty} \dfrac{(2n+1)\pi}{\sinh((2n+1)\pi)} +  \dfrac{1}{4}\sum_{n=1}^{+\infty} (-1)^{n+1} n \pi \left(1-\tanh\left(\dfrac{n \pi}{2}\right)\right) - \dfrac{\pi}{16} \right) \dfrac{1}{r^2}.
\end{equation}
Once again, the sums involved are deeply related to elliptic functions singular values, with the results:
\begin{equation}
    \sum_{n=0}^{+\infty} \dfrac{(2n+1)\pi}{\sinh((2n+1)\pi)} = \dfrac{\Gamma\left(\dfrac{1}{4}\right)^4}{64 \pi^2},
\end{equation}
and:
\begin{equation}
    \sum_{n=1}^{+\infty} (-1)^{n+1} n \pi \left(1-\tanh\left(\dfrac{n \pi}{2}\right)\right) = \dfrac{\pi}{4} - \dfrac{\Gamma\left(\frac{1}{4}\right)^4}{32 \pi^2},
\end{equation}
such that:
\begin{equation}
    c(r) \underset{r \rightarrow +\infty}{\sim} \dfrac{5 \Gamma\left(\frac{1}{4}\right)^4}{384 \pi^2 r^2}.
\end{equation}
As the scaling functions for the marginal case $d_s=2$ were already decaying as $r^{-2}$ for large $r$, the $r^{-2}$ decay of the scaling function for correlations in discovery order might hold in full generality for recurrent universality classes.

\subsection{From correlations to the variance}

We now develop how to recover results for variance scaling and prefactors from the structure of discovery order covariances. This was already presented for recurrent walks in the case of $1$-dimensional Brownian motion because summing the correlations to obtain the variance was the natural approach in this case, but such an approach is absolutely general. We proved the following scaling form for discovery order correlations in the regime $k,k' \rightarrow +\infty$, $r=l/k$ fixed:
\begin{equation}
\mathrm{Cov}(I_k,I_{k+l}) \sim L_{d_s}(l) c\left(\dfrac{l}{k}\right),
\end{equation}
with:
\begin{equation}
\begin{cases}
L_{d_s}(l) = 1 \; & , \, d_s < 2 \\
L_{d_s}(l) = (\log{l})^{-2} \; & , \, d_s = 2 \\
L_{d_s}(l) = l^{-(d_s-2)} \; & , \, 2 < d_s < 4.
\end{cases}
\end{equation}
As we can write $\Var[X_n] = \dfrac{2}{n^2}\sum_{k=1}^n \sum_{l=1}^{n-k} \mathrm{Cov}(I_k,I_{k+l})$, we can directly express the variance.

In the recurrent case $d_s <2$, the variance is dominated by terms with $k$ and $l$ large and one gets:
\begin{equation} \begin{split}
\Var[X_n] & \sim \dfrac{2}{n^2} \sum_{k=1}^n \sum_{l=1}^{n-k} c\left(\dfrac{l}{k}\right) \sim 2\sum_{k=1}^n \sum_{l=1}^{n-k} \dfrac{1}{n^2}c\left(\dfrac{l}{k}\right) \\
& \underset{u=k/n , v=l/n}{\sim} 2\int_0^1 \dd u \int_0^{1-u} \dd v \, c\left(\dfrac{v}{u}\right).
\end{split} \end{equation}
Therefore $\Var[X_n]$ has a finite non zero limit when $n \rightarrow +\infty$, with value:
\begin{equation}
\Var[X_\infty] = 2\int_0^1 \dd u \int_0^{1-u} \dd v c\left(\dfrac{v}{u}\right) = \int_0^{+\infty} \dd r\dfrac{c(r)}{(1+r)^2}.
\end{equation}
This is exactly the expression we used for $1$-dimensional Brownian motion.

The marginal case behaves exactly in the same way as $\log{l} = \log(n v) \sim \log{n}$ at leading order. Thus the integral part does not change but an extra factor $(\log{n})^{-2}$ appears in front, giving:
\begin{equation}
\Var[X_n] \underset{n \rightarrow +\infty}{\sim} \dfrac{1}{(\log{n})^2}\int_0^{+\infty} \dd r\dfrac{c(r)}{(1+r)^2}.
\end{equation}

For transient walks, the prefactor in front of the scaling function strongly modifies the behavior of the variance. Naively taking the scaling limit in the sum gives:
\begin{equation} \begin{split}
\Var[X_n] \sim 2 \sum_{k=1}^n \sum_{l=1}^{n-k} \dfrac{1}{n^2 l^{d_s-2}} c\left(\dfrac{l}{k}\right)& \underset{u=k/n , v=l/n}{\sim} \dfrac{2}{n^{d_s-2}} \int_0^1 \dd u \int_0^{1-u} \dfrac{\dd v}{v^{d_s-2}} c\left(\dfrac{v}{u}\right) \\
& \sim \dfrac{2}{(4-d_s)n^{d_s-2}} \int_0^{+\infty} \dd r \dfrac{c(r)}{r^{d_s-2}} \dfrac{1}{(1+r)^{4-d_s}},
\end{split} \end{equation}
and one sees that the integral over $r$ defining the prefactor for the decay of the variance diverges for $d_s \geq 3$ because of the region $r \sim 0$. This signals the second transition for variance decay and the breakdown of the scaling approximation for the sum of correlations. We therefore have the following complete picture. For $2<d_s<3$, the variance decays anomalously with an amplitude fixed by the scaling function for correlations:
\begin{equation}
\Var[X_n] \underset{n \rightarrow +\infty}{\sim} \dfrac{2}{(4-d_s)n^{d_s-2}} \int_0^{+\infty} \dd r \dfrac{c(r)}{r^{d_s-2}(1+r)^{4-d_s}}.
\end{equation}
For $d_s =3$ we obtain a supplementary logarithmic correction by keeping the lower cutoff $r_{\text{min}} \sim 1/n$ in the integral. This gives:
\begin{equation}
\Var[X_n] \underset{n \rightarrow +\infty}{\sim} \dfrac{2 c(0) \log{n}}{n}.
\end{equation}
For transient walks, $c(0)$ has a quite complicated expression involving microscopic details of the walk. For $1$-dimensional Lévy walks, we computed this value and plugging in $d_s=3$ directly gives the prefactor.

For strongly enough transient walks with $d_s > 3$, the variance amplitude is not related anymore with the scaling function for correlations $c(r)$, but depends quite heavily on the microscopic details of the walk, as shown in section \ref{sec:quant}. The sum over $k,l$ needed to go from correlations to the variance is dominated by terms with $l$ of order $1$ instead of scaling terms with $k$ and $l$ large, and the decay of the variance becomes the classic $1/n$ decay of the central limit theorem.

 This behavior of the sum also explains the transition from non-Gaussian fluctuations to Gaussian fluctuations for the discovery-share. For $2< d_s <3$, covariances with arbitrarily large label separation $l$ play a role at leading order, and correlations are truly long range causing the breakdown of Gaussian CLT. For $d_s > 3$, the variance is dominated by short-range correlations where $l \sim 1$ and the discovery-share decomposes as a sum of effectively decorrelated random variables because of this short-range nature of correlations. The limiting fluctuations of $X_n$ are then Gaussian.

\section{Robustness of theoretical predictions for variance and correlations}

In this section, we discuss extensions of the minimal competitive-exploration model for which the fluctuation structure described in Eq.~\eqref{eq:VarXn_main} persists. We consider unequal mobilities, more than two competitors, heterogeneous substrates and non-Markovian exploration. In the first two cases, the Markovian first-passage framework developed above can be adapted, at least in principle, to obtain variance scalings and prefactors. For heterogeneous substrates and non-Markovian dynamics, key ingredients of the quantitative derivation, in particular translation invariance and Markovian path decompositions, no longer apply. In these cases, evidence for the persistence of the fluctuation structure is mainly provided by numerical simulations. 

\subsection{Unequal mobilities}

We start with the important case of unequal mobilities. Until now, we always assumed the $2$ walkers were identical: same universality class for the dynamics (this will not be relaxed) and same hopping speed. Our results, however, still hold even in the case of unequal mobilities with different hopping speeds. Although the mean discovery share now shifts away from the symmetric value $1/2$, fluctuations around this mean retain the structure described in Eq.~\eqref{eq:VarXn_main} in the main text. This is seen numerically (Figure \ref{fig:illu_decay_robustness} in the main text) and can also be proven using the tools from section \ref{sec:quant} for marginal and transient walks.

In the recurrent case, we already showed a generalization to unequal mobilities in section \ref{sec:one-sided} and we present here the generalization of Eq.~\eqref{eq 230} to Brownian walkers in one dimension with different diffusion coefficients. The calculation is exactly the same as for the symmetric case, but the building block is now:
\begin{equation}
\rho^{+,(1,\delta)}_{x,y}(z) = 2 \sum_{n=1}^{+\infty} \dfrac{\sin(n \pi z) \sin(n \pi x) \sinh\left(\frac{n \pi y}{\sqrt{\delta}}\right)}{\sinh\left(\frac{n \pi}{\sqrt{\delta}}\right)},
\end{equation} 
where $\delta=\frac{D_y}{D_x}$ is the diffusion ratio between the walker starting in $y$ and the one starting in $x$. Denoting $\delta=\frac{D_1}{D_2}$, $c(r)$ is now expressed as:
\begin{multline}
    c(r) = \dfrac{\langle X_\infty \rangle}{(1+r)} \int_0^1 \dd \zeta \, 2 \zeta \int_0^r \dd x  \int_0^1 \dd u \left\{ \left[ \rho^{+,(1,\delta)}_{\frac{x+\zeta}{1+r},\frac{x}{1+r}}\left(u\right) +  \rho^{+,(1,\delta)}_{1-\frac{x+\zeta}{1+r},1-\frac{x}{1+r}}\left(u\right)\right] \partial_2 \pi^{+,(1,\delta)}_{1-u,0}\right.\\
    \left. + \left[ \rho^{+,(1,\delta^{-1})}_{\frac{x}{1+r},\frac{x+\zeta}{1+r}}\left(u\right)+  \rho^{+,(1,\delta^{-1})}_{1-\frac{x}{1+r},1-\frac{x+\zeta}{1+r}}\left(u\right)\right] \partial_1 \pi^{+,(1,\delta)}_{0,1-u}\right\} + \frac{1}{2}\int_0^1 \dd \zeta \, 2 \zeta \, \pi^{+,(1,\delta)}_{\frac{1-\zeta}{1+r},\frac{1}{1+r}} - \langle X_\infty \rangle^2,
\end{multline}
since $\langle X_\infty \rangle = \mathbb{P}(I_k = 1)$. Factors with indices $(1,\delta)$ (respectively $(1,\delta^{-1})$) are associated to discoveries by walker $1$ (respectively walker $2$).
If $X_n$ is the discovery-share of walker $1$ with diffusion coefficient $D_1$ against walker $2$ with diffusion coefficient $D_2$, then:
\begin{tcolorbox}[
colback=white,
colframe=black,
arc=2mm]
\begin{varwidth}{\textwidth}
\begin{multline}
\Var[X_\infty] = \sum_{n=0}^{+\infty} \dfrac{2}{(1+\delta)\pi^2(2n+1)^2}\left[\dfrac{\delta f_1^\delta((2n+1)\pi)}{\sinh\left(\frac{(2n+1)\pi}{\sqrt{\delta}}\right)} - \dfrac{\sqrt{\delta}f_2^\delta((2n+1)\pi)}{1+\cosh\left(\frac{(2n+1)\pi}{\sqrt{\delta}}\right)}\right] - \dfrac{\delta^2}{(1+\delta)^2} \\
+ \sum_{n=1}^{+\infty} \dfrac{\sqrt{\delta}}{(1+\delta)n^2 \pi^2 (1+\cosh(n \pi\sqrt{\delta}))}\left[-f_2^{\delta^{-1}}(n\pi) + (-1)^{n+1} f_3^{\delta^{-1}}(n\pi)\right],
\end{multline}
where the auxiliary functions now account for rescaling of the arguments of the hyperbolic functions:
\begin{multline}
    \begin{cases}
    f_1^\delta(x) = 4 \int_0^x \dfrac{\dd u}{u} [1 - \cos(u)]\sinh\left(\frac{u}{\sqrt{\delta}}\right),  \\
    f_2^\delta(x) = 4 \int_0^x \dfrac{\dd u}{u} \sin(u)\sinh\left(\frac{u}{\sqrt{\delta}}\right), \\
    f_3^\delta(x) = 4 \int_0^x \dfrac{\dd u}{u} \sin(u) [\sinh\left(\frac{x}{\sqrt{\delta}}\right) - \sinh\left(\frac{x-u}{\sqrt{\delta}}\right)].
    \end{cases}
\end{multline}
\end{varwidth}
\end{tcolorbox}
This expression for the variance satisfies $\Var[X_\infty](\delta)= \Var[X_\infty](\delta^{-1})$, since $X_\infty$ and $1-X_\infty$ have the same variance.
As an intermediate result in the calculation, we obtained $\langle X_\infty \rangle = \frac{D_1}{D_1+D_2}$ to ensure $c(r) \rightarrow 0$ for $r \rightarrow 0$. Obtaining such a compact expression for the mean can be considered a surprise here since computing it is much harder than in the one-sided case, but the final value has lower complexity:
\begin{equation}
    \langle X_\infty \rangle = \dfrac{D_1}{D_1+D_2} \text{ vs. } \langle X_\infty \rangle = \dfrac{2}{\pi}\arctan\left(\sqrt{\dfrac{D_1}{D_2}}\right),
\end{equation}
which is the value of the mean in the one-sided case. Of course, when going to second moments, we saw that the complexity of the two-sided case becomes clear, and we obtain quite an intricate expression for the variance, instead of:
\begin{equation}
    \Var[X_\infty] = \dfrac{1}{\pi}\arctan\left(\sqrt{\dfrac{D_1}{D_2}}\right)\left[1-\dfrac{2}{\pi}\arctan\left(\sqrt{\dfrac{D_1}{D_2}}\right)\right],
\end{equation}
the variance in the one-sided case.

\subsection{Multiple competitors}

A second natural extension is competition between \(M>2\) walkers. For \(M\)
identical competitors, exchange symmetry gives:
\begin{equation}
    \langle X_n\rangle=\frac{1}{M}
\end{equation}
for the discovery-share of any tagged competitor. Numerical simulations shown in
Fig.~\ref{fig:illu_decay_robustness} of the main text indicate that fluctuations
around this mean follow the same scaling regimes as in the two-competitor case.
The approaches of Sections~\ref{sec:fluc} and~\ref{sec:quant} remain
structurally applicable, but explicit calculations become combinatorially more
involved. In particular, the analogue of Eq.~\eqref{eq 43}, which is the basis
of the quantitative calculation in Section~\ref{sec:quant}, involves a product
of \(2M\) first-passage quantities and \(M+1\) spatial sums. This expression
still provides, in principle, access to the scaling of the discovery-share
variance and to its prefactors.

\subsection{Heterogeneous substrates}

We next consider diffusion on substrates whose geometry is not translationally
invariant. Deterministic fractals, such as the Sierpinski triangle, and random fractals, such as
critical percolation clusters, provide standard examples of geometrically
constrained, crowded or disordered environments. Random walks on these
structures are standard examples of recurrent exploration, with spectral
dimension $d_s<2$. The prediction of the general theory is therefore that the
discovery-share should remain broadly fluctuating. This is precisely what is
observed in the numerical simulations shown in
Fig.~\ref{fig:illu_decay_robustness} of the main text.
\\
Although the exact quantitative machinery of Section~\ref{sec:quant} relies on
translation invariance and therefore does not apply directly to such substrates,
the scaling mechanism leading to the phase diagram is expected to be robust. The
reason is that the central input of the scaling argument is the large-scale form
of the one-walker propagator, controlled by the spectral dimension. For
deterministic and random fractal substrates, this propagator scaling is a
standard property, at least in an averaged sense for random substrates and up to
possible oscillatory corrections associated with discrete scale invariance for
some deterministic fractals. We therefore expect the same overlap and
label-correlation scaling arguments to apply once the Euclidean dimension is
replaced by the appropriate spectral dimension.
\\
On heterogeneous substrates, the detailed amplitudes and scaling functions may
depend on geometric features of the lattice or on disorder averaging. The point
tested here is therefore the universal part of the prediction: recurrence, or
$d_s<2$, prevents self-averaging and leaves a broad limiting discovery-share
distribution. The numerical results support this spectral-dimension mechanism
beyond translationally invariant lattices.

\subsection{Fractional Brownian motion as a non-Markovian test case }

Finally, we consider fractional Brownian motion as a representative
non-Markovian dynamics. Numerical simulations, shown in
Fig.~\ref{fig:illu_decay_robustness} of the main text and in
Section~\ref{sec:SI_num}, indicate that the scaling predictions for both the
discovery-share variance and the two-point label correlations remain valid for
fractional Brownian exploration.
\\
An analytical treatment is substantially harder in this case. Non-Markovian
walks, and fractional Brownian motion in particular, are notoriously difficult
to handle at the level of first-passage observables. The few available results
on first-passage properties of such processes are mostly restricted to one
dimension~\cite{wiese_perturbation_2011,bremont_persistence_2025}. Nevertheless,
universal first-passage and exploration scalings have been observed in several
fractional-Brownian settings~\cite{regnier_universal_2023,regnier_visitation_2025},
and the present results provide an additional example in a genuinely competitive
observable.
\\
At the technical level, Markovianity enters the derivation at several essential
steps. Eq.~\eqref{eq 43}, which underlies the quantitative formulas of this
work, follows from path-decomposition identities that immediately fail for non-Markovian walks. Similarly, the scaling arguments of
Section~\ref{sec:fluc} rely on many classical results on spatial exploration by
Markovian random walks, only a small subset of which has been extended to
non-Markovian dynamics. Fractional Brownian motion should therefore be viewed
here as a robust numerical test of the predicted scaling structure, rather than
as a case for which the full analytical machinery is presently available.

\section{Summary of important formulas for the variance and scaling functions}

In this section, we regroup all the relevant formulas for competitive exploration, namely the variance decay for all marginal and transient walks and the expression of the scaling function $c(r)$ quantifying correlations in discovery labels. We also include explicit formula for $1$-dimensional nearest-neighbor diffusion.

\subsection{Variance amplitudes}

Defining the overlap function for the scaling function $p$ of the propagator as:
\begin{equation}
\Theta(u,v,w) = \int \dd^d x \, u^{-\frac{d_s}{2}}p\left(x u^{-\frac{d_s}{2d}}\right) v^{-\frac{d_s}{2}}p\left(x v^{-\frac{d_s}{2d}}\right) w^{-\frac{d_s}{2}}p\left(x w^{-\frac{d_s}{2d}}\right),
\end{equation}
we have the following formulas for the variance of the discovery-share $X_n$:
\\
\\

\textbf{Variance of the discovery-share for marginal random walks:}

For $d_s=2$, we have:
\begin{multline}
    \Var[X_n] \underset{n \rightarrow +\infty}{\sim} \dfrac{1}{4 p(0)^2 (\log{n})^2} \int_0^1 \dd u \int_0^{1-u} \dd v \int_0^u \dd a \left\{ \int_0^a \dd b \left[\Theta(a-b,u-a,v) - \Theta(a-b,u-a,u+v+b) \right. \right.\\
    \left. \left. + \Theta(a-b,u+b,v) - \Theta(a-b,u+b,u+v-a) \right] + \int_0^u \dd b \left[\Theta(u-a,a+b,v) - \Theta(u-a,a+b,v+u-b)\right]\right\}.
\end{multline}
\\
\\

\textbf{Variance of the discovery-share for weakly transient random walks:}

In the regime $2<d_s<3$, we have:
\begin{multline}
    \Var[X_n] \underset{n \rightarrow +\infty}{\sim} \dfrac{2^{d_s}}{16 g_0^{d_s}} n^{2-d_s} \int_0^1 \dd u \int_0^{1-u} \dd v \int_0^u \dd a \left\{ \int_0^a \dd b \left[\Theta(a-b,u-a,v) - \Theta(a-b,u-a,u+v+b) \right. \right.\\
    \left. \left. + \Theta(a-b,u+b,v) - \Theta(a-b,u+b,u+v-a) \right] + \int_0^u \dd b \left[\Theta(u-a,a+b,v) - \Theta(u-a,a+b,v+u-b)\right]\right\},
\end{multline}
where $g_0$ is the averaged total number of visits to the origin:
\begin{equation}
g_0 = \sum_{n=0}^{+\infty} P_n(0).
\end{equation}
\\
\\

\textbf{Variance of the discovery-share for strongly transient random walks:}

We have the following formulas depending on the spectral dimension $d_s$. For $d_s=3$, the decay is:
\begin{equation}
\Var[X_n] \sim \dfrac{3\beta(d)C^3}{8dg_0^3}\dfrac{\log{n}}{n},
\end{equation}
where $\beta(d)$ is the area of the $d-1$-dimensional hypersphere with the convention $\beta(1)=2$ and $C$ describes the divergence of the scaling function $\mathcal{G}$ at the origin:
\begin{equation}
\mathcal{G}(x) \underset{x\rightarrow 0}{\sim} C |x|^{-\frac{d}{3}}, C=\int_0^{+\infty} \dd u \, u^{-\frac{3}{2}} p\left(u^{-\frac{3}{2d}}\right).
\end{equation}

For $3<d_s\leq4$, the decay is:
\begin{equation}
\Var[X_n] \sim \displaystyle \left\{1 - \dfrac{2}{g_0} +2 \sum_{x\neq 0} G(x,1) \left[\dfrac{g_0 - G(-x,1)}{g_0^2 - G(x,1)G(-x,1)} - \dfrac{1}{g_0} + \dfrac{G(-x,1)}{g_0^2} \right]\right\} \dfrac{1}{8 n},
\end{equation}
and depends strongly on the microscopic details of the walk.

For $d_s > 4$, the expression is:
\begin{equation}
\Var[X_n] \sim \displaystyle \left\{ 2 \sum_{x\neq 0} G(x,1) \left[\dfrac{g_0 - G(-x,1)}{g_0^2 - G(x,1)G(-x,1)} - \dfrac{1}{g_0}\right]- 1 +2 \dfrac{c}{g_0^2}\right\} \dfrac{1}{8 n},
\end{equation}
where we introduced a new microscopic constant:
\begin{equation}
c= \sum_{n=1}^{+\infty} n P_n(0) = G_0'(1).
\end{equation}
\\
\\

We now explicit the formulas above for specific universality classes.

\textbf{Variance of the discovery-share for $1$-dimensional Lévy walks:}

In the regime $2<d_s<3$ which corresponds to $\frac{2}{3}<\alpha<1$ for the Lévy exponent in $1$D ($\alpha = \frac{2}{d_s}$), the variance decays as:
\begin{multline}
\Var[X_n] \sim \dfrac{2^{d_s}}{64 \pi^2 D^{d_s} g_0^{d_s}} \dfrac{8 \pi d_s \csc(\pi d_s)}{\Gamma(5-d_s) n^{d_s-2}} \int_0^{+\infty} \dd r \left\{\frac{r^{\frac{4}{d_s}-2}}{\left(r^{\frac{2}{d_s}}-2\right) \left(r^{\frac{2}{d_s}}-2 (r+1)^{\frac{2}{d_s}}\right)} \right. \\
   +\frac{2^{-d_s} \left((r
   (r+1))^{\frac{2}{d_s}}-4\right)}{\left(r^{\frac{2}{d_s}}-2\right) \left(r^{\frac{2}{d_s}}-1\right)
   \left((r+1)^{\frac{2}{d_s}}-2\right)
   \left((r+1)^{\frac{2}{d_s}}-1\right)}+\frac{(r+1)^{\frac{4}{d_s}-2}}{\left((r+1)^{\frac{2}{d_s}}-2\right)
   \left((r+1)^{\frac{2}{d_s}}-2 r^{\frac{2}{d_s}}\right)} \\
   +\frac{1}{\left(2 r^{\frac{2}{d_s}}-1\right)
   \left(2 (r+1)^{\frac{2}{d_s}}-1\right)}+\frac{2^{-d_s} \left((r+1)^{\frac{2}{d_s}}-4
   r^{\frac{4}{d_s}}\right) r^{\frac{4}{d_s}-2}}{\left(r^{\frac{2}{d_s}}-1\right) \left(2
   r^{\frac{2}{d_s}}-1\right) \left(r^{\frac{2}{d_s}}-(r+1)^{\frac{2}{d_s}}\right) \left(2
   r^{\frac{2}{d_s}}-(r+1)^{\frac{2}{d_s}}\right)}\\
   \left.+ \frac{2^{-d_s} (r+1)^{\frac{4}{d_s}-2} \left(r^{\frac{2}{d_s}}-4
   (r+1)^{\frac{4}{d_s}}\right)}{\left(r^{\frac{2}{d_s}}-2 (r+1)^{\frac{2}{d_s}}\right)
   \left(r^{\frac{2}{d_s}}-(r+1)^{\frac{2}{d_s}}\right) \left((r+1)^{\frac{2}{d_s}}-1\right) \left(2
   (r+1)^{\frac{2}{d_s}}-1\right)} \right\}.
   \label{eq 257}
\end{multline}
This expression involves the generalized diffusion coefficient $D$, defined as:
\begin{equation}
\tilde{p}(k) = \int_{-\infty}^{+\infty} e^{i k x}p(x) \underset{k \rightarrow 0}{\sim} 1-D |k|^\alpha.
\end{equation}
\\
\\

\textbf{Variance of the discovery-share for the $1$-dimensional Cauchy walk:}

For the Cauchy walk ($\alpha=1$), which is marginal in $1$D, we have the very compact formula:
\begin{equation}
\Var[X_n] \sim \dfrac{\log{2}}{2 (\log{n})^2}.
\label{eq 259}
\end{equation}
\\
\\

\textbf{Variance of the discovery-share for $2$-dimensional Brownian motion:}

The variance in the case of $2$-dimensional Brownian motion is expressed using Clausen function $\Cl_2(\theta) = \sum_{n=1}^{+\infty} \frac{\sin(n\theta)}{n^2}$:
\begin{equation}
\Var[X_n] \sim \dfrac{\Cl_2\left(\frac{\pi}{2}\right) - \frac{1}{\sqrt{3}} \Cl_2\left(\frac{\pi}{3}\right)}{(\log{n})^2}
\label{eq 260}
\end{equation}
\\
\\

\textbf{Variance of the discovery-share for $3$-dimensional Brownian motion:}

\begin{equation}
\Var[X_n] \sim \dfrac{3 \sqrt{6} \; 2^9 \pi^7}{\Gamma^3\left(\dfrac{1}{24}\right)\Gamma^3\left(\dfrac{5}{24}\right)\Gamma^3\left(\dfrac{7}{24}\right)\Gamma^3\left(\dfrac{11}{24}\right)} \dfrac{\log{n}}{n}
\end{equation}
\\
\\

We end this section with the case of $1$-dimensional nearest neighbour diffusion, which is the only recurrent universality class where the value of $\Var[X_\infty]$ can be computed analytically.

\textbf{Variance of the discovery-share for $1$-dimensional Brownian motion:}

Defining the following functions:
\begin{multline}
    \begin{cases}
    f_1(x) = 4 \int_0^x \dfrac{\dd u}{u} [1 - \cos{u}]\sinh{u},  \\[10pt]
    f_2(x) = 4 \int_0^x \dfrac{\dd u}{u} \sin{u}\sinh{u}, \\[10pt]
    f_3(x) = 4 \int_0^x \dfrac{\dd u}{u} \sin{u} [\sinh{x} - \sinh(x-u)],
    \end{cases}
\end{multline}
the variance behaves as:
\begin{equation} \begin{split}
    \Var[X_n] \sim \Var[X_\infty] & =  \sum_{n=0}^{+\infty} \dfrac{1}{\pi^2(2n+1)^2}\left[\dfrac{f_1((2n+1)\pi)}{\sinh((2n+1)\pi)} - \dfrac{f_2((2n+1)\pi)}{1+\cosh((2n+1)\pi)}\right]- \frac{1}{4} \\
    & + \sum_{n=1}^{+\infty} \dfrac{1}{2 n^2 \pi^2 (1+\cosh(n \pi))}\left[-f_2(n\pi) + (-1)^{n+1} f_3(n\pi)\right] \approx 0.03433.
\end{split} \end{equation}

\subsection{Scaling functions $c(r)$}

\textbf{Expression of $c(r)$ for marginal walks:}

\begin{multline}
    c(r) = \dfrac{1}{8 p(0)^2} \int_0^1 \dd a \left\{ \int_0^a \dd b \left[ \Theta(a-b,1-a,r) - \Theta(a-b,1-a,1+r+b) \right. \right. \\
    \left. \left. + \Theta(a-b,1+b,r) - \Theta(a-b,1+b,1+r-a) \right] + \int_0^1 \dd b \left[ \Theta(1-a,a+b,r) - \Theta(1-a,a+b,1+r-b) \right] \right \}.
\end{multline}
\\
\\

\textbf{Expression of $c(r)$ for weakly transient walks:}

In the regime $2<d_s<4$, we have:
\begin{multline}
    c(r) = \dfrac{2^{d_s} r^{d_s-2}}{32 g_0^{d_s}} \int_0^1 \dd a \left\{ \int_0^a \dd b \left[ \Theta(a-b,1-a,r) - \Theta(a-b,1-a,1+r+b) \right. \right. \\
    \left. \left. + \Theta(a-b,1+b,r) - \Theta(a-b,1+b,1+r-a) \right] + \int_0^1 \dd b \left[ \Theta(1-a,a+b,r) - \Theta(1-a,a+b,1+r-b) \right] \right \}.
\end{multline}
\\
\\

\textbf{Expression of $c(r)$ for $1$-dimensional Lévy walks in the intermediate region:}

For $2<d_s<4$ which corresponds to $\frac{1}{2}<\alpha<1$, the scaling function is expressed as:
\begin{equation}
    c(r) = \dfrac{2^{d_s} r^{d_s-2}}{32 (2 \pi)^2 g_0^{d_s} D^{d_s}} \int_0^{+\infty} \dd u K\left(u^{\frac{2}{d_s}},(1+u)^{\frac{2}{d_s}},r\right),
    \label{eq 266}
\end{equation}
with:
\begin{multline}
    K(x,y,r) = \frac{d_s \Gamma \left(d_s-2\right)}{(x-1) x (y-1) y (x-y)} \left(x^2 y r^{2-d_s}-x^2 r^{2-d_s}-x y^2 r^{2-d_s}+x r^{2-d_s}+y^2
   r^{2-d_s}-y r^{2-d_s} \right. \\
   +r^2 x^5 y (r x)^{-d_s}-r^2 x^5 (r x)^{-d_s}-r^2 x^4 y^2 (r x)^{-d_s}+r^2 x^4 (r
   x)^{-d_s}+r^2 x^3 y^2 (r x)^{-d_s}-r^2 x^3 y (r x)^{-d_s}+r^2 x^2 y^4 (r y)^{-d_s} \\
   -r^2 x^2 y^3 (r
   y)^{-d_s}-r^2 x y^5 (r y)^{-d_s}+r^2 x y^3 (r y)^{-d_s}+r^2 y^5 (r y)^{-d_s}-r^2 y^4 (r y)^{-d_s}-(r+2)^2
   x^3 y^2 ((r+2) x)^{-d_s} \\
   +x^3 (r x+2 y)^{2-d_s}-x^3 y (r x+2)^{2-d_s}+(r+2)^2 x^3 y ((r+2)
   x)^{-d_s} +(r+2)^2 x^2 y^3 ((r+2) y)^{-d_s}-x^2 y^2 (r y+2)^{2-d_s}\\
   +x^2 y^2 (r x+2)^{2-d_s}+x^2 (r+2
   y)^{2-d_s}-x^2 (r x+2 y)^{2-d_s}-x^2 y (r+2)^{2-d_s}+x y^3 (r y+2)^{2-d_s}-(r+2)^2 x y^3 ((r+2)
   y)^{-d_s}\\
   -(y-1) y^2 (r y+2 x)^{2-d_s}-y^2 (r+2 x)^{2-d_s}+x y^2 (r+2)^{2-d_s}-x (r+2 y)^{2-d_s}+y (r+2
   x)^{2-d_s}\left. \right).
\end{multline}
\\
\\

\textbf{Expression of $c(r)$ for the Cauchy walk:}

\begin{equation}
c(r) = \dfrac{1}{4}\Li_2\left(\dfrac{r}{2(1+r)}\right) + \dfrac{1}{8}\log\left(\dfrac{r}{2(1+r)}\right)^2 - \dfrac{\pi^2}{48}.
\end{equation}
\\
\\

\textbf{Expression of $c(r)$ for $2$-dimensional Brownian motion:}

Defining:
\begin{equation}
\begin{split}
	K(r) & = \log\left(\dfrac{r}{1+r}\right)^2 -\log\left(1+r+\sqrt{2r+1}\right)		\log\left(\dfrac{r}{1+r}\right)\\
	& + \Li_2\left(\dfrac{1}{1+r}\right) - \Li_2\left(\dfrac{1+r}{1+r+\sqrt{1+2r}}\right)+\Li_2\left(1-\dfrac{\sqrt{1+2r}}{1+r}\right),
	\end{split}
\end{equation}
\begin{equation}
	C(r) = \Cl_2\left(2r+\dfrac{2\pi}{3}\right) - \Cl_2\left(2r-\dfrac{2\pi}{3}\right)
\end{equation}
with $\Cl_2(\theta) = \sum_{n=1}^{+\infty} \frac{\sin(n \theta)}{n^2}$, and:
\begin{equation}
	H(x,y) = (\log{y})^2 - \log{x}\log{y} - \Li_2(x y) + \Li_2\left(\dfrac{y}{x}\right),
\end{equation}
$c(r)$ is expressed as:
\begin{multline}
	c(r) = \dfrac{1}{8}\left[K(r) + H\left(\dfrac{r}{1+r},\dfrac{\sqrt{2r+1}-1}{\sqrt{2r+1}+1}\right) - H\left(\dfrac{r}{1+r},\dfrac{\sqrt{4r+1}-1}{\sqrt{4r+1}+1}\right)\right] -\dfrac{1}{8\sqrt{3}}\left[C\left(\arctan\left(\sqrt{\dfrac{3(1+r)}{r+5}}\right)\right) \right. \\
	\left.+\dfrac{2}{3}\Cl_2\left(\dfrac{\pi}{3}\right) - 2 C\left(\arctan\left(\sqrt{\dfrac{3}{2r+1}}\right)\right) - C\left(\arctan\left(\sqrt{\dfrac{3(1+r)}{r+5}}\right)+\arctan\left(\sqrt{\dfrac{3}{4r+5}}\right)\right) \right. \\
	\left.  - C\left(\arctan\left(\sqrt{\dfrac{3(1+r)}{r+5}}\right)-\arctan\left(\sqrt{\dfrac{3}{4r+5}}\right)\right) + C\left(\arctan\left(\sqrt{\dfrac{3}{4r+5}}\right)\right)\right].
\end{multline}

\textbf{Expression of $c(r)$ for $3$-dimensional Brownian motion:}

A direct application of the general formula with $\Theta(u,v,w) = \frac{27}{8 \pi^3 (u v+v w + u w)^{3/2}}$ gives:
\begin{equation}
c(r) = \dfrac{6144 \sqrt{6} \pi^6}{ \Gamma\left(\dfrac{1}{24}\right)^3 \Gamma\left(\dfrac{5}{24}\right)^3 \Gamma\left(\dfrac{7}{24}\right)^3 \Gamma\left(\dfrac{11}{24}\right)^3} \dfrac{\arctan\left(\frac{1}{\sqrt{2r}}\right)}{(2+r)}.
\end{equation}
\\
\\

\textbf{Expression of $c(r)$ for $1$-dimensional Brownian motion:}

Defining:
\begin{multline}
\begin{cases}
U(x) = (\cosh{x} -1)(x-\sin{x}) + (1-\cos{x})(x-\sinh{x}), \\
V(x) = x(\sin{x}+\sinh{x}) - \sin{x}\sinh{x} + (1-\cos{x})(\cosh{x}-1), \\
W(x) = (1-\cos{x})(\cosh{x}-1) - x(\sin{x}+\sinh{x}) + \sin{x}\sinh{x},
\end{cases}
\end{multline}
we have:
\begin{multline}
c(r) = \dfrac{(1+r)^2}{\pi^3}\left[2 \sum_{n=0}^{+\infty} \dfrac{1}{(2n+1)^3} \left(\dfrac{V\left(\dfrac{(2n+1)\pi}{1+r}\right)}{\sinh((2n+1)\pi)} + \dfrac{U\left(\dfrac{(2n+1)\pi}{1+r}\right)}{1+\cosh((2n+1)\pi)}\right) \right. \\
\left. + \sum_{n=1}^{+\infty} \dfrac{(1+(-1)^n \cosh(n\pi))U\left(\dfrac{n\pi}{1+r}\right) + (-1)^n \sinh(n \pi) W\left(\dfrac{n \pi}{1+r}\right)}{n^3(1+\cosh(n\pi))}\right] - \dfrac{1}{4}.
\end{multline}

\section{Numerical validation}\label{sec:SI_num}

In this section, we analyze in a detailed way simulation data to show agreement between theoretical predictions and direct simulation of the walkers on various lattices. This concerns both predictions for the variance (decay amplitudes or limiting variance for $1$-dimensional Brownian motion) and for discovery order covariances (scaling forms and predictions for $c(r)$).

\subsection{Details for simulations shown in the main text}

We begin by providing additional details on the simulation data presented in the main text, proceeding figure by figure, starting with Fig.~\ref{fig:variance+correlations_decay}. In this figure, blue curves and symbols correspond to the universality class of one-dimensional Brownian motion, simulated as a nearest-neighbor simple random walk. For panel (b), the walks were simulated until they had collectively visited $10^6$ sites. For the discovery-label correlations shown in panels (c) and (d), simulations were instead stopped after $2000$ visited sites in order to obtain sufficient statistical accuracy.\\
Similarly, red curves and symbols correspond to one-dimensional symmetric Levy walks with exponent $\alpha = 3/4$, simulated as Riemann walks (discrete space and discrete time). In panel (b), simulations were run until the walkers had collectively visited $10^7$ sites, whereas the discovery--label correlation measurements were again limited to the first $2000$ visited sites.\\
Finally, the green curve in panel (b) corresponds to a one-dimensional Levy walk with exponent $\alpha = 1/2$, while in panel (c) the corresponding Levy walk has exponent $\alpha = 2/3$. In both cases, the walks were simulated as Riemann walks, with simulations stopped after $10^6$ collectively visited sites for panel (b) and after $2000$ visited sites for the discovery-label correlations shown in panel (c).

In Fig.~\ref{fig:illu_1D_limiting_distributions}, we present numerically simulated distributions of the discovery share for one-dimensional Brownian motion. In both the left panel (where discoveries are counted only on positive integers) and the right panel (where discoveries are counted on both sides), the data are obtained by simulating nearest-neighbor simple random walks until the walkers have collectively discovered $10^4$ sites, and recording the value of the discovery share at the end of the simulation.

In Figure~\ref{fig:illu_decay_robustness}, we illustrate the robustness of the variance decay of the discovery share across several generalizations of the minimal model. In panel (a) (unequal mobilities), we simulate two one-dimensional Levy walks with identical Levy exponent $\alpha$ but different characteristic jump lengths, one walker having jumps twice as long as the other. The walks are simulated as Riemann walks (discrete space and discrete time). The recurrent regime is illustrated by $\alpha=3/2$ (blue), the intermediate regime $2<d_s<3$ by $\alpha=3/4$ (red), and the strongly transient regime $d_s>3$ by $\alpha=1/2$ (green).\\
In panel (b) (multiple competitors), we consider $M>2$ identical Lévy walkers. The recurrent regime corresponds to $\alpha=3/2$ with $M=7$ (blue), the intermediate regime $2<d_s<3$ to $\alpha=3/4$ with $M=4$ (red), and the strongly transient regime $d_s>3$ to $\alpha=1/2$ with $M=5$ (green).\\
In panel (c) (heterogeneous substrates), we simulate nearest-neighbor diffusion on complex substrates. Deterministic fractals are represented by Sierpinski triangles (blue), on which the walkers evolve directly on the infinite recursive structure as new ramifications are generated during the simulation. Random fractals are represented by critical percolation clusters (purple). These clusters are generated on a $200\times200$ square lattice by retaining each bond independently with probability $1/2$. The largest connected component is then extracted, and both walkers are initialized from the same site chosen uniformly at random within the cluster.\\
In panel (d) (memory-driven transport), we simulate fractional Brownian motion (fBm) in two and three dimensions. To recover the discrete-space setting relevant to competitive exploration, the continuous trajectories are discretized by generating standard fBm paths with the Davies--Harte algorithm and replacing each coordinate by its integer part at every time step. The recurrent regime (blue) corresponds to two-dimensional fBm with Hurst exponent $H=1/3$. The weakly transient regime ($2<d_s<3$, red) is obtained using two-dimensional fBm with $H=2/3$, whereas the strongly transient regime ($d_s>3$, green) is represented by three-dimensional fBm with $H=2/3$.

\subsection{$1$-dimensional nearest neighbor diffusion}

\begin{figure}[hb]
\centering
\includegraphics[scale=1]{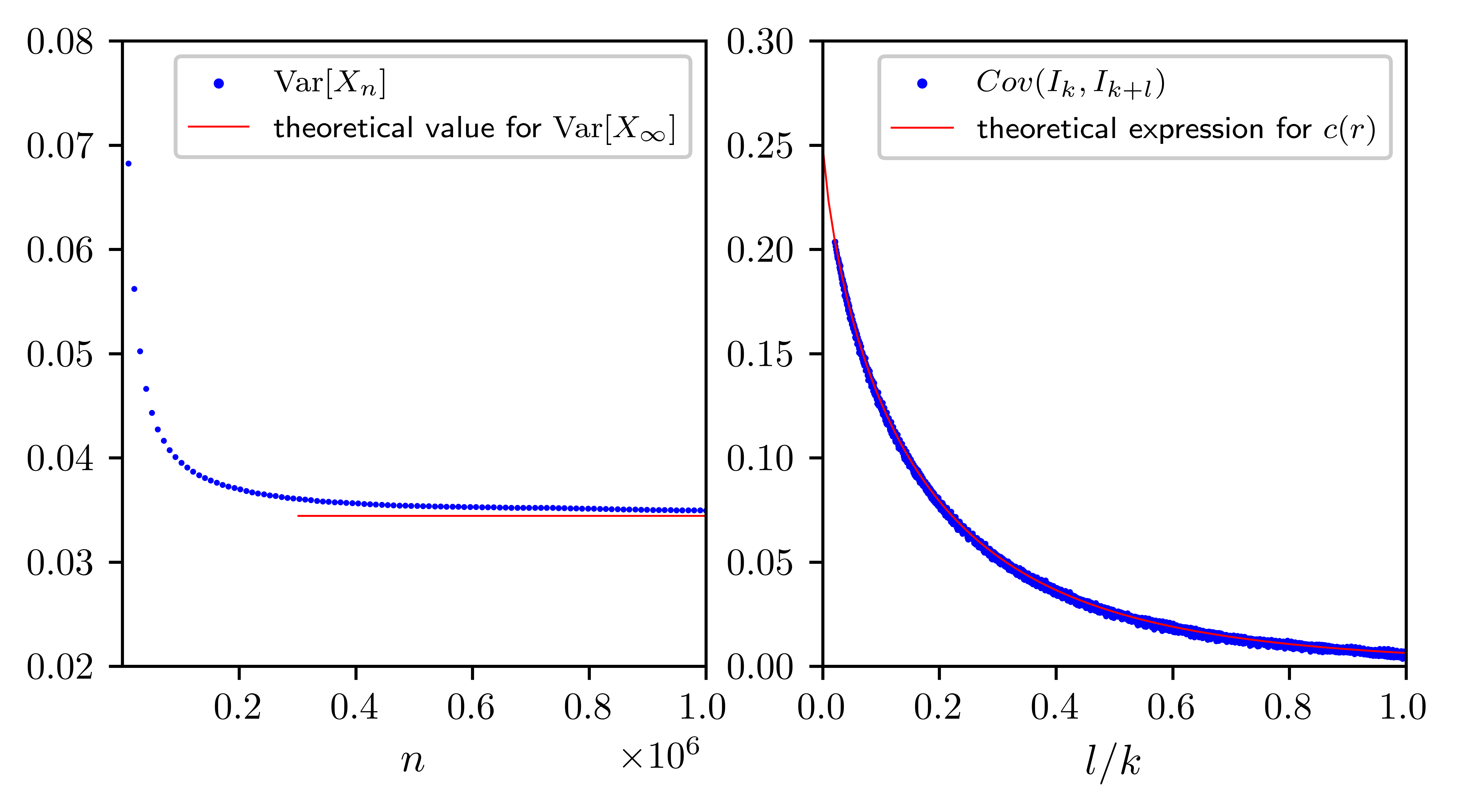}
\caption{Numerical validation of predictions for $1$-dimensional brownian motion. Left: Convergence of $\Var[X_n]$ towards the limiting value $\Var[X_\infty]$. The finite $n$ correction to the variance is numerically obtained to decay as $1/n$. Right: Collapse of the discovery order covariances for $1$-dimensional Brownian walkers. Agreement with the theoretical prediction for the scaling function $c(r)$ is very good. Deviations are caused by finite-size corrections to the dominant scaling term and finite sampling.}
\label{fig:num_1DBM}
\end{figure}
This is the only recurrent case where we have a fully quantitative theory, predicting the asymptotic variance and the scaling function for covariances. For the variance, the prediction is:
\begin{equation}
    \Var[X_n] \underset{n \rightarrow +\infty}{\rightarrow} \Var[X_\infty] \approx 0.03433.
\end{equation}
For $c(r)$, the theoretical expression is given by Eq.~\eqref{eq 179}. Agreement with numerical simulations is shown on Fig.~\ref{fig:num_1DBM}.

\subsection{Exclusion of Beta shape for the limiting discovery-share distribution}

\begin{figure}[hbtp]
\centering
\includegraphics[scale=1]{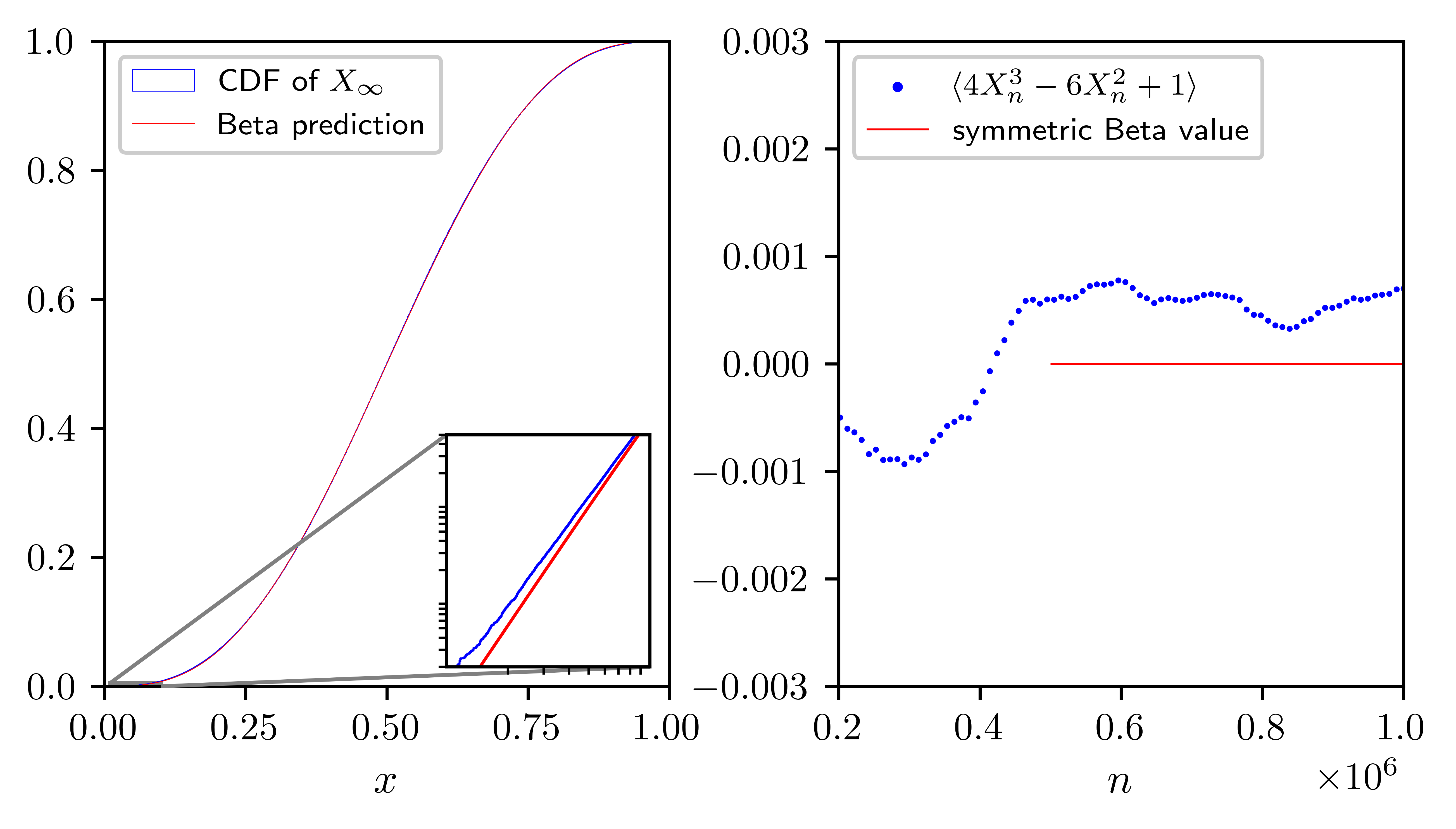}
\caption{Numerical analysis of the limiting shape for the distribution of $X_\infty$. Left: Simulated distribution of $X_\infty$ compared with the centered Beta distribution with matching variance. The two curves are extremely close, but deviations most notably appear in the tails $x\sim 0$ and $x\sim 1$. The zoom around the region $x,y \sim 0$ reveals that the numerical distribution and the Beta curve do not approach $0$ with the same exponent. Right: The higher moment combination $\langle 4 X_n^3 - 6 X_n^2 +1 \rangle$ does not converge towards $0$ as $n$ increases, as it should for any centered Beta distribution. This proves that the limiting-shape of $X_\infty$ is non-Beta, although agreement with the correctly chosen Beta candidate is nearly perfect in the bulk of the distribution.}
\label{fig:nonBeta}
\end{figure}

We also obtain numerically the limiting distribution of $X_\infty$. To compare it with centered Beta distributions of shape $\rho(x) = \frac{x^{\alpha-1}(1-x)^{\alpha-1}}{\Beta(\alpha,\alpha)}$, we compute the value of $\alpha$ that reproduces the correct variance for $X_\infty$. This gives $\alpha \approx 3.14$. The corresponding centered Beta distribution then matches extremely well the simulated distribution for $X_\infty$, but one can see deviations in the region $x\sim 0$ and $x\sim 1$. We also test the matching of higher order moments. For all centered Beta distributions $X\sim \Beta(\alpha,\alpha)$, we have:
\begin{equation}
\langle 4 X^3 - 6 X^2 +1 \rangle = 0.
\end{equation}
Testing the convergence towards $0$ of this quantity for $X_\infty$ then gives a way of excluding all centered Beta distributions, whatever the $\alpha$ value. This analysis, resulting in the validation of the non-Beta shape for the limiting random variable $X_\infty$, is presented in Fig.~\ref{fig:nonBeta}.

\subsection{Other recurrent walks}

\begin{figure}[hbtp]
\centering
\includegraphics[scale=1]{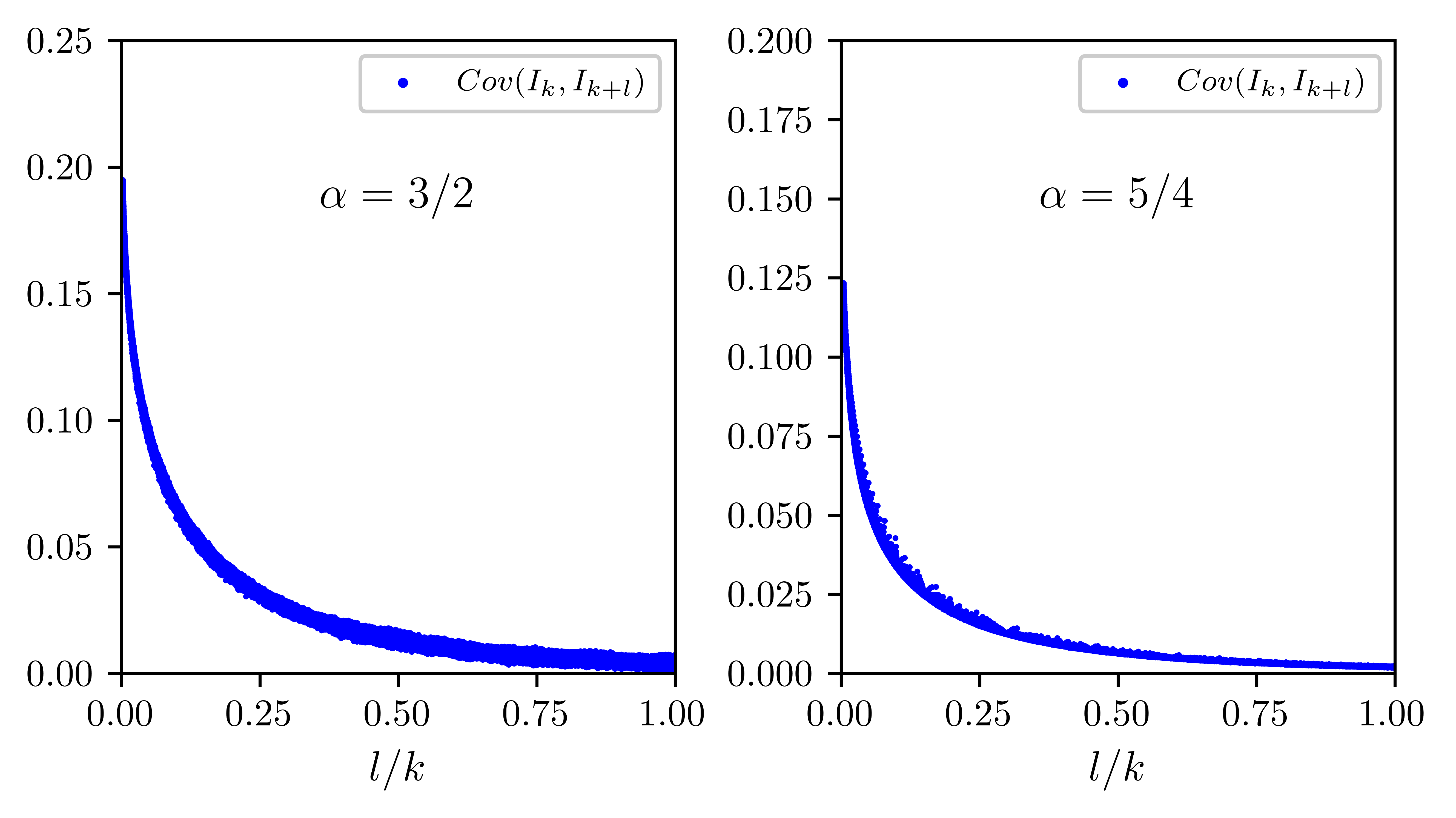}
\caption{Numerical validation of the aging scaling behavior of discovery order correlations for general recurrent walks. As shown here for one-dimensional Levy walks with exponents $3/2$ and $5/4$, numerically simulated covariances collapse onto a single function of the aging ratio $l/k$.}
\label{fig:CollapseRecu}
\end{figure}

For the other recurrent universality classes, we do not have any quantitative predictions for the variance or scaling functions, but we check numerically that the scaling form for discovery order covariances hold, by collapsing $\mathrm{Cov}(I_k,I_{k+l})$ onto a function of $l/k$. This is done for Lévy walks in $1$ dimension with exponents $\alpha = 3/2$ and $5/4$ and shown on Fig.~\ref{fig:CollapseRecu}.

\subsection{The marginal case - Cauchy walk and $2$-dimensional Brownian motion}

\begin{figure}[hbtp]
\centering
\includegraphics[scale=1]{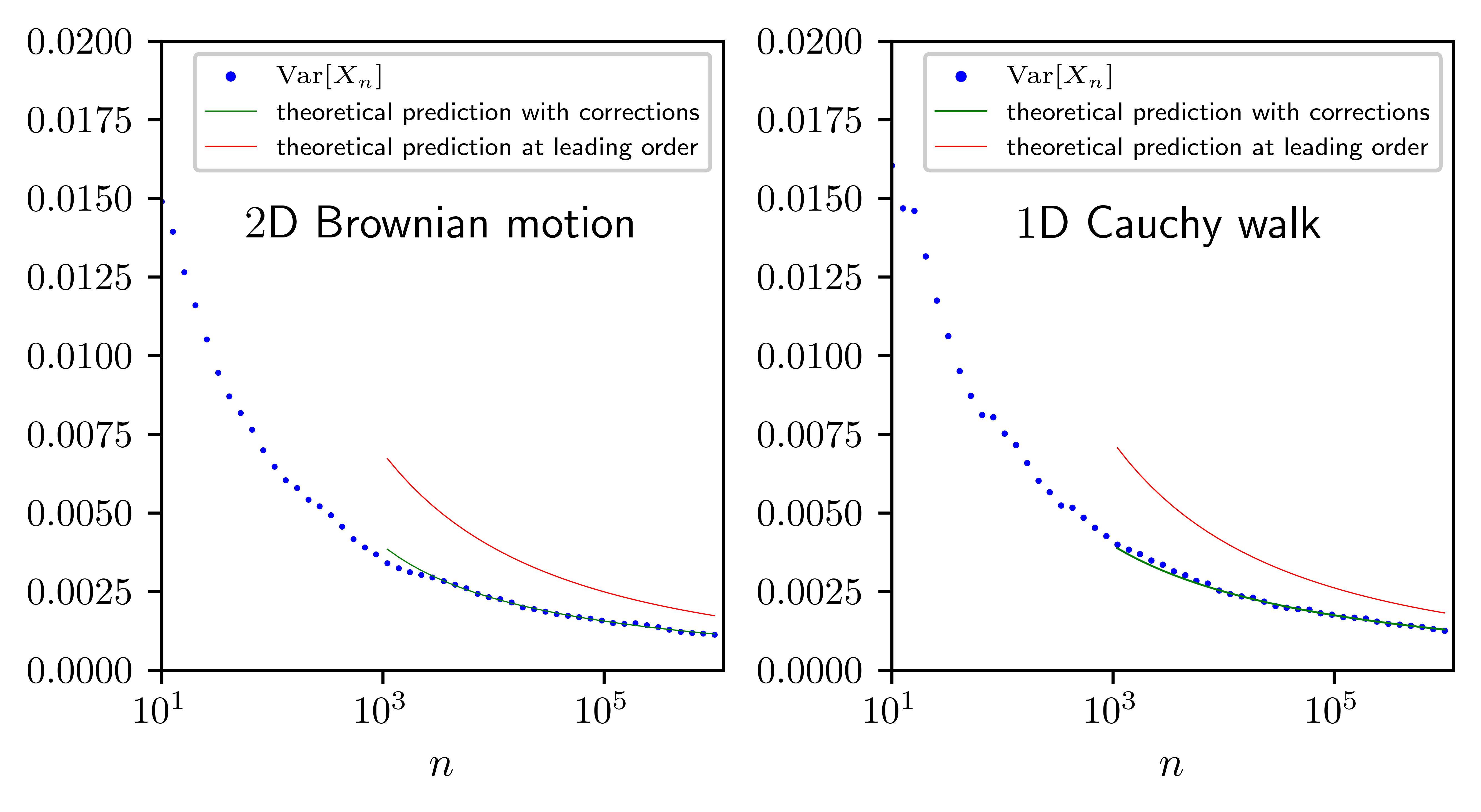}
\caption{Numerical validation of the variance decay for marginal walks. Taking into account higher order corrections to the leading behavior, the simulated variance values match the predicted decay and amplitudes for $2$-dimensional Brownian motion and $1$-dimensional Cauchy walks. Fitted coefficients are $a=2.09,b=7.69$ in the Brownian case, and $a=1.69,b=4.15$ in the Cauchy case.}
\label{fig:variance_marginal}
\end{figure}
We now check predictions Eq.~\eqref{eq 259} and Eq.~\eqref{eq 260} for the variance of $1$-dimensional Cauchy walks and $2$-dimensional Brownian motion, which are the two most relevant universality classes among marginal walks. Because the variance is actually expressed as an infinite series in powers of $1/\log{n}$, convergence towards the leading behavior predicted by the theory is extremely slow, and higher order corrections still play an important role even for $n=10^6$. We therefore use an ansatz of the form:
\begin{equation}
\Var[X_n] \sim \dfrac{A}{(\log{n})^2}-\dfrac{a}{(\log{n})^3}+\dfrac{b}{(\log{n})^4},
\end{equation}
where $A$ is fixed by the theoretical prediction ($A=\Cl_2(\pi/2) - \Cl_2(\pi/3)/\sqrt{3} \approx 0.33$ for $2$-dimensional Brownian motion, $A = \log{2}/2 \approx 0.35$ for $1$-dimensional Cauchy walk) whereas $a$ and $b$ are fitted numerically. Including these higher order corrections leads to an excellent agreement with simulations, despite the leading order term being quite far off, as shown on Fig.~\ref{fig:variance_marginal}.

\subsection{$1$-dimensional transient Levy walks}

\begin{figure}[hbtp]
\centering
\includegraphics[scale=1]{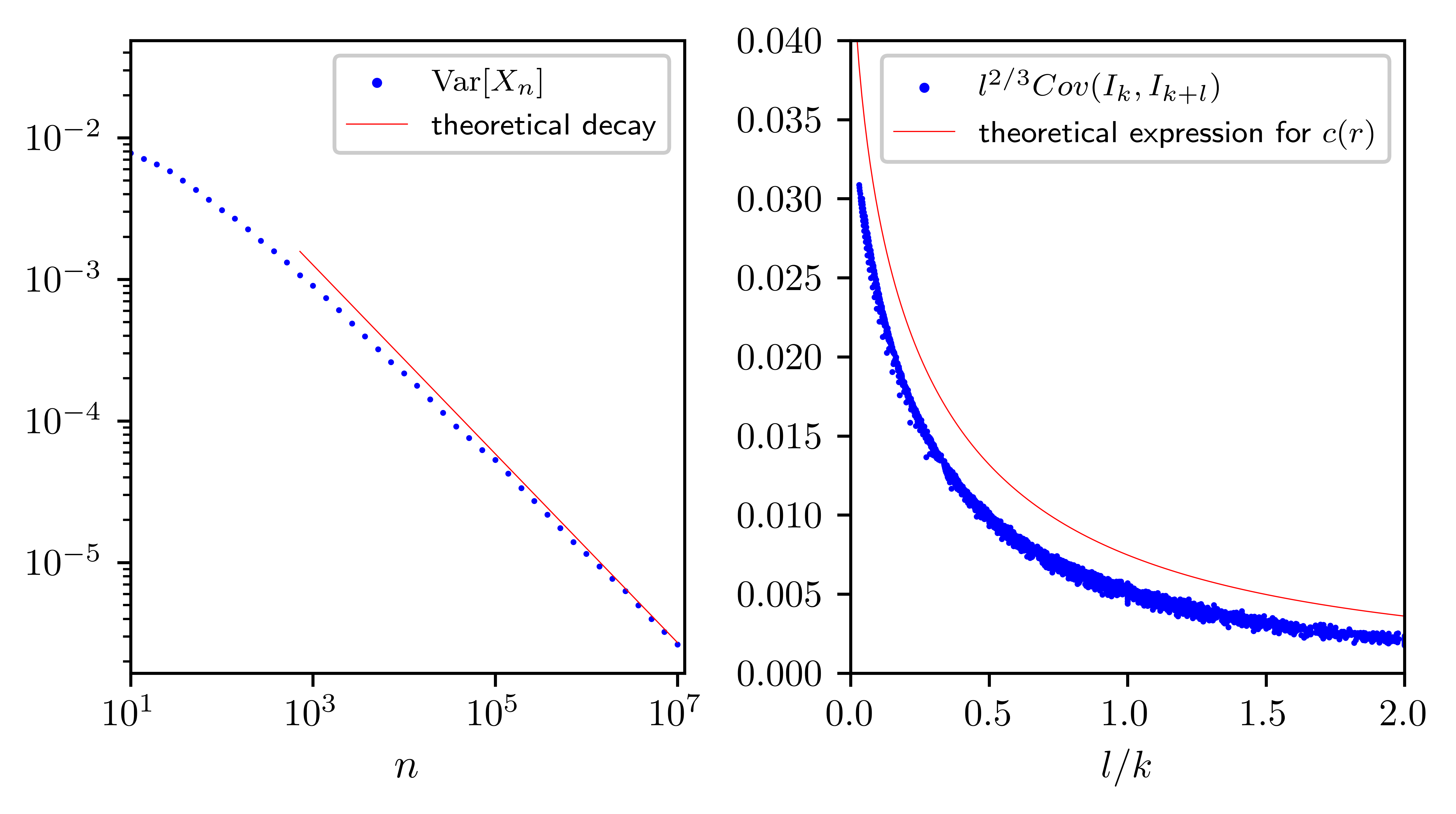}
\caption{Numerical validation of theoretical predictions for walks in the intermediate region $2<d_s<3$. For the $1$-dimensional Levy walk with exponent $\alpha=3/4$ ($d_s=8/3$), the variance converges towards the predicted leading behavior $0.1266 n^{-2/3}$. The gap between numerically simulated correlations and the theoretical expression for $c(r)$ is expected because of important finite-size effects for low values of $k,l$.}
\label{fig:comparison alpha 0.75}
\end{figure}

As an example for walks belonging to the intermediate region $2<d_s<3$, we focus on the Lévy walk in $1$ dimension with exponent $\alpha = 3/4$ ($d_s = 8/3$). We compare both the theoretical expression for the variance amplitudes and for the scaling function $c(r)$ of discovery order covariances. Starting from Eq.~\eqref{eq 257}, one simply needs to compute the generalized diffusion coefficient and the constant $g_0$. For this particular Lévy walk, we have:
\begin{equation}
\tilde{p}(k) = \dfrac{\Li_{7/4}(e^{ik}) + \Li_{7/4}(e^{-ik})}{2\zeta(7/4)},
\end{equation}
from which we extract:
\begin{equation}
D = \dfrac{2 \sqrt{2-\sqrt{2}}\Gamma\left(\frac{1}{4}\right)}{3\zeta\left(\frac{7}{4}\right)},
\end{equation}
and:
\begin{equation}
g_0 = \int_{-\pi}^{\pi} \dfrac{\dd k}{2\pi}\dfrac{1}{1-\tilde{p}(k)} \approx 1.937.
\end{equation}
This gives $\Var[X_n] \sim \dfrac{0.1266}{n^{2/3}}$. The theoretical prediction for the scaling function $c(r)$ quantifying correlations in discovery order is given by Eq.~\eqref{eq 266} and is also compared with rescaled correlations obtained numerically on Fig.~\ref{fig:comparison alpha 0.75}. The collapse of the correlations onto a single curve is excellent but there is a significative deviation compared to the theoretical prediction for $c(r)$. This is expected because of finite size effects. The collapse uses values of $k,l$ of the order of $10^2$ where the corrections are not at all negligible compared with the leading scaling behavior. Good agreement would require going to values of $k,l$ of order $10^4$ at least, where good enough statistics are extremely long to obtain.

\subsection{$3$-dimensional Brownian motion}

\begin{figure}[hbtp]
\centering
\includegraphics[scale=1]{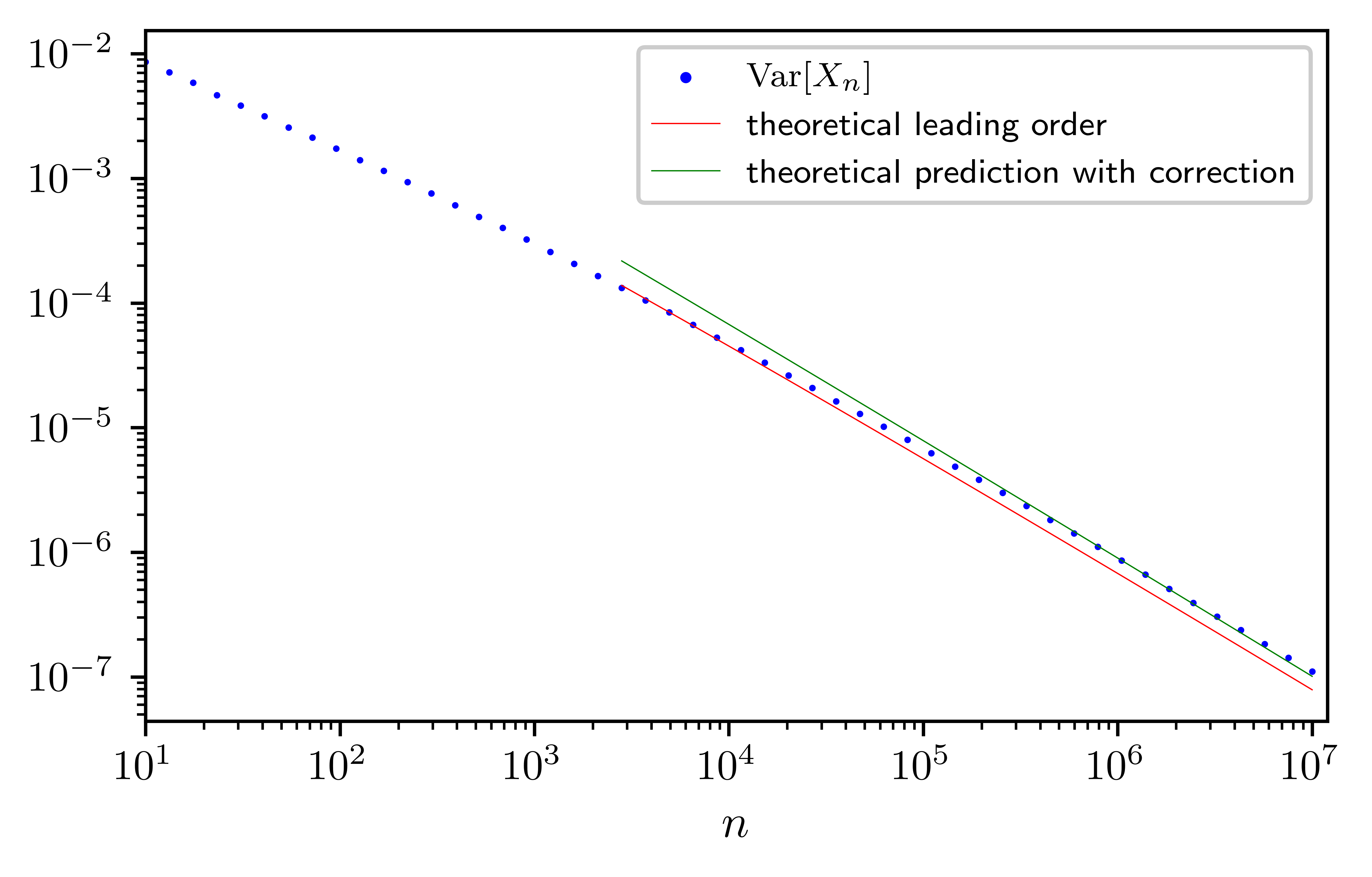}
\caption{Numerical validation of the predicted decay of the variance for $3$-dimensional Brownian motion. Agreement is already decent at leading order, and becomes excellent for large $n$ when including a higher order correction. The fitted coefficient is $a\approx 0.22$.}
\label{fig:variance_3DBM}
\end{figure}

For $3$-dimensional Brownian motion, we check the prediction of Eq.~\eqref{eq 153} for the decay of the variance. As for marginal walks ($d_s=3$ can be considered "marginal" in that logarithmic corrections are present), one has to include corrections for good agreement with simulations as they are smaller only by a logarithmic factor. We thus fit the variance by a form:
\begin{equation}
\Var[X_n] \sim \dfrac{0.049 \log{n}}{n} + \dfrac{a}{n},
\end{equation}
allowing for good matching of theoretical predictions with numerics. This is shown on Fig.~\ref{fig:variance_3DBM}.

In the last four subsections expanding on the robustness discussion present in the main text, all relevant variance decays are shown in the main text on Fig.~\ref{fig:illu_decay_robustness}, but we provide numerical validation of discovery-label correlation scaling and obtain the scaling functions $c(r)$ numerically. We also provide simulation details for these generalizations of the minimal model.

\subsection{Unequal mobilities}

\begin{figure}
    \centering
    \includegraphics[width = 0.65\linewidth]{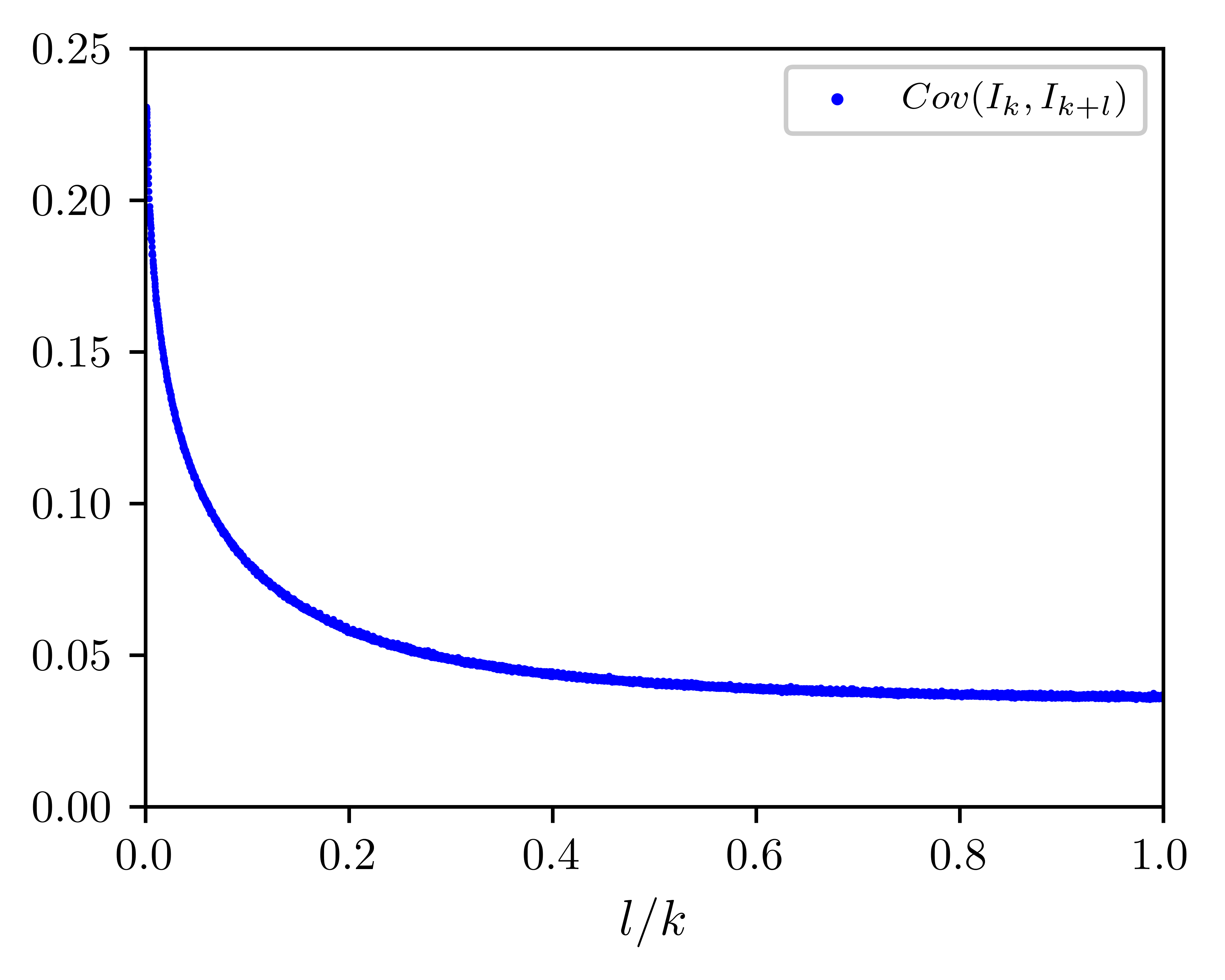}
    \caption{Numerical validation of the collapse for discovery-label correlations for recurrent, non-identical one-dimensional Levy walks. Specifically, we simulate two Levy walks with exponent $\alpha=3/2$ in one dimension, where one walker has jump lengths that are twice as large as those of the other.}
    \label{fig:collapseRecuHetero}
\end{figure}

To represent the case of unequal mobilities, we simulate $2$ Levy walks in one dimension, with exponent $\alpha = 3/2$, but where one of the walkers is performing bigger jumps than the other one (the jumps are taken twice as big). Because $3/2 > 1$, these walks are recurrent and the correlations $\mathrm{Cov}(I_k,I_{k+l})$ collapse onto a single function of $l/k$, as shown on Fig.~\ref{fig:collapseRecuHetero}.

\subsection{Multiple competitors}

\begin{figure}
    \centering
    \includegraphics[width = 0.65\linewidth]{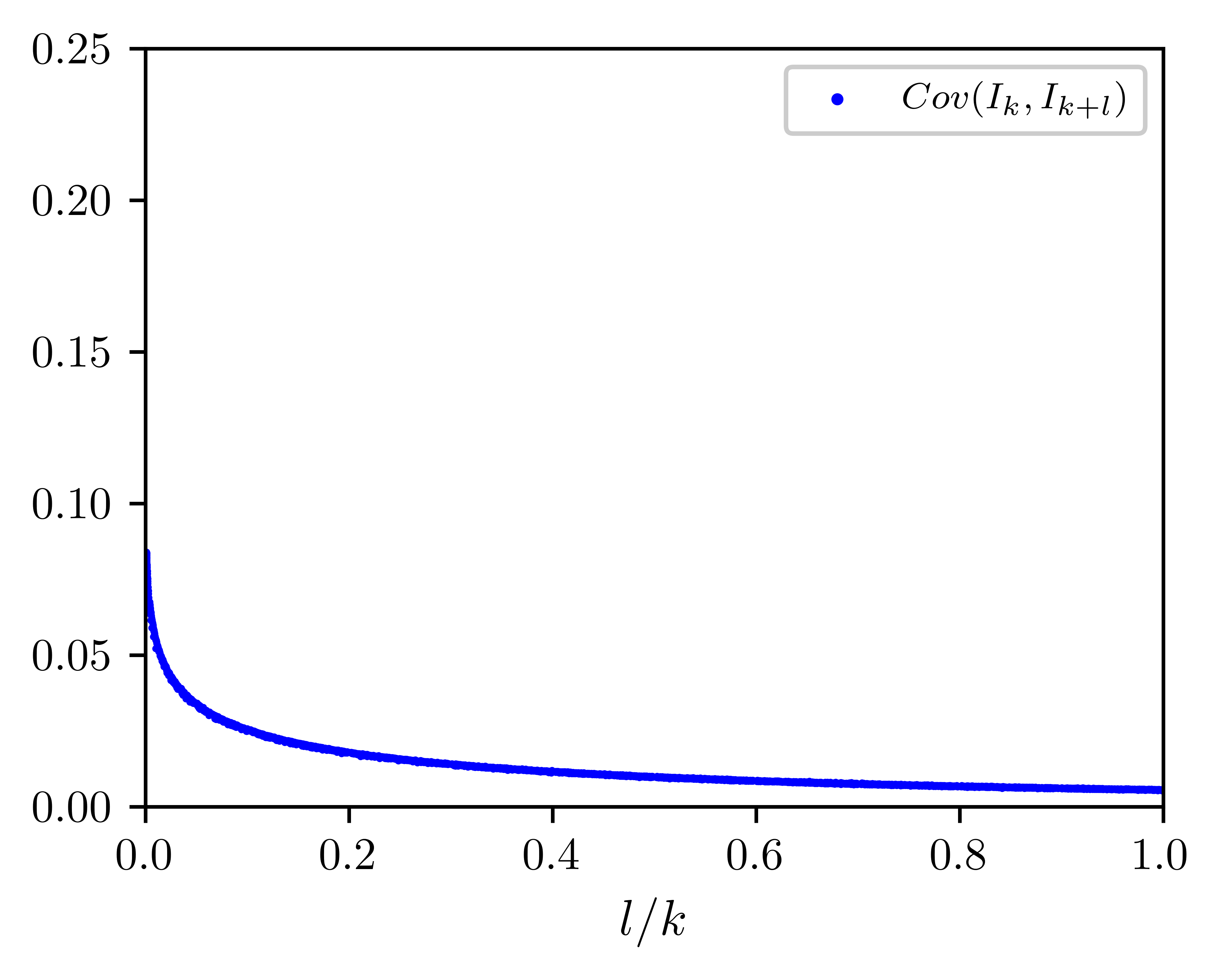}
    \caption{Numerical validation of the collapse for discovery-label correlations in many walkers competitive exploration. Data is obtained simulating $7$ Levy walks with exponent $\alpha=3/2$ in one dimension. The value $c(0)$ is shifted down compared to previous plots for scaling functions $c(r)$ since now $\langle I_k \rangle = 1/7 < 1/2$.}
    \label{fig:collapseRecuMany}
\end{figure}

As a generalization to multiple competitors, we simulate $7$ Levy walks in one dimension, with exponent $\alpha = 3/2$. Defining $I_k$ to be one if walker $1$ discovers the $k$-th site and $0$ if it is any of the other $6$ walkers, the correlations $\mathrm{Cov}(I_k,I_{k+l})$ collapse onto a single function of $l/k$, as shown on Fig.~\ref{fig:collapseRecuMany}.

\subsection{Heterogeneous substrates}

\begin{figure}
    \centering
    \includegraphics[width=0.65\linewidth]{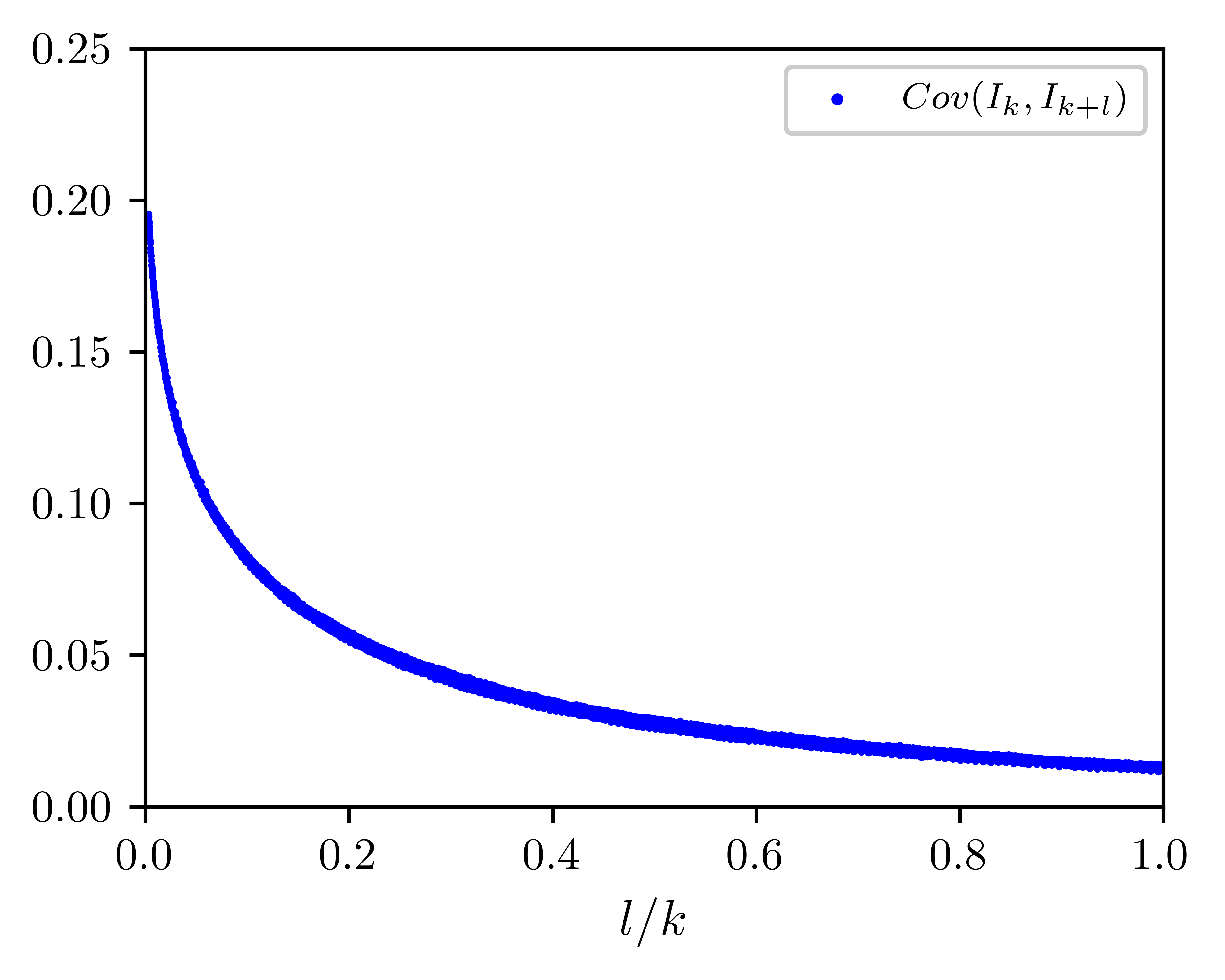}
    \caption{Numerical validation of the collapse for discovery-label correlations on critical percolation clusters. The walkers perform nearest-neighbor diffusion on clusters generated inside a $200 \times 200$ two-dimensional grid.}
    \label{fig:CollapseFractal}
\end{figure}

The aging structure of correlations for recurrent walks persist on heterogeneous substrates with complex spatial structure. In the case of critical percolation clusters, correlations $\mathrm{Cov}(I_k,I_{k+l})$ collapse onto a single function of $l/k$, as shown on Fig.~\ref{fig:CollapseFractal}. It is however required to perform simulations on many randomly generated clusters with random starting position to reach scale invariance, and correlations do not collapse as functions of $l/k$ when simulations are limited to a single cluster.

\subsection{$2$-dimensional fractional Brownian motion}

\begin{figure}
    \centering
    \includegraphics[width=0.65\linewidth]{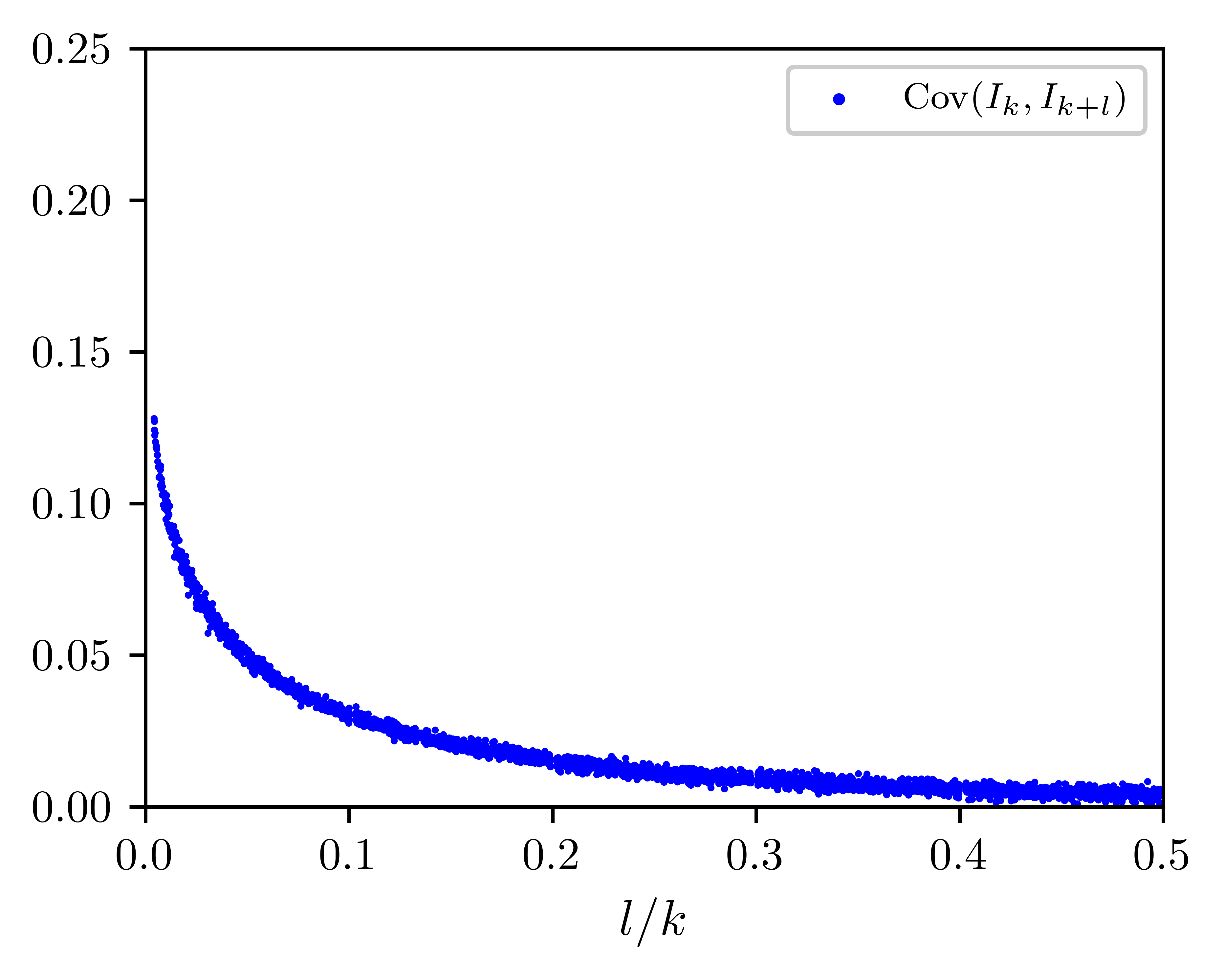}
    \caption{Numerical validation of the collapse for discovery-label correlations for two-dimensional fractional Brownian motion with Hurst exponent $H=1/3$. The correlations are typically small among recurrent walks considered in this work, smaller than for one-dimensional Brownian motion and percolation clusters and comparable to one-dimensional Levy walks with $\alpha=5/4$.}
    \label{fig:collapseFBM}
\end{figure}

Lastly, we obtain the scaling function $c(r)$ for discovery-label correlations in the case of recurrent fBm. More precisely, we simulate two-dimensional fBm with Hurst exponent $H=1/3$ ($d_s=4/3<2$), generated using the Davies-Harte algorithm and then discretized spatially by taking the integer part. The collapse and the resulting numerical curve for $c(r)$ are shown on Fig.~\ref{fig:collapseFBM}.
